\documentclass[11pt]{article}
\usepackage{natbib}
\usepackage{amsmath}
\usepackage[margin=1.4in]{geometry}
\usepackage{graphicx}
\usepackage{caption}
\usepackage{hyperref}

\title{\textbf{Why Do We Want a Theory of Quantum Gravity?}}
\author{Karen Crowther -- \textit{University of Oslo}}
\date{January 2025}

\begin{document}

\maketitle

\bibliographystyle{newapa}

\begin{abstract}

\begin{normalsize} 

The search for a new scientific theory is typically prompted by an encounter with something in the world that cannot be explained by current theories. This is not the case for the search for a theory of quantum gravity, which has been primarily motivated by theoretical and philosophical concerns. This Element introduces some of the motivations for seeking a theory of quantum gravity, with the aim of instigating a more critical perspective on how they are used in defining and constraining the theory sought. These motivations include unification, incompatibilities between general relativity and quantum field theory, consistency, singularity resolution, and results from black hole thermodynamics.

\paragraph{}
Preprint draft for a forthcoming publication in \textit{Cambridge Elements in the Philosophy of Physics}.
\end{normalsize}

\end{abstract}

\pagebreak
\tableofcontents

\pagebreak
\section{Introduction}\label{Intro}

Usually, when physicists search for a new theory it's because they have encountered something in the world that cannot be explained by the current theories. Such was the case, for example, with the development of quantum mechanics, which began with the anomalous experimental phenomena of the spectrum of blackbody radiation and the photoelectric effect; and also for the development of special relativity, prompted by asymmetries in the predictions of Maxwell's electrodynamics which---famously---did not seem to accord with nature. In these instances, observations precede, and go hand-in-hand along with, the theoretical work: motivating, guiding, and constraining it. Yet, this is not so for the current search for a new fundamental theory of physics, known as \textit{quantum gravity}, which---although potentially related to some empirical problems---has been primarily motivated, guided, and constrained by theoretical and philosophical considerations (but cf. \S\ref{empirical}). 

The search for a theory of QG has grown into one of the greatest challenges in modern physics. It is my contention that one of the reasons for the difficulty in finding a theory of QG is that it is not clear what the theory is supposed to achieve, and what, in fact, it \textit{should} be expected to achieve. Thus, a better understanding of the motivations for seeking the new theory can help us to more precisely frame, and eventually answer, the question `What is the problem of QG?' The hope is that critical investigation of the theoretical and conceptual problems which \textit{motivate} the theory can reveal some of the principles (or prejudices) which may underlie these problems. We can then asses whether these principles are well-founded, as well as the consequences of adopting them in various roles in theory-development, especially in conjunction with other principles and constraints. It is an aim of this \textit{Element} to introduce and instigate this new, critical perspective upon the problem of QG. As such, this \textit{Element} will not cover all of the important developments and philosophically interesting aspects of QG; in particular, discussion of spacetime emergence, spacetime functionalism, and the AdS/CFT duality are conspicuously absent.

So, why do we want a theory of QG? The usual answer refers to `two pillars' of fundamental physics: \textit{general relativity} (GR), providing our best understanding of spacetime, and \textit{quantum mechanics} (QM), our best understanding of matter.\footnote{\textit{Thermodynamics}, a third pillar of modern physics, is typically neglected in the Primary Motivation for QG, but c.f., \S\ref{threepillars}; \S\ref{BHT}.} Both these frameworks are supposed to be \textit{universal}: unrestricted in their domains of applicability, i.e., both are supposed to describe everything in the universe. In practice, however, we typically only \textit{need} to use GR to describe `big stuff' (the universe at large distance scales), and QM to describe `small stuff' (matter and forces at short distance scales, or, equivalently, high-energy scales). Yet, there are domains of the universe (`parts of the world') where both GR and QM are thought to be necessary---where we cannot get away with just using one or the other theory, or any known combination of both. These domains are characterised by extreme densities or temperatures (potentially as high as $10^{93}$ grams per cubic centimetre, or $10^{32}$ degrees Celsius), and include the cores of black holes (within the Planck length $10^{-35}$ m), cosmological singularities such as the `big bang', and the first instants of early universe cosmology. The desire for a description of these domains is part of the primary motivation for seeking the new theory of QG. 

In line with my aims of reframing the motivations for seeking QG into positive constraints defining the objectives of the new theory, we might suppose then, `that the theory describe these particular domains' could be the minimal constraint upon the new theory, serving to characterise it. Such a constraint is insufficient, however. GR, on its own terms, \textit{does} describe these domains.\footnote{On this view, the black hole and cosmological GR singularities are seen as predictions of the theory. This is discussed in \S\ref{sing}.} Quantum theory, as well, is supposed to be unrestricted in its domain of applicability. The problem, then, is apparently to find a new theory that describes the domains where both GR and quantum theory are supposed to be necessary, \textit{and} which somehow captures (or `takes into account') the lessons of \textit{both} GR and quantum theory. Let us call this the \textbf{Primary Motivation} upon any acceptable theory of QG (although it is very imprecisely defined as stated).

The most straightforward attempts at constructing a theory that fulfils the Primary Motivation---by quantizing GR, treating gravity in the framework of \textit{quantum field theory} (QFT), or otherwise `combining' GR and quantum theory---are beset by various conceptual and technical problems, which have led to their dismissal as unable to fulfil the expectations of a theory of QG. For example, \textit{canonical quantization} of GR leads to the infamous `problem of time' (\S\ref{time}); the treatment of gravity in the framework of perturbative QFT is divergent, and thus typically viewed as internally inconsistent (\S\ref{UVC}); while \textit{semiclassical gravity}, which couples classical gravity to quantum matter fields, also leads to divergences, and has been accused of paradox in various thought-experiments (\S\ref{quantiz}). Each of these problems could be referred to as `the problem of QG', since they are obstacles that prevent the acceptance of each approach as a contender for QG. I maintain, however, that these problems are damning only to the extent that they prevent these approaches from satisfying the Primary Motivation on QG. Beyond this, the criticisms of these approaches to QG reveal interesting prejudices we have about the nature of physical theories, and expectations about what QG is to achieve. 
\paragraph{}
\textbf{The main general motivations for seeking QG} that will be explored include:

\begin{itemize}
    \item \textit{The Primary Motivation}: To have a theory that describes the domains where both GR and QFT are supposed to be necessary, \textit{and} which somehow `takes into account' the lessons of both GR and quantum theory;
       \item \textit{Incompatibilities between GR and QFT}: To have a theory that provides a coherent picture of the world;
         \item \textit{Unification}: i. [Minimally] to have GR and quantum theory accounted for by a common framework; ii. [Full] to have a unified theory of all forces, including gravity, as stemming from a single interaction;
                 \item \textit{Putative inconsistency of semiclassical and other `hybrid' approaches to QG}: i. Claims that a theory with quantum matter coupled to a classical spacetime is inconsistent, and thus unable to serve in place of a full theory of QG; ii. Claims that perturbative GR is inconsistent because it is non-renormalizable;
    \item \textit{Singularity resolution or explanation}: i. To have a theory that addresses the singularity theorems of GR, particularly to describe black holes and the cosmological `big bang' singularity, ii. To have a theory that `cures' the divergences and other mathematical inconsistencies in QFT;
    \item \textit{Black hole thermodynamics}: To have a theory that describes the evolution of black holes, particularly black hole evaporation as suggested by theoretical work on \textit{black hole thermodynamics};

\end{itemize}

However, note that there are (at least) four \textbf{other main general motivations for QG} which will not be explored in this book, namely:

\begin{itemize}
    \item \textit{Complete cosmology}: i. To have a theory that describes the initial conditions of the universe; ii. To have a theory that solves the cosmological constant problem;
\item \textit{Heirarchy problem}: To have an explanation for why the gravitational force is so weak compared to the other fundamental forces.
    \item \textit{The measurement problem of QM}: To have a more fundamental framework that explains the origin of this problem, and is free of it;
    \item \textit{The problem of becoming in GR}: To have a more fundamental framework that explains the passage of time (this is briefly mentioned in \S\ref{time}).

\end{itemize}

Each of these represents a number of considerable philosophical challenges in their own right---several have already spawned their own healthily growing philosophical literature. This \textit{Element} will briefly introduce these related philosophical issues and current debates. Additionally, this \textit{Element} aims to motivate interest in exploring how these are, could be, and should be translated into positive heuristics to guide theory-development.

\subsection{Motivations, constraints, and desiderata}\label{principles}

The motivations for QG are the reasons for seeking a theory that satisfies the Primary Motivation: answers to the question `why do we want this theory in the first place?' The project that this \textit{Element} aims to promote starts as a descriptive one, considering \textit{any} motivations cited by physicists, but also looking for other potential motivations. The next stage is an evaluative and prescriptive one. The motivations are framed as principles, and their utility, and implications, assessed in terms of their potential roles in theory-development in various approaches to QG. Both of these stages are merely sketched in this \textit{Element}, with the aim of inspiring the reader to engage with this project in their own research. Ultimately, the goal is to better understand the theory being sought, and determine whether it is actually what we want or require of QG.

A principle may be used in different ways in different research programs, and at different stages of theory-development within a given program (see \cite{Crowther2021} for a discussion of these, with examples from QG). Since I am concerned with motivations here, it is useful to distinguish them from those non-motivational principles and features that share some of the same roles.

\paragraph{}
Broadly, motivations can be translated into either:

\begin{enumerate}
    \item \textit{Heuristics} or \textit{guiding principles}: these are desirable or useful features that can aid in the discovery of the theory by leading to new insights, but which may or may not actually be retained in the resulting theory; they are non-necessary \textit{desiderata}: features that it would be nice if the theory possessed, and which would make us more inclined to accept the theory; or,
    \item \textit{(Strong) constraints} or \textit{criteria of theory acceptance}: the new theory should not be accepted if it is incompatible with the principle (unless there is strong evidence in favour of the theory, and/or the principle is shown to be violated under the relevant conditions); when a motivation is adopted as a constraint, it typically is taken to form part of the definition of the new theory.
\end{enumerate}

Not all of the constraints on QG come from motivations, however, and nor do all desiderata motivate the need for a new theory. Typically, the non-motivational constraints are not taken to define the theory, as the motivational ones are. Some examples of constraints on QG that are not motivations include those principles coming from current theories within their domains of applicability (e.g., QG predictions should not violate Lorentz invariance in the domains where we know this symmetry holds). There is also the requirement that the new theory \textit{explain} the success of the theories it replaces, in the domains where we know those theories work well. This constraint is related to the Generalised Correspondence Principle, introduced below (\S\ref{Fundamentality}), which also plays a significant role in ensuring the new theory satisfies certain empirical constraints, namely those observations that are explained by current theories. Some of the non-motivational constraints, in spite of being non-motivational, can nevertheless still also be heuristically useful. Some possible examples of desiderata that may not be motivational include the `theoretical virtues' (if these are viewed as guides in developing a new theory rather than motivations to seek a new theory), as well as unexpected explanation of (theoretical or empirical) problems that are not hitherto thought to be quantum-gravitational. 

\begin{figure}[h]

  \centering
  
      \includegraphics[width=\textwidth]{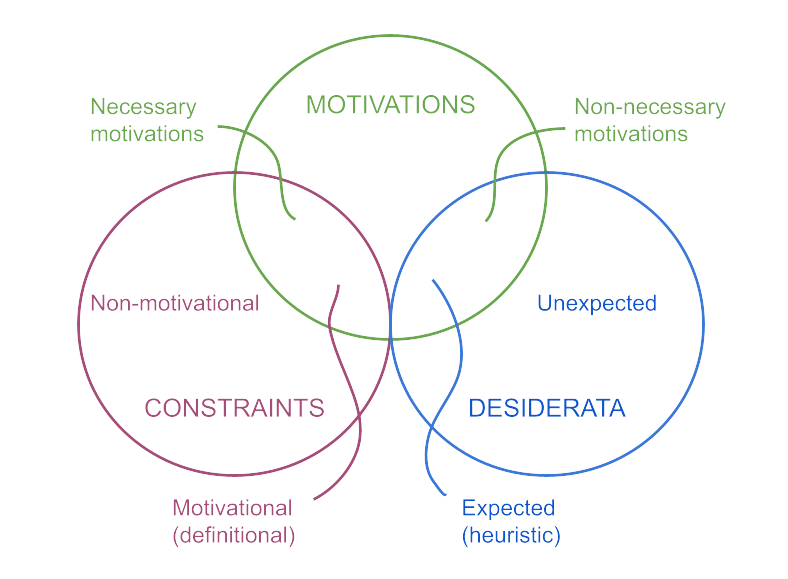}

  \caption[figure]{\textit{Motivations, constraints, and desiderata:} Motivations may be necessary (constraints), or non-necessary (desiderata).}
  \label{motivations}

\end{figure}

The relationship between motivations, constraints (necessary features), and desiderata (non-necessary features), is illustrated in Fig. \ref{motivations}

Motivations, constraints, and desiderata can all play important roles as \textit{means of confirmation} or as indicators of \textit{pursuit-worthiness} of a program: compatibility with the principle (or empirical result) serving to increase credence in the theory, or as being suggestive of the theory's potential future success, if it were to be developed (`pursued'). I believe this holds regardless of whether the principle has been used in the development of the theory, or whether the theory's possession of the feature in question is unexpected (`prediction').\footnote{Some philosophers accuse this of illegitimate `double counting', but I've never understood this argument. If the theory satisfies various constraints and desiderata (including the theoretical virtues), then this is evidence in its favour; though, it can perhaps be argued that the surprise factor adds more weight.} Keep in mind, too, that principles are not always purely theoretical, but can be based on empirical results. Empirical results can serve in various roles, though different results may feature in different roles, as discussed below (\S\ref{empirical}).

\begin{figure}[h]

  \centering
  
      \includegraphics[width=\textwidth]{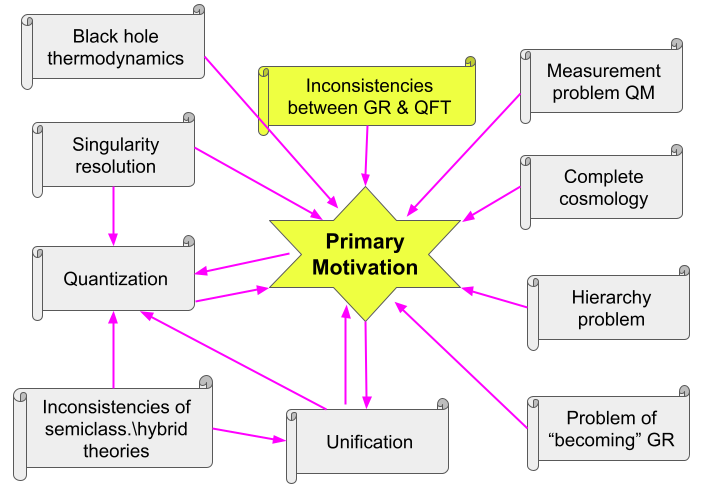}

  \caption[figure]{\textit{Various motivations as related to the Primary Motivation}. Arrows represent `motivates'; unification and quantization are possible routes towards fulfilling the Primary Motivation, so can also be thought of as motivated by the Primary Motivation (as well as independently motivated).}
  \label{primarymap}

\end{figure}

So, how are the motivations listed above used, and how are they related to the Primary Motivation? These questions will be explored in the rest of this \textit{Element}, though my conclusions here are certainly more provisional than conclusive, and I invite further engagement. A suggestion is illustrated in Figure \ref{primarymap}.

\begin{itemize}
    \item \textit{Incompatibilities between GR and QFT} motivate the Primary Motivation, representing the constraint of \textit{consistency} (which itself requires philosophical assessment in this context), but also frustrate attempts at satisfying it;

         \item \textit{Unification} motivates the Primary Motivation, but also represents different (i-ii) ways of satisfying the Primary Motivation; it is not a constraint, but a desideratum;
        \item \textit{Putative inconsistency of semiclassical and other `hybrid' approaches to QG} arise from attempts to satisfy the Primary Motivation,  underpinned by the constraint of consistency; they are not motivations for the Primary Motivation, but are motivations for particular ways of satisfying it (through unification, typically taken to indicate the need for quantization), which I argue are not necessary ways;
    \item \textit{Singularity resolution or explanation} motivates the Primary Motivation, through the constraint of consistency; can serve as a guiding principle and desideratum, but not all singularities in GR and QFT require resolution by QG;
    \item \textit{Black hole thermodynamics} stems from attempts to bring together GR and QFT (plus thermodynamics), it motivates the Primary Motivation; leads to various desiderata and potential constraints;
    \item \textit{Complete cosmology}: these are problems related to empirical observations which QG is expected to solve, and so they motivate the Primary Motivation; arguably, they represent constraints on QG;
\item \textit{Heirarchy problem}: stems from empirical concerns; motivates the Primary Motivation, but is (arguably) a desideratum rather than a constraint;
    \item \textit{The measurement problem of QM}: as above;
    \item \textit{The problem of becoming in GR}: as above.

\end{itemize}

\subsection{Relative fundamentality}\label{Fundamentality}

The general motivations for QG listed above may be seen as reasons for not treating GR and QFT as fundamental. There are different notions of, or requirements for, fundamentality in physics\footnote{\cite{Morganti2020b, CrowtherDigging}.}, but the key notion here concerns \textit{completeness} (or \textit{maximality}) of description. In this \textit{Element}, I take it that QG need not be a fundamental theory, in the sense of being a \textit{final theory} or \textit{theory of everything}. All that is required is fulfilment of the Primary Motivation. As such, we should not require as strong constraints on the QG all those features that are expected of a final theory. 

However, QG is expected to be more fundamental than GR and QFT. The notion of relative fundamentality here can be understood simply as an asymmetric relation of non-causal dependence \citep{Crowther2018}. 
QG is more fundamental than GR and QFT by virtue of its having a broader domain of applicability, one that subsumes the domains of success of the previous theories---QG describes `more stuff' than either of these frameworks. GR and QFT are thus expected to be \textit{effective} theories (or frameworks): valid as approximations to QG in the restricted domains where they are known to be successful.

Asymmetric dependence between an older, `predecessor' theory $O$, and a newer, `successor' theory, $N$, may be demonstrated through the establishment of particular \textit{correspondence relations} between them, such that $O$ is taken to be \textit{derivable} from $N$. There are various types of inter-theory correspondence relations\footnote{\cite{Radder1991}, identifies three types of correspondence relations, and \citet{Hartmann2002}, describes seven.}, though the most familiar are characteristic limiting relations (a standard example is the transition from relativistic to classical mechanics in the \textit{Newtonian limit} of small velocities compared to the speed of light, i.e., $v/c \rightarrow 0$). The idea is essentially what \citet{Nickles1973} calls ``reduction\textsubscript{2}'', where $N$ is said to `reduce to' or \textit{recover} $O$ in the domains where $O$ is known to be successful. This is standardly taken as a constraint on any new theory of physics that is more fundamental than its predecessor. This is also a way of satisfying the \textit{Generalised Correspondence Principle}, which forms an important constraint on QG, and states that any two theories whose domains of applicability overlap share approximately the same results within these domains.\footnote{Here, I follow \citet[][p. 79]{Butterfield2001} in taking ``results'' to include theoretical propositions as well as observational ones, and even ``larger structures'' such as derivations and explanations. The Generalised Correspondence Principle is discussed by \cite{Post1971}, but the statement here comes from \cite{Crowther2020}.} 

Given that QG is taken to be more fundamental than GR and QFT, we expect that GR and QFT be derivable from QG in this way. (In spite of this, the \textit{recovery} of QFT has not been appealed to as a constraint on QG to the same extent as has the recovery of GR or spacetime, except within the context of string theory). On the other hand, QG is not standardly expected to be more fundamental than the framework of quantum theory: most approaches to QG attempt to satisfy the Primary Motivation by simply providing a quantum description of spacetime (or, more precisely, the quantum physics underlying classical spacetime). According to this `standard perspective', QG is itself a quantum theory, rather than a more fundamental framework supposed to explain quantum mechanics (\S\ref{primary}). This said, however, it is a significant, and underexplored, question as to whether the explanation of quantum mechanics \textit{should} be understood as part of the motivation for QG---for instance, if we expect QG to explain the measurement problem of QM. There are some approaches that do take this as a goal, most notably the work by Penrose \citeyearpar{Penrose2002, Penrose2014}. If standard quantum theory is modified in the construction of QG, or QG were to represent a new framework, of which quantum mechanics were a special case, then QG would be considered more fundamental than quantum theory.

\subsection{Empirical constraints}\label{empirical}

Although QG is primarily motivated by non-empirical concerns, the need to be consistent with empirical observations is still a principle that is used extensively in the search for QG. Three different types are employed: empirical observations that \textit{are not} explained by current theories; empirical observations that \textit{are} explained by current theories; future empirical observations. I briefly describe each in turn.

\paragraph{}
\textbf{Empirical results not explained by current theories}

These are motivations, but may or may not be constraints. Both GR and quantum theory are remarkably successful in describing the world: there are scant observations that they leave unexplained. Nevertheless, there are some open problems related to observations. The two biggest ones are the problems of dark matter and dark energy. There are various astrophysical observations that cannot be explained by GR given the known amount, and type, of matter in the universe---these can be explained by positing an abundant amount of a new form of matter, known as \textit{dark matter}, with behaviour very different from known matter. Additionally, the expansion of the universe is observed (via, e.g., measurements of supernovas) to be accelerating in a way that cannot be accounted for according to the main currently accepted cosmological model based on GR and the known matter in the universe; this underlies a problem known as \textit{dark energy}. 

Although the problems of dark matter and dark energy may be related to QG---and to each other---they are not standardly treated as such. The observations associated with dark matter, for instance, are instead are generally taken to motivate searches for new particles beyond the Standard Model---though other candidates, such as primordial black holes, do not require new physics. Additionally, most approaches to QG have not taken these anomalies as empirical evidence motivating the need for a more-fundamental theory of QG. There are some exceptions to this `traditional' position, however---and, indeed, there are a growing number of papers related to QG that treat the solution to these problems as a guiding principle, means of confirmation, or indicator of pursuit-worthiness for a theory of QG.\footnote{The most prominent is \cite{Verlinde2017}, but see also \cite{Calmet2018, Kastner2018, Oriti2021}.} This raises interesting questions as to how, or whether, the solutions to the problems of dark energy and dark matter should be used in the search for QG. Perhaps these empirical anomalies \textit{should} motivate the search for a more fundamental theory of matter and spacetime, and be counted as part of `the problem of QG'?

\paragraph{}
\textbf{Empirical results explained by current theories}

These are not motivations, but are constraints. It is a criterion of acceptance that QG reproduce the empirical results explained by current theories, in accordance with the Generalised Correspondence Principle (\S\ref{Fundamentality}). The Generalised Correspondence Principle plays many roles in theory development and assessment: Its wide use is reflective of its status as a `shortcut' to empirical results, through the demonstration the new theory is consistent with the empirical results already covered by the older theory. In QG, the Generalised Correspondence Principle takes a more specific form, as the requirement that GR be appropriately \textit{derivable} from QG in the domain where GR is known to be successful. This correspondence between QG and GR has overwhelmingly dominated as a constraint of interest---\citet[][p. 927]{Carlip2001}, for instance, refers to this as the ``zeroth test'' of QG.

\paragraph{}
\textbf{Novel empirical results}

These are motivations and constraints. Novel empirical results come from  \textit{QG phenomenology}: a unique field of research that aims to reciprocally connect QG to observable phenomena, by building models that bridge the considerable gap (many orders of magnitude) between them. So far, there has not been any philosophical work dedicated to exploring QG phenomenology in general, in spite of the importance of the field, and its offering intriguing connections to philosophy of scientific modelling and experiment.\footnote{The reader may take this as a hint of an interesting research project to explore.} QG phenomenology has led to numerous results, including tight empirical constraints on any possible violation of Lorentz invariance (this is a key symmetry of GR that some approaches to QG suggest may be broken under some circumstances, but QG phenomenology has revealed no evidence of its being violated).

While QG phenomenology typically connects with cosmological and astrophysical observations, it can also potentially connect with laboratory experiments, such as `tabletop' Gravitationally Induced Entanglement experiments, which may provide a `witness' of the underlying quantum nature of gravity in the non-relativistic limit, using superpositions of Planck-mass bodies.\footnote{\label{fntt}These include \cite{Bose2017, MV2017}; philosophical exploration in \citet{Adlam2022, huggettQGLab}.}

Other novel empirical tests of QG phenomena may (arguably) come from \textit{analogue experiments}, which are experiments whose Source system (the particular system being experimented on) is thought to be analogous to the Target system (the general class of systems that the Source system is supposed to represent) \textit{and} where the Target system itself is inaccessible (the inaccessibility of the Target system is part of the definition of analogue experiments, following \citeyearpar{CLW}). An example is the case of acoustic `dumb hole' models, in which sound waves in fluids encounter a horizon, and which are described by equations that are isomorphic to those of black holes. These were originally proposed by \citet{Unruh1981}, using \textit{analogical reasoning}, since the dumb holes are supposed to be analogous to black holes. There has been interest in whether or not analogue models of black holes in the laboratory are able to confirm the existence of \textit{Hawking radiation}---which, as discussed in \S\ref{BHT}, would give us insight into properties of black hole spacetimes that are potentially interesting for QG.\footnote{\label{fnanalogue}\citet{Steinhauer2016} has reported to have detected Hawking radiation using Bose-Einstein condensates in the laboratory. See also \citet{Barcelo2011, Weinfurtner2013}. Philosophical discussion: \citet{CLW, Dardashti2017, Dardashti2019}.}

\subsection{Approaches to QG}\label{approaches}

There are various different research programs, or \textit{approaches}, towards QG, with different motivations, assumptions, formalism, and of varying degrees of completeness. Each of these have their own problems, which, from their own perspective could be called `the problem of QG'. This \textit{Element}, however, is written from a general perspective, discussing issues relevant to \textit{any} possible theory of QG: indeed, it aims, specifically, at better understanding the common features by virtue of which we can even conceptualise these different approaches to QG as being `approaches to QG'. The idea, as stated, is to clarify, and begin to assess, the most basic motivations for seeking QG in the first place. The different approaches to QG each assign varying levels of importance to the various principles and motivations, and utilise them in different ways. As I go through the motivations and principles, I will mention the ways they have been used in some approaches to QG as illustration. But, given the limited space of this \textit{Element}, I will not introduce or explain what these approaches are in any detail.

We can broadly classify the approaches of three types, owing to their historical development as stemming from the different methods of attempting to quantize gravity \citep[see, e.g.,][]{Carlip2001, Rovelli2000}. As described in \S\ref{UVC}, one natural approach to quantizing GR is to use familiar techniques from QFT to treat gravity, which broadly captures the `particle physics perspective' (or `high-energy physics perspective') towards QG. An early attempt utilises quantum perturbation theory, which is known as the perturbative \textit{covariant quantization} approach to QG. This approach to quantization splits the GR metric into a background part plus small perturbations; the perturbations are quantized, while the background remains classical and non-dynamical. [Splitting apart the dual role played by the metric field in GR, and restoring spacetime as a fixed, static, `background', goes against the principle of \textit{background independence} in QG (\S\ref{BGInd}), and was thus objectionable from the `GR perspective' on QG]. This approach was shown to suffer from gravity being non-renormalizable (see \S\ref{UVC}). In response, the years 1980-2000 saw the development of research into \textit{non-perturbative} QG, and the majority of research since has focused on this \citep{QG30questions}. The main approach to QG that historically stems from the covariant quantization methods is \textit{string theory} \citep[see, e.g.,][]{Zwiebach2009}.

\textit{Asymptotic safety} is an approach based on the Wilsonian idea of renormalization, and the conjecture that quantized gravity is non-perturbatively renormalizable (discussed in \S\ref{UVC}).

Closely related to the covariant approach is \textit{Feynman quantization}, which applies Feynman's path integral (sum over histories) quantization techniques to the GR metric. Some important current examples of approaches that aim to implement these methods include \textit{causal dynamical triangulations} \citep{Loll2019}, and \textit{causal set theory} \citep{Surya2019}. 

Alternatively, the \textit{canonical} or \textit{Hamiltonian} quantization of GR aims to construct QG as a theory of the fluctuations of the metric as a whole, so no fixed metric is involved. This approach represents the `GR perspective' on QG. It involves first casting GR into canonical form, and then using canonical quantization methods. This requires splitting spacetime into space and time, with a fixed topological structure. The metric information is represented, in full, by operators on a Hilbert space. The goal is then to find the eigenvalues of the operators along with their transition probabilities. While the canonical QG approaches aim at preserving the principle of background independence, they suffer more manifestly from the \textit{problem of time}, \S\ref{time} (though problems of time also arise in other approaches to QG). The main approach to QG stemming from the canonical quantization procedure is \textit{loop quantum gravity} (although now loop quantum gravity also is being developed using a covariant formalism, see, \cite{Rovelli2004, Rovelli2014}). 

Two other classes of approaches mentioned in this \textit{Element} are \textit{hybrid theories} and \textit{emergent gravity} approaches (\S\ref{takes},\ref{quantiz}). Note, however, that this general classification by no means exhausts the range of approaches pursued in QG.

\pagebreak
\section{Incompatibilities between GR and QFT}\label{incom}

QM is not a theory, but a framework within which particular theories are formulated; similarly, QFT is also a framework---one which combines QM and SR. Basically, it extends QM to to fields, i.e., systems with an infinite number of degrees of freedom, defined on a fixed, flat, continuum spacetime of SR. The framework uses spacetime, but it does not \textit{describe} spacetime. Theories formulated in this framework are said to be \textit{background dependent}, because the spacetime serves as a `background' in the theory (see below, \S\ref{BGInd}). The Standard Model of particle physics is a theory formulated in the framework which we might call `conventional QFT' (CQFT) \cite{Wallace2006, Wallace2011}. The framework treats all matter as composed of particles, which are understood as local excitations of quantum fields; the fundamental forces are themselves represented by quantum fields, whose corresponding excitations interact locally with the other particles, depending on their type. Any dynamical field, according to QFT, is quantized. Incorporating gravity into this framework would entail treating gravity as a field whose force is mediated by a particle called the graviton.

GR is a theory \textit{of} spacetime; the equivalence of gravitational and inertial mass means we can understand gravity as a property of spacetime itself, rather than a field propagating on a fixed spacetime background. It stands in contrast to QFT for several reasons. Most basically, GR says that spacetime is a dynamical field rather than a fixed, static background structure. It is a \textit{nonlinear} theory (while QM is a linear theory), in that spacetime `reacts' to matter and energy, and in turn, the behaviour of matter and energy are affected. To use the famous 12-word summary of Wheeler, ``Spacetime tells matter how to move; matter tells spacetime how to curve'' \citep[][p. 235]{Wheeler2000}.\footnote{Cf. \cite[][p. 5]{MTW2017}.} But, from the perspective of GR, we could say that gravity is not really a force at all--it does not `act' on objects \citep{Maudlin2012}. Objects simply exist in spacetime, and GR tells us that an object's inertial path is determined by the curvature of spacetime (rather than, e.g., conceiving of gravity as a field that deflects objects from their inertial paths).\footnote{For other formal compatibility issues, see \S3.6.3 of \cite{Rickles2006}.} 

The formalism of GR can be defined by the Einstein-Hilbert action\footnote{I'm here following the presentation in \cite{Kiefer2007a}.},
\begin{equation}\label{EHaction}
   S_{EH}=\frac{c_4}{16\pi G}\int_{\mathcal{M}}d^4x\sqrt{-g}(R-2\Lambda)-\frac{c^4}{8\pi G}\int_{\partial\mathcal{M}}d^3x\sqrt{h}K
\end{equation}
where $g$ is the determinant of the metric , $R$ is the Ricci scalar, and $\Lambda$ is the cosmological constant. The first integral is over a spacetime region $\mathcal{M}$ (a four-dimensional manifold of spacetime points, encoding the topology and differentiable structure), and the second integral is defined on the boundary, $\partial\mathcal{M}$, of this region. This term is required for a consistent variational principle; here, $h$ is the determinant of the three-dimensional metric, and $K$ is the trace of the second fundamental form.
We also consider a `matter action', $S_m$, for non-gravitational fields, which give rise to the energy-momentum tensor,
\begin{equation}
	T_{\mu\nu}=\frac{2}{\sqrt{-g}}\frac{\delta S_m}{\delta g^{\mu\nu}}
\end{equation}
which acts as a `source' of the gravitational field. From the variation of $S_{EH}+S_m$, we obtain the Einstein Field Equations (EFE),
\begin{equation}\label{EFE}
G_{\mu\nu}\equiv R_{\mu\nu}-\frac{1}{2}g_{\mu\nu}R+\Lambda g_{\mu\nu} = \frac{8\pi G}{c^4}T_{\mu\nu}
\end{equation}
where $G_{\mu\nu}$ is the Einstein tensor describing the curvature of spacetime, $g_{\mu\nu}$ the the Lorentzian metric tensor, encoding the geometry, $R_{\mu\nu}$ is the Ricci curvature tensor. A model (i.e., a spacetime) of GR is specified as $\text{M}=\langle \mathcal{M} , g_{\mu\nu}, T \rangle$ where the two tensors $g_{\mu\nu}$ and $T_{\mu\nu}$ satisfy the EFE (\ref{EFE}).

So, on the one hand, we have GR describing classical geometry as a dynamical field coupled to classical matter, and on the other hand, we have QFT which uses a fixed, non-dynamical classical geometry, and which describes quantum matter, saying that all dynamical fields should be quantized. How can they be combined? One straightforward option would be to modify (\ref{EFE}) to take into account the quantum nature of matter, by replacing $T_{\mu\nu}$ with the expectation value of the \textit{quantum energy-momentum tensor operator}, $\left\langle \hat{T}_{\mu\nu}  \right\rangle$, to obtain,

\begin{equation}\label{eqsemiclass}
G_{\mu\nu} = \frac{8\pi G}{c^4}\left\langle \hat{T}_{\mu\nu}  \right\rangle 
\end{equation}

These are the \textit{semiclassical Einstein equations}, where gravity stays classical while the other fields are quantum. Note that the quantum energy-momentum tensor operator is not only difficult to compute mathematically, but its expectation value is also difficult to understand physically.\footnote{Cf. \cite{huggettQGLab, Kiefer2007a, Wald1994}.} Nevertheless, it is an operator that acts on states $|\psi\rangle$ of a material quantum system, and thus obeys the Schr\"{o}dinger equation, with Hamiltonians describing both the dynamics of matter with itself, and with gravity,
\begin{equation}\label{eqsemiclass2}
    i\partial_t|\psi\rangle = \hat{H}_{\text{matter + gravity}}|\psi\rangle
\end{equation}
A system described by equations (\ref{eqsemiclass}-\ref{eqsemiclass2}) is referred to as \textit{semiclassical gravity}. While the semiclassical Einstein equations may be of value as approximations, leading to insights into the low-energy regime of QG, they face several serious conceptual and theoretical difficulties if treated as exact equations at the fundamental level of QG.\footnote{\label{fnCuriel}While \cite{Grossardt, Grossardt2} discusses ``three little paradoxes'' of semiclassical gravity, Erik Curiel describes a ``panopticon of problems'' with semiclassical gravity and black hole thermodynamics; see his talks linked at http://strangebeautiful.com/phil-phys.html. See also \cite{Kiefer2007a}.}

More generally, the basic intuition that semiclassical gravity is inconsistent can be appreciated by analogy with the case of electromagnetism: \cite{BohrRosenfeld} analysed an equation akin to the `semiclassical' equation for electromagnetism and demonstrated that the electromagnetic field had to be quantized in order to be consistent with the quantized matter it couples to.\footnote{\citet[][\S3.1.2.]{Butterfield2001}, \cite{Huggett2001}.} The basic idea is usually taken as the uncertainty relations in the quantized system spread to (`infect') the coupled non-quantized system. In the case of semiclassical gravity, the uncertainty in the position of a quantized gravitating object would lead to quantum uncertainty in the gravitational field, so the gravitational field itself should be quantized. Although such arguments are heuristically very powerful, the consensus is they cannot by themselves compel the quantization of gravity (see below, \S\ref{quantiz}).

Nevertheless, this does suggest another natural way to combine GR and quantum theory: attempt to quantize the gravitational field, similar to the way in which the electromagnetic field was quantized. This approach, too, faces extraordinary conceptual and theoretical difficulties. The quantization of the electromagnetic field in quantum electrodynamics (QED) results in a theory of quantum fluctuations of the electromagnetic field against a well-defined classical spacetime background. The attempt to quantize gravity, however, means subjecting some of the properties of spacetime to quantum fluctuations. We thus run into trouble in giving a mathematical characterisation of the quantization procedure itself without a well-defined background spacetime. Even if we are able to achieve this, we then face trouble in giving a physical account of the theory that results---for example, a fluctuating metric would seem to imply a fluctuating causal structure and spatiotemporal ordering of events, which makes it difficult to define equal-time commutation relations in the quantum theory.

One of the most prominent clash of principles between GR and QM has to do with the superposition principle of QM. It seems, firstly, that superpositions conflict with general relativity's fixed causal structures---it is difficult to see how we could understand superposed causal structures of GR. Possibly relevant for this, however, are recent `quantum switch' experiments, which show quantum superposition of causal structure \citep{Chiribella2020, Goswami2020}. Furthermore, as \cite[][\S30.10-30.11]{Penrose2004} argues, using a thought experiment called `Schr\"{o}dinger's lump', there may be a conflict between the superposition principle of QM and the principle of general covariance of GR.\footnote{Cf. \cite{Penrose2014, RicklesSEP}.} \cite{Penrose2014} uses this as part of an argument for the ``gravitization of quantum theory'', as opposed to the quantization of gravity.

The superposition principle has also, however, been used to argue for the quantization of gravity. This argument comes from Feynman, who considers a Stern-Gerlach type experiment in which a spin-1/2 particle is guided to two counters. The counters are connected to an indicator which is either up when the particle arrives at counter 1 or down when it arrives at counter 2. The indicator itself is a ball of macroscopic dimensions (1 cm), which would then be in a superposition of being in two positions. Since the ball is macroscopic, its gravitational field would also be in a superposition. ``We would then use that gravitational field to move another ball, and amplify that, and use the connections to the second ball as the measuring equipment. We would then have to analyze through the channel provided by the gravitational field itself via the quantum mechanical amplitudes. Therefore, there must be an amplitude for the gravitational field.'' (Feynman, quoted in \cite{Zeh2011}, p. 66).\footnote{The original report is from 1957 and republished in \cite{DewittRickles}.} The idea is that the gravitational field itself must be described by quantum states subject to the superposition principle. As \cite{Kiefer2013}, states, this is not an argument that demonstrates the necessity of quantizing gravity, but rather ``is an argument based on conservative heuristic ideas that proceed from the extrapolation of established and empirically confirmed concepts (here, the superposition principle) beyond their present range of application. It is in this way that physics usually evolves'' (p. 2). \cite{Kiefer2013} points out that while there are several such heuristic arguments, there are none that \textit{logically demand} the quantization of gravity.

In the rest of this chapter, I discuss three other notorious issues that arise related to the attempt to quantize GR: the problem of \textit{background independence}, \S\ref{BGInd}, the \textit{problem of time}, \S\ref{time}, and the \textit{perturbative non-renormalizability} of quantum GR, together with the principle of \textit{UV-completeness}\S\ref{UVC}, \S\ref{UV}.

\subsection{Background independence}\label{BGInd}

Background independence, and the principle of \textit{relationalism} are used widely in the search for QG. Several authors view the history of physics in terms of the increasing relationalism of its theories, and thus motivate the quest for QG along these lines. \cite{Smolin2020}, for instance, adopts relationalism as a methodological imperative to ``make our theories more relational by eliminating background structure''.\footnote{\label{fndiscrete}Others who motivate QG along these lines include \cite{Rovelli2004}, and proponents of the `discrete approaches' to QG, e.g., \cite{Caravelli2011, Ambjorn2006, Loll1998, Markopoulou2009, Williams2006}.} One might wonder, however, whether it actually represents a \textit{motivation} for QG, rather than just being, e.g., a constraint or desideratum upon a new theory of QG. On the one hand, we might argue that the goal of having a background independent theory is not a reason to seek QG, since we already have a theory of gravity that is background independent, namely GR. But, while background independence can certainly be viewed as part of the motivation that drove the development of GR, it is not straightforward to identify what exactly is meant by background independence in GR, as I discuss below. On the other hand, though, QM and QFT are straightforwardly not background independent theories, so a drive for increasing relationalism through the elimination of background structure could promote bringing these theories in closer connection to GR, and thus to motivate the Primary Motivation. 

There are four general, but distinct, principles associated with background independence:
\begin{enumerate}
    \item[a.] That the theory not feature an absolute, fixed, `background' structure (typically: spacetime);
       \item[b.] That the theory be compatible with \textit{relationalism};
        \item[c.] That the theory be \textit{generally covariant};
        \item[d.] That the theory be \textit{non-perturbative}.
\end{enumerate}

The first of these (a.)\ is most properly called `background independence': basically, a theory that is background independent does not posit the existence of a fixed background object, or require the existence of such an object in order to define the properties of other objects in the theory. A well-known example of a background \textit{dependent} theory is Newtonian mechanics, which features absolute space and time as fixed background structures used to define position and motion. For Newton, space was not only \textit{absolute} and \textit{immutable} (meaning that it is fixed, existing unchanged throughout all time, totally unaffected by matter, but also responsible for affecting matter, being used to define the properties of---or itself conferring properties upon---matter) but also \textit{substantival} (meaning that space itself is a `substance', as real as matter, and which can exist independently of matter).

For QG, background independence standardly means that the theory not feature a fixed (background) spacetime, but there are several meanings, corresponding to different notions of `fixity' \citep[][\S 4.1]{Butterfield1999}, as well as the different (layers of) structures that could be identified as `spacetime' (e.g., metric, manifold, affine structure, etc.). There are (at least) four proposals that attempt to formally capture the idea of background independence in the sense of (a.), above. Note that each is a separate proposal: 
\begin{enumerate}
    \item[i.] A theory is background independent if it features no absolute objects (i.e., structures that are the same in all models of the theory); 
    \item[ii.] A theory is background independent if it has no formulation which features fixed fields (i.e., there is no formulation of the theory that is not diffeomorphism invariant);
    \item[iii.] A theory is background independent if its solution space is determined by a generally-covariant action (all of whose dependent variables are subject to Hamilton's principle, and represent physical fields);
    \item[iv.] A theory is background independent if all of its physical degrees of freedom correspond to geometrical degrees of freedom.
\end{enumerate}
 For details on (i--iii), \cite{Pooley2017}, and for (iv), \citet{Belot2011} (a comprehensive discussion of  background independence in classical and quantum gravity can be found in \cite{Read2023}). Each of these attempted characterisations has its own difficulties, however---and, arguably, none are adequate to do the job.\footnote{\label{BGI}Further details: \cite{Giulini2007, Pitts2006, Pooley2017, Read2023}.} 

The idea of (b.)\ relationalism in regards to background independence comes from the thought that a theory that does not rely on a fixed background structure will have to define the properties of its objects purely in terms of the relations between these objects, rather than relying on the background structure. In this sense, relationalism is opposed to absolutism (that there be a structure that affects objects without itself being affected), as well as substantivalism (that space itself has an independent existence, rather than just being a means of representing the relations between objects). Also, an absolute object is typically understood as being fundamental, in the sense that it not ontologically dependent on any further, more basic, underlying objects. So, when it is claimed that a background independent theory is relational, it means that the properties in the theory are defined only with respect to one another, and that these relationships are not fixed (in the sense of being non-dynamical), but evolve according to the equations of motion of the theory. While we might for this reason identify `relationalism' and `background independence' as synonymous,\footnote{As does \citet[][p. 204]{Smolin2006}.} \citet{Rickles2008} argues against the identification on the grounds that a substantival view of spacetime is also compatible with background independence.

A major driving force for seeking a background independent theory is the desire to preserve, or continue, what is seen as one of the most important insights of GR. Yet, while GR is taken as the prime exemplar of a relational or background independent theory, it is difficult to precisely identify what it is that `makes GR special' in this regard. The intuitive notion we have is that GR is special because---owing to the theory's \textit{diffeomorphism invariance} or \textit{general covariance}---we cannot define the physical observables of the theory without solving the dynamics, or, in other words, that ``there is no kinematics independent of dynamics'' \citep{Stachel2006}. However, it is not easy to define explicitly the idea of a dynamical object along the lines of `being solved for'. The main problem, following \citet{Kretschmann1917}, is that we can turn \textit{any} background field into a dynamical object by making it satisfy some equations of motion, however physically vacuous they might happen to be. Even the metric in special relativity is able to be made dynamical in the sense that it satisfies some equations of motion---hence special relativity can be made background independent, which then conflicts with our basic intuitions of what background independence is supposed to be \citep{Rickles2008}.\footnote{There is a long-running debate on the question of `what makes GR special', e.g. \citet{Earman2006, Pooley2010}.}

It is not impossible that general covariance (c.)\ be used in defining `what is special' about GR,\footnote{Cf. \citet{Norton2003}; also \citet[][\S 5.3.1]{Brown2005}.} and thus feature in a principle of theory selection, though general covariance can not \textit{by itself} be the principle of background independence. However, most of the recent literature has instead moved on in to principle (a.)\ in their attempts to salvage the general intuition that GR is special by virtue of possessing some particular conception of background independence. In this context, the proposal (i) above, fails for GR (and any theory with massive fields),\footnote{See fn. \ref{BGI}.} and while proposal (ii) arguably works for GR, it is difficult to implement as a guiding principle in the search for a new theory \citep{Read2023}. Nevertheless, the success of GR as a theory of spacetime, together with the significant heuristic role of the principle in its discovery, mean, however, that background independence (in some sense, which may, arguably, be any of (ii-iv)) continues to be viewed as a desirable principle in QG, in all roles.

Background independence is not standardly considered a \textit{motivation} for QG: it is not a reason for seeking a new theory in the first place. One way of framing it as a motivation would be to argue (1) QM/QFT are not fundamental because they are not background independent, and (2) that fundamental physics, as it progresses through theory-change, trends towards increasingly relational, or less-background-dependent theories. Such an argument by induction is not very strong, however, without additional reasons for expecting a fundamental theory should be background independent, or an argument as to why background independent theories are better or more successful than background dependent ones. The main argument is that background independent theories are \textit{more explanatory} than background dependent ones because they do not require structures to be fixed in advance, i.e. put in `by hand', but instead have their structures follow from the dynamical laws and principles specified by the theory itself. In this case, \textit{if} background independence is a \textit{motivation} for QG, then it would mean that QG is motivated by the need to explain the background elements of QM/QFT (and potentially also those in GR), and itself to not feature these elements. If we were to reject the principle as a motivation for QG, then, arguably, the principle need not be treated as a \textit{constraint} on QG, unless QG were conceived of as a final theory---if the theory required background structures, not themselves explained by the theory, then we would be motivated to search for a more fundamental theory that did explain these structures.

Finally, there is the distinct notion, (d.), that the theory be \textit{non-perturbative}. Perturbative treatments of gravity involve fixing a background metric (e.g., Minkowski metric) and studying gravitons propagating on this background. The desire that a theory be non-perturbative is a natural one, given that perturbation theory is a set of techniques used to construct approximations when the exact theory is unknown or unsolvable. These techniques are of limited applicability, and can easily lead to problems when misapplied. Some problems with perturbative gravity are considered in \S\ref{UVC}.

I close this section with a brief mention of background independence in string theory and some other approaches to QG. From early on, string theory has been intensely criticised for being a background \textit{dependent} approach---on account of its being based on perturbative methods, and describing strings moving on a background spacetime---it apparently fails on all (a.--d.).\footnote{See, e.g., \cite{Smolin2006}.} In recent years, however, the criticism has died down, as string theorists came to appreciate the significance of background independence, and the theory itself shows hints of being background independent in various ways. (For this reason, the philosophical debate concerning the importance of background independence in QG is no longer so active, having achieved its goal, in a sense). As \citet[][\S5.2]{Read2023} demonstrates (building on a claim made by \citet{HuggettVistarini}), there are in fact several senses in which string theory (considered both at the level of spacetime fields, and at the level of worldsheet fields) may be considered background independent, on some of the definitions (ii--iv) above, given that the background fields in the theory are required to be dynamically coupled together in the same dynamical equations of GR (the Einstein field equations, plus higher order corrections), and thus are not `fixed' after all, but possess background independence in ways similar to GR. 

Additionally, there are attempts to construct a non-perturbative version of string theory; the most celebrated development being the \textit{AdS/CFT duality}\label{AdS} (also known as the \textit{Maldacena conjecture}, after \citet{Maldacena1998}). This is a relationship---a conjectured exact equivalence---between a string theory featuring gravity, describing closed strings propagating on a spacetime (anti-de Sitter space (AdS)), known as the `bulk', and a gauge theory without gravity (a conformal field theory (CFT)), defined on the boundary that contains the bulk spacetime. In the regime where the string theory is \textit{strongly coupled}, and computations difficult---the CFT is \textit{weakly coupled}, meaning that perturbative techniques can be used to make calculations. Since the theories are supposed to be equivalent (describe the same physics), the weak coupling regime of the CFT can be used to shed light on the strong coupling regime of the string theory.\footnote{See, e.g., \cite{Polchinski2017}.} This is closely related to the holographic principle, \S\ref{Holog}. While string theorists hope to ultimately arrive at a theory that is background independent, the approach began as a background dependent one (perturbative string theory). Credence in the approach increases by its being shown to be background independent, it may also feature as a criterion of theory-acceptance. 

These comments also apply, broadly, to the program of \textit{asymptotic safety}, which is an approach to QG that treats gravity in the framework of QFT, but avoids the problems associated with perturbative non-renormalizablity (see \S\ref{UVC}). While the approach is premised on non-perturbative QFT techniques, it still relies on splitting the full metric $g_{\mu\nu}$ into a fixed, but arbitrary, background metric $\bar{g}_{\mu\nu}$ plus a dynamical fluctuation field $h_{\mu\nu}$. The principles of diffeomorphism invariance and background dependence (in the sense of (a.)) pose challenges---both technical and conceptual---for the theory, associated with its use of the \textit{functional renormalization group} techniques and the \textit{background field method} \citep{Criticalasysafety}.

Other approaches to QG have placed heavy emphasis on background independence from their inception, including LQG \citep{Rovelli2014}, causal set theory \citep{Surya2019}, causal dynamical triangulations \citep{Ambjorn2009}, group field theory \citep{Oriti2009b}, and other `pre-geometric' approaches \citep{Caravelli2011}. These approaches strive for a fully background independent theory, and sell themselves as upholding what they see as a central insight of GR (and of any acceptable theory). These approaches argue that the key to finding a successful theory of QG is in complete (as much as possible) elimination of any background structure or geometry. Adopting Smolin's (2006) slogan, the idea is ``the more relational the better''. These approaches use the principle of background independence in all roles.

The principle of background independence is now widely accepted as an important feature for QG to possess: it is presented as the main candidate for satisfying the condition that QG `take into account' the lessons of GR in the Primary Motivation (\S\ref{takes}). Nevertheless, it is an open question whether background independence is a sufficient means for satisfying this aspect of the Primary Motivation: there may be other principles from GR that could serve this role. Background independence is furthermore desirable since it functions as means of principle correspondence, connecting QG with GR, and thus helping, in part, to satisfy the Generalised Correspondence Principle and establish a connection between QG and known empirical results.

\subsection{Problem of time}\label{time}

There is an incompatibility between the treatment of time quantum theory and in GR. According to quantum theory, time is external to the system being studied: it is \textit{fixed} background in the sense that it is specified from the outset and is the same in all models of the theory. In GR, however, time forms part of what is being described by the theory. Time in GR is subject to dynamical evolution, and it is not `given once and for all' in the sense that it is the same across all models \citep{Butterfield1999}. This clash between the external time in quantum theory and the dynamical time in GR leads to problems in combining the theories, and is often taken as a suggestion---along with other heuristics, such as the suggestion of `quantum foam', or the Planck length as a minimal length---that time `disappears', or `does not exist' in QG.

The \textit{problem of time} is a cluster of problems, usually framed along the lines just suggested, but specifically as manifesting in the context of \textit{canonical QG}. This is a class of approaches to QG that attempts to quantize the canonical (or Hamilitonian) formulation of GR, via familiar canonical quantization techniques. The oldest and most straightforward of such approaches is known as \textit{quantum geometrodynamics}, which proceeds in a spirit similar to Schr\"{o}dinger's original heuristic approach to quantization (via the Hamiltonian-Jacobi formalism). In this context, one finds the primary form of the `problem of time': the supposed `dynamical' equation of the resulting quantum theory (the Wheeler-de Witt equation) does not feature a time variable, and so presents us with a `frozen formalism'.\footnote{\label{fn:time}It is important to emphasise that there are several difficulties for interpreting the Wheeler-de Witt equation. Detailed reviews of the problem of time: \citet{Isham1993, Kuchar1999}. Recent philosophical treatments: \citet{AndersonTime, Thebault2021}.} Yet, as \cite{Thebault2021} emphasises, the problem as I have just set it up, is misleading in the sense that it does not in fact arise from ``from forcing a background independent theory of spacetime onto the Procrustean bed of quantization with respect to a background time''. Instead, \cite{Thebault2021} argues, ``the problem arises generically from the manipulation of a particular class of classical theories, that includes general relativity, according to the standard formal steps that are preparatory for quantization. That is, the problem of time becomes apparent in the process of preparing any reparametrization invariant theory for quantization'' (p. 390).

Thus, the problem of time (or, at least, the \textit{global} problem of time that \cite{Thebault2021} focuses on) may not be specific to GR, and may arise even before quantization. This does not ameliorate the fact, however, that there are numerous problems related to the treatment of time in QG. The relevant question for this \textit{Element} is whether these problems represent \textit{motivations} for seeking the new theory, or rather challenges that crop up once we embark on this quest. As mentioned in \S\ref{Intro}, one motivation for seeking QG (which is not discussed in this \textit{Element}) is the `problem of becoming': if we are dissatisfied with the lack of `real change', or temporal passage, in the standard `block universe' picture of spacetime as presented by SR and GR, a possible option is to see if we can find a more hospitable notion of time in the more-fundamental theory beyond GR. This type of motivation is based on a perceived inconsistency between time in SR/GR, and our own human experience of time (i.e., of time flowing, and the experience of events occurring `Now'). One \textit{prima facie} difficulty with this strategy is, of course, analogous to that faced by attempts to overcome the determinism of GR (i.e., in the hopes of finding `free will') by looking to more fundamental theory: given that relativity does describe our macroscopic world, and that any more-fundamental theory needs to be compatible with this in the regimes where GR is successful, we may question what relevance any concept of time in QG has to our everyday experiences of `change' and `becoming'. Plausibly, though, the connection with becoming could be relevant in the context of QG cosmology, where the theory is expected to replace the Standard Model of cosmology (based on GR), including the idea of the `big bang', in describing the initial conditions of the universe.

More generally, we might take the disparate treatments of time in GR and QM as motivation for seeking a new theory that has a single, coherent account of time. The exploration of time in QG has spawned a huge amount of discussion, with all sorts of different solutions being proposed: from those that feature space without time (e.g., \cite{Barbour2000, Gomes2020}), time without space (e.g., \cite{MarkopoulouSpace, Smolin2013, Smolin2020}, neither time nor space \citep[e.g.,][]{Rovelli2004}, and disagreements over the implications of QG for the notion of `becoming' \citep[e.g.,][]{Dowker2020, Rovelli2020}. So, is `the problem of QG' about finding a fundamental description of time, or of timeless fundamental physics that can nevertheless explain the description of spacetime in GR, or of providing an account of change and/or becoming? Arguably, none of these are necessitated by the Primary Motivation. In order to translate these into positive constraints or heuristics for the new theory, we need to evaluate the arguments (including metaphysical arguments) for one solution over another---assuming, of course, we follow \cite{Huggett2013}, in not outright rejecting such theories as unphysical or `empirically incoherent' on account of not featuring spacetime fundamentally.

\subsection{Perturbative non-renormalizability of quantum GR}\label{UVC}

A natural way of approaching QG is to utilise familiar techniques from QFT---for example, using quantum perturbation theory to treat gravity. The framework of QFT within which the Standard Model of particle physics is formulated relies upon perturbative techniques in order to make predictions in most cases. A perturbative calculation of any particular physical process involves a summation over all possible intermediate states, and this is done at all orders of perturbation theory (though, in practice, often only the first few terms are taken, and it is hoped that the higher-orders decay rapidly). The dynamics of QFT is local, utilising point sources and interactions---it features arbitrarily small distances and, thus, arbitrarily large momenta. This, as well as the integration over all the momentum-energy states, implies that there are an infinite number of intermediate states. Consequently, perturbative calculations within the theory lead to divergent integrals (if the terms are not sufficiently suppressed). These infinities are dealt with in particular QFTs by the process of \textit{renormalization}, which first imposes a finite-valued cutoff in momenta (so that the integral is not taken over all possible high values of momenta), and then the parameters of the theory are re-defined in terms of `physical' coupling constants that absorb the divergences, and whose values are then fixed by experiment. A theory is \textit{renormalizable} if only a finite number of counterterms are required to absorb the divergences, whereas a \textit{non-renormalizable} theory requires an infinite number of such terms. Because an infinite number of terms cannot be determined experimentally, a non-renormalizable theory is not predictive.\footnote{See, e.g., \cite{Weinberg1995}. In the context of QG, see \cite[][\S2.2.2]{Kiefer2007a}.}

The strong and electroweak interactions are renormalizable, and thus the Standard Model of particle physics is a renormalizable theory. Gravity, however, is non-renormalizable: the Einstein-Hilbert action (\ref{EHaction}), treated in the perturbative framework, has divergences appear in loop diagrams at first order (in the matter case, or second order in the matter-free case), and there is an expectation that infinitely more infinities appear at higher orders.\footnote{The one-loop divergences were shown by \cite{Oneloop}; cf. \cite{Bern2002}.} The divergences occur at high energy scales, near the Planck scale, and we are prevented from calculating anything in this regime. 

In standard QFT, however, even non-renormalizable theories can be predictive (and impressively so), if they are treated as \textit{effective field theories}. This is thanks to the development of \textit{renormalization group} (RG) techniques, which led to a framework for studying the way in which the physical coupling constants change depending on the energy scale at which the theory is being applied (the couplings are said to `run' or `flow' with the RG).\footnote{\label{fnRG} The RG and associated philosophy of effective field theory is rich in philosophical interest, see, e.g., \cite{Batterman2017, Butterfield2015, Cao1993, Crowther2015, Williams2021}, and \cite{Kaplan2005, Manohar1997, Weinberg2009} for the physics.} Effective field theories are valid at low energies compared to some high energy valued cutoff scale: they are not formally predictive to arbitrarily high energies, i.e., they are not thought to be \textit{UV-complete} (\S\ref{UV}). 

Perturbative quantum GR can be treated as an effective field theory which makes predictions for low energies compared to the Planck scale. It has proven to be an extremely good approximation for all the present experimental tests of gravity \citep{Burgess2004, Wallace2022}. However, the low-energy nature of its predictions is taken to imply that quantum effects are important where gravitational fields are very strong, such as inside black holes or cosmological singularities. In other words, its failure here is taken to indicate the domains of necessity of QG. It also means that this approach towards finding QG fails to satisfy the Primary Motivation, and thus motivates the search for a new theory.

However, it is possible that the proliferation of infinities in the ultraviolet (UV) regime does not signal the failure of GR treated in the framework of QFT, but instead signals the limitations of the perturbative approach in this regime. The \textit{asymptotic safety scenario} claims that the physical couplings, in the \textit{non-perturbative} RG flow, do not actually diverge, but instead flow to a finite value: a `fixed point' in the UV.\footnote{This was proposed by \cite{Weinberg1979}; cf. \cite{Donoghue2020, Eichhorn2019, Niedermaier2006}.} This is similar to QCD, where the couplings flow to a fixed value of 0 in the UV, and the theory is said to be asymptotically free (the theory is non-interacting in this regime). In the case of gravity, however, the fixed point is supposed to be non-zero (the theory is interacting in at least one of the couplings), and the theory is said to be \textit{asymptotically safe}, since the physical quantities are `safe' from divergences as the cutoff is removed (taken to infinity). If there is a fixed point, then following the RG trajectory (almost) to it, one can in principle extract unambiguous answers for physical quantities on all energy scales \cite{Niedermaier2006}. At the fixed point, the dependence on the UV cutoff is lost, and the theory is \textit{scale invariant}: it does not change as smaller length scales are probed. While this approach may satisfy the Primary Motivation for QG, it does suffer other problems, including a conflict with the principle of unitarity. This \textit{problem of unitarity}, means---roughly--that probabilities will not be conserved in time.\footnote{Another approach, known as \textit{conformal gravity}, which is perturbatively renormalizable, also suffers this problem of unitarity \citep[][p. 179]{Gomes2020}.}

\subsection{UV completeness}\label{UV}

The failure of perturbative quantum GR to provide a predictive theory at the Planck scale is motivation for the Primary Motivation. The problem, from this perspective, is that the theory fails in the regime where we expect quantum effects to be significant for gravity. Nevertheless, the problem is often misdiagnosed as the failure of the theory to be UV-complete, and as indicating the need for a UV-completion of this theory. Thus, UV-completeness is often taken as a criterion of theory acceptance for QG. Here, following \cite{CrowtherLinnemann}, I argue that it need not be.


A theory is UV-complete if it is formally predictive to all possible high energy scales (or, equivalently, all short distance scales). In other words, the theory can be used to generate predictions at all possible short distance scales. Being UV-complete does not guarantee that all of these predictions are \textit{correct}, however (Newtonian mechanics is UV complete, for instance, since it does not formally `break down' at any short distance scale, and yet we know that it is not the correct theory to use at all short distance scales). There are several ways of obtaining a UV-complete theory, including having a QFT that is non-perturbatively renormalizable, such as quantum chromodynamics.\footnote{\textit{Perturbative} renormalizability is insufficient to establish UV-completion, since a perturbatively renormalizable theory may still face a Landau Pole, as is the case in quantum electrodynamics \S\ref{Land}.} Yet, the idea of UV-completion is \textit{not restricted to QFT}, and there are other ways in which a theory may be UV-complete, apart from being renormalizable.\footnote{\citet{Butterfield2015, CrowtherLinnemann,Dvali,Zee2010}.}

In the heyday of QFT, UV-completion was usefully employed as a criterion of theory success in QFT: i.e., a means of selecting theories that were not only viable, but \textit{physical}. The development of the RG and effective field theory techniques (and associated philosophy) led to a change in this perspective, where we recognise that theories can be physical without being fundamental---effective field theories correctly describe the world within their limited domains, and are consistent when applied within these domains. Nevertheless, it seems there remains a sort of hangover, where the general perception of UV-incomplete theories as being faulty, or mathematically inconsistent, still lingers.

Although approaches such as asymptotic safety are guided by the principle of UV-completion, QG need not be a UV-complete theory, \textit{unless QG is conceived of as a fundamental (final) theory}. If we conceive of QG in a more minimal sense, then all that is required to satisfy the Primary Motivation is a `UV-better' theory, predictive at the Planck scale \citep{CrowtherLinnemann}. Following Fig. \ref{Flowchart}, we may thus distinguish between two options: either the `amalgamation' of QG (as not ToE)\footnote{If one includes the criterion that QG be a ToE, then QG is the ToE in (ii), rather than part of the amalgam (i).} plus the Standard Model (i.e., a non-unified set of theories) are together fundamental (i), or they are not. If they are not, then (some of) these theories are emergent, and they can either emerge from a ToE (ii), or they emerge from another amalgam (iii). And so on. Ultimately, there are three possible scenarios: (1) some non-unified set of theories is fundamental, (2) there is a ToE, or (3) there is no fundamental level, but a never-ending tower of theories.

\begin{figure}[h]

  \centering
  
      \includegraphics[width=\textwidth]{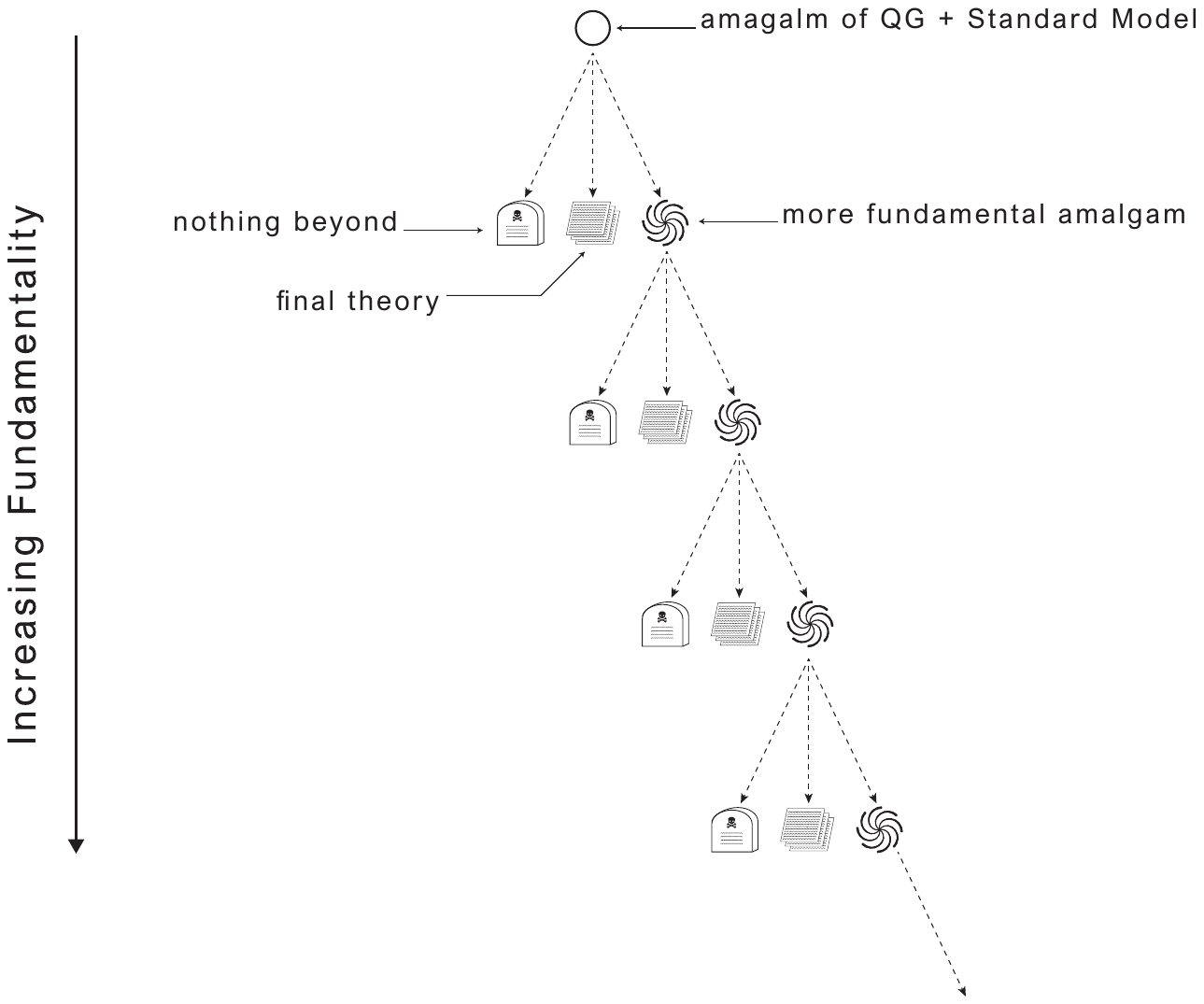}

  \caption[figure]{\textit{Options on the road towards higher-energy physics:} 
  Suppose we have a non-unified set of theories `amalgam' of QG and the Standard Model. It is possible that there is no deeper theory beyond, or that this amalgam emerges from a unified `final theory', or that it emerges from another amalgam. This could happen at each `level' of inquiry as we seek increasingly fundamental theories.}
  \label{Flowchart}

\end{figure}

It is only in scenario (i) in which QG is necessarily UV-complete. This scenario is distasteful, however, in that it represents a disunified amalgamation of several theories as standing in place of a final ToE. The disunified nature of the amalgam would prevent us from believing it fundamental\footnote{Other concerns, such as arbitrariness of the values of the parameters within each theory, the lack of explanatory power, etc. would also contribute to this thought, \citep{CrowtherDigging}.}, and we would push on in the search for a more unified theory beyond. This would be the case in spite of all the theories in the amalgam being UV-complete. The lesson from this is that UV-completion should only be used as a constraint on QG if you accept that, at the fundamental level, there is no final theory, but rather a disunified collection of several theories.\footnote{In some approaches to QG, UV-completion may conflict with other principles that are desirable and otherwise apparently viable, aside from unification. For instance, if UV-completion is obtained by cutoff, it may pose problems for Lorentz invariance, and, in the `higher-derivative approaches' and asymptotic safety scenario, UV-completion apparently conflicts with the principle of unitarity.}

String theory is supposed to be a ToE, rather than just QG---as such, it must be UV-complete. The extendedness of the theory's basic entities is very likely to ensure that it is UV-complete, although no proof has yet been found that it is (see \citet[][\S 7.2]{Hagar2014}, \citet[][Ch. 1]{Dawid2013a}). Textbooks on string theory often present an analogy between the perturbative non-renormalizability of quantum GR and that of 4-Fermi theory, which is a non-renormalizable theory that was revealed to be the low-energy limit of the \textit{renormalizable} theory of electroweak interactions. Similarly, proponents of string theory claim that string theory is the renormalizable theory underlying perturbative quantum GR. Thus, the alleged UV-completeness of string theory is treated as as a means of confirmation or indicator of pursuit-worthiness.

Other approaches to QG use UV-completion in different roles, particularly as a guiding principle, constraint, and means of confirmation. In some cases, including in LQG, UV-completion is obtained automatically in virtue of the theory naturally describing a minimal length scale \citep{Rovelli2004}. Here, without taking UV-completeness as a constraint on QG, it may make sense to treat the principle as a means of confirmation, especially given the consilience of the idea of minimal length with other motivations for QG, such as singularity resolution, \S\ref{sing}. 

\pagebreak

\section{The Primary Motivation: Reconciling the two pillars}\label{primary}

I've characterised the Primary Motivation for QG very roughly as a theory that describes the domains where both GR and QFT are supposed to be \textit{necessary}, and which somehow `takes into account' the lessons of both GR and quantum theory. Three points require more elaboration: `describes', `domains', and `taking into account'. 

\subsection{`Domains'}\label{domains}

\cite{Smeenk2013} describes the ``overlapping domain'' argument for QG, arguing that the early universe falls within the domains of applicability of both QFT and GR, and thus requires a theory of QG for its description. As stated above, however, both GR and quantum theory are supposed to be universal theories, so technically their domains of applicability overlap completely! Thus, QG should also be a universal theory, not restricted to particular scales or parts of the world. Properly, the motivation for QG is to describe domains where QG effects cannot be neglected---where the predictions of QG are thought to differ from those of GR and QFT. In other words, it should describe the overlapping domains of \textit{necessity}---rather than applicability---of GR and quantum theory, as viewed from the current paradigm.\footnote{Of course, though, from the perspective of the imagined future paradigm, QG is supposed to replace GR and QFT in these domains, so---strictly---the presently incumbent theories are not supposed to be necessary, nor even correct, in these domains.}

\cite{Wallace2022}, as well as the physicists interested in QG phenomenology, emphasises that not all the domains of QG are exotic and inaccessible. Also, in fact, we have a working theory that satisfies the second condition of the Primary Motivation as a combination of GR and quantum theory: this is just QFT applied to GR\footnote{\label{LEQG} This is not just semiclassical gravity (eq. \ref{eqsemiclass}), but also a QFT treatment of fluctuations using the \textit{background field method}, such that GR is evaluated via the path integral formalism of QFT \citep{Wallace2022}.}, \textit{and} which describes QG phenomena and observations---that is, observations of QG phenomena that we in fact already have access to---\textit{and} the theory is relatively well-confirmed (depending on the different domains where it is applied, its support ranges from strong, to moderate, to weak). This is a \textit{low energy theory of QG} (LEQG); it is an effective field theory, meaning that it is valid in a particular, restricted domain. This domain includes QG phenomena---being self-gravitating objects that also require QM to describe---in the semiclassical regime, both non-relativistic (describing planets and white dwarfs), and relativistic (describing stars, neutron stars, supernovas, and cosmology), and also the perturbative regime (having some support from observations of the solar system and cosmic microwave background).

The domain of LEQG, however, is low-energy\footnote{`Low-energy' here refers to anything below the Planck energy scale, however, so it includes all known physics!}---the theory is not thought to describe the Planck scale (explained below), the interior of black holes, or very early universe cosmology. It is these domains for which we need a full theory of QG, and which properly serve to characterise the Primary Motivation for QG. Thus, `the problem of QG' from the perspective of high-energy physics, is that LEQG does not describe the high-energy (UV) regime. (The theory also faces particular theoretical and conceptual problems, but these are not what prevents it from satisfying the Primary Motivation). 

The Planck scale is characterised by combining the fundamental constants from our current theories: the speed of light, $c$, the gravitational constant (Newton's constant), $G$, and the quantum of action (the reduced Planck's constant), $\hbar$. It was shown by Planck in 1899 that there is a unique way (apart from numerical factors) to do so. We arrive at the Planck length, $l_\text{P}$, the Planck time, $t_\text{P}$ and the Planck mass (equivalently, the Planck energy), $m_\text{P}$, respectively:\footnote{See, e.g., \cite{Kiefer2007a}, \S1.1.3.}

\begin{equation}
l_\text{P}=\sqrt{\frac{\hbar G}{c^3}}\approx 1.62\times 10^{-35}\text{m}
\end{equation}

\begin{equation}
t_\text{P}=\sqrt{\frac{\hbar G}{c^5}}\approx 5.40\times 10^{-44}\text{s}
\end{equation}

\begin{equation}
m_\text{P}=\sqrt{\frac{\hbar c}{G}}\approx 1.22\times 10^{19} \text{GeV}
\end{equation}

In cosmology, we also refer to the Planck temperature $T_\text{P}$, and Planck density,  $\rho_\text{P}$,

\begin{equation}
T_\text{P}=\frac{m_\text{P}c^2}{k_\text{B}}\approx 1.41\times 10^{-32}\text{K}
\end{equation}

\begin{equation}
\rho_\text{P}=\frac{m_\text{P}}{{l^3}_{\text{P}}}\approx 5\times 10^{93} \text{g/cm}^3
\end{equation}

where $k_\text{B}$ is Boltzmann's constant.

These are obtained by dimensional analysis, a familiar and useful technique in physics. Theories are typically characterised by their own fundamental constants, and because this dimensional analysis combines both the characteristic constants of GR and QM in a unique way to form a new set of characteristic constants, it is standardly thought that these characterise the regime where QG is necessary. 

 In spite of the popularity of the Planck scale as characterising the domains of QG, the dimensional analysis that gives us the Planck scale is heuristic: it does not establish that this is the characteristic scale of QG. The relevance of the Planck scale can be criticised as a naive general estimate. Nevertheless, it is possible to make the case that the Planck scale is the relevant scale at which to expect new physics, based on arguments from both particle- and gravitational-physics which may be taken to suggest that our current picture of familiar `large scale' (i.e., energies lower than the Planck energy) physics is complete \citep{Held2019}. Such a conclusion, however, neglects the problems of dark matter and dark energy (the cosmological constant Lambda), which may be considered as part of QG phenomenology. 
 
 Perhaps stronger than the heuristic arguments for the relevance of the Planck scale is the argument that the Planck scale is where LEQG---as standardly conceived, perturbatively---breaks down, so necessarily a new theory of QG is required here (\S\ref{UVC}).

The other domains that characterise QG are the interior of black holes and the very early universe. This is motivated by the idea that GR `breaks down' in the vicinity of curvature singularities because of the neglect of expected quantum effects in these domains (\S\ref{sing}). Another motivation comes from the theoretical predictions of black hole thermodynamics (\S\ref{BHT}). 

\subsection{`Describes'}

Ideally, for a theory to \textit{describe} particular domains, the theory gives predictions in these domains that are tested and not falsified. This is a challenge given the domains in question, however. More minimally, then, we might require that the theory give predictions that are physically reasonable when applied in these domains---although, what `physically reasonable' means is a non-trivial question in this context! More work needs to be done to explore this, particularly in connection to discussions in general philosophy of science.

Of course, a theory being formally predictive, and that the predictions are taken as `physically reasonable', in these domains does not guarantee that what it says is correct when applied here, but we can usefully appeal to the empirical constraints described in \S\ref{empirical}, including correspondence relations with GR and QFT to provide at least some guidance.

\subsection{`Takes into account both QM and GR'}\label{takes}

Finally, consider the second main condition of the Primary Motivation: that the theory \textit{take into account both QM and GR}. This condition is motivated by the success of GR and QM, and also the fact that both are expected to be necessary in the particular domains indicated above---and yet, each are thought to be inaccurate here. There are several ways of understanding this condition, representing different viewpoints on the problem of QG. Ashtekar expresses some of these,

 \begin{quote}
Everything in our past experience tells us that the two descriptions of Nature we currently use must be approximations, special cases which arise as suitable limits of a single, universal theory. That theory must be based on a synthesis of the basic principles of general relativity and quantum mechanics. This would be the quantum theory of gravity that we are seeking. \citep[][p. 186]{Ashtekar1995} (p. 186).\footnote{Cf. \cite{Ashtekar2005}.}
  \end{quote}
 
This quote demonstrates two ways in which QG could take into account GR and quantum theory: by recovering both theories in their respective domains of applicability (through correspondence relations, including limiting relations), and combining the principles of both theories. These two conditions represent what we might call the `new framework perspective' of the problem of QG. It is also described in \cite{Ashtekar1974}, in a passage where the authors state that GR and QM each can be understood as a distinct ``body of universal rules", which means that the quantization of gravity is essentially different from quantizing the electromagnetic field: ``From this viewpoint, then, the problem is to obtain a new body of rules which suitably encompasses the essential features of those of quantum mechanics and of general relativity." (p. 1214).
 
 This `new framework perspective' on the problem of QG, however, is not the standard one adopted by physicists working on QG.\footnote{Note that the labels new framework perspective' and `standard perspective' are my inventions, and that Ashtekar's work is ambiguous between the two views.} Rather than seeking a new theory that replaces GR and QM, and recovers both theories as approximations in their respective domains, most physicists tend to treat QM as fundamental and GR as not.\footnote{Which is not to say that GR fails to embody some principles, e.g., background independence, that are thought to be fundamental.} For instance, \citet[][p. 16]{QG30questions} expresses the sentiment that the `universality of quantum theory' is one of the fundamental principles of modern physics: the physical world at the fundamental level is governed by quantum laws, and that the classical picture is only an approximation, valid at sufficiently low energies and on sufficiently large scales. 

 Thus, we find at least two different ways of satisfying the condition that QG `take into account' the lessons of GR and QM. What I have called the \textbf{\textit{New Framework Perspective}} holds (1) we need a \textit{new body of rules} that express insights of both GR and QM, and (2) we need to recover \textit{both} GR and QM as approximations in their respective domains. The \textbf{\textit{Standard Perspective}}, however, holds (1) we need a \textit{quantum theory}, and (2) from this we can recover GR as an approximation in the domains where we know GR is successful (on such a view, the recovery of QM is not required). While I will not here argue for one view or the other, I would at least like to suggest that the fundamentality of QM is an assumption that has not been questioned seriously enough in the search for QG, and that the problem of QG may well be that we require a whole new framework that could `explain' the success of quantum theory just as it would explain the success of GR.
 
There are more-specific ways of satisfying the second main condition of the Primary Motivation. \cite{Butterfield2001}, \S3.1.3. describes four different types of approach to QG: 1. quantize GR, 2. GR as the low-energy limit of a quantization of a different classical theory, 3. GR as the low-energy limit of a theory that is not a quantization of a classical theory, 4. start \textit{ab initio} with a radically new theory. Approaches 1-2 represent the Standard Perspective, approach 4 represents the New Framework Perspective, while approach 3 could represent either perspective, depending on whether or not the theory is formulated in the framework of quantum theory as we currently understand it. Any of these approaches could produce a theory that accords with the condition of `taking into account' both GR and QM, if it satisfies the two necessary aspects described above. Nevertheless, the by-far dominant approach is (1), where not only is QG supposed to be a quantum theory, but one in which gravity is quantized.
 
It is possible that QG take into account QM without itself being a (purely) quantum theory. The most familiar way of thinking about this is along the lines of semiclassical gravity (\S \ref{incom}) or other `hybrid' approaches (\S \ref{quantiz}), which attempt to couple the classical and the quantum as a sort of `amalgamation' (as opposed to a unification) or `mongrel gravity' theory (a term borrowed from \cite{Mattingly2009}; cf. \cite{Tilloy2018} for discussion of some other hybrid theories). The New Framework Perspective countenances the possibility that the framework of QM---as it stands---is not applicable in QG, and gets modified somehow (in which case QG would be more fundamental than both GR and QM, \S\ref{Fundamentality}, but need not be a full unification of GR and QM \S\ref{unif}).

As stated above, Penrose believes that the measurement problem of QM points to the need for QG, and that the unitarity of QM will be modified, so that QM will become a non-linear theory at the Planck length. More recently, the unitarity of QM has been called into question in the context of quantum cosmology, where \citet{cotler2022} replace linear unitary time evolution with linear isometric time evolution (note: this is very different from the quite radical proposal of Penrose). Other recent work in the New Framework perspective is the hybrid, ``post-quantum classical gravity'' theory which modifies both GR and QM \cite{Oppenheim2023}. This approach is motivated by problems with semiclassical gravity and as well as perceived inadequacies to mainstream responses to the black hole information loss paradox.

QG can `take into account' the lessons of GR and QG by means of \textit{principle correspondence}: the new theory features one or more principles from each of the older theories. As part of taking into account GR, it is popular to utilise background independence as a principle-correspondence, featuring in both GR and QG (\S\ref{BGInd}). For principle correspondence in QM, the favourite principle to use is quantization (\S\ref{quantiz}), which accords with the Standard Perspective. These principles are naturally, and conservatively, motivated by the universality of both GR and QM.

 \subsection{Three pillars perspective}\label{threepillars}
 
 An increasingly significant perspective upon the problem of QG is that it involves bringing together not just GR and QM, but also the third pillar of modern physics: thermodynamics (also including statistical mechanics). There are several approaches to QG that tie gravity to the thermodynamic concept of entropy. For instance, \cite{Jacobson1995} argues that the Einstein Field Equations can be derived from the proportionality of entropy and black hole horizon area together with the first law of thermodynamics. \cite{Padmanabhan2004, Padmanabhan2010} and \cite{Verlinde2011, Verlinde2017} also present arguments for gravity being an emergent phenomenon of entropic origin. According to the `emergent gravity' approaches, discussed in \S\ref{quantiz}, spacetime is an effective thermodynamic entity; as such, a quantization of gravity (the metric field) would not lead us to the `micro' degrees of freedom which we seek to describe by a more fundamental theory of QG. (These micro degrees of freedom, however, could themselves be described by a quantum theory, and so these approaches need not represent the New Framework perspective). The link between GR, QFT and thermodynamics is motivated, and further revealed, through the theoretical results of black hole thermodynamics, discussed in \S\ref{BHT}.
 
\subsection{Delimitations}
The Primary Motivation is supposed to represent the minimal definition of QG: any approach that does not (attempt to) satisfy it would not be accepted, by \textit{the current mainstream consensus}, as an approach to QG. The definition thus serves to delimit what would `count' as an acceptable theory, and necessarily excludes alternative possibilities. We should bear in mind that there may be other solutions to the problems discussed here, which are overshadowed if we take them exclusively as problems that motivate QG (as a theory that satisfies the Primary Motivation). Examining the particular motivations in more detail can help us also to explore the other possibilities for their resolution, or reconsider the need to resolve them at all. 

This is nicely illustrated by the example of the differing attitudes one can take towards spacetime singularities (§6): not all cases of singularities motivate the need for a new theory, and in the cases where the (particular) singularities do motivate a new theory, it is not certain that these point to the need for QG rather than a different theory formulated `at the level of current theory' (i.e., a new classical theory of gravity such as a modification of GR). There is the possibility that other motivations for QG may also be addressed by reconsidering our current best theories, including their principles and formulation, rather than moving to a theory of QG. "

 \pagebreak
\section{Unification}\label{unif}

\makebox[0pt][l]{%
\begin{minipage}{\textwidth}
\centering
    \includegraphics[width=\textwidth]{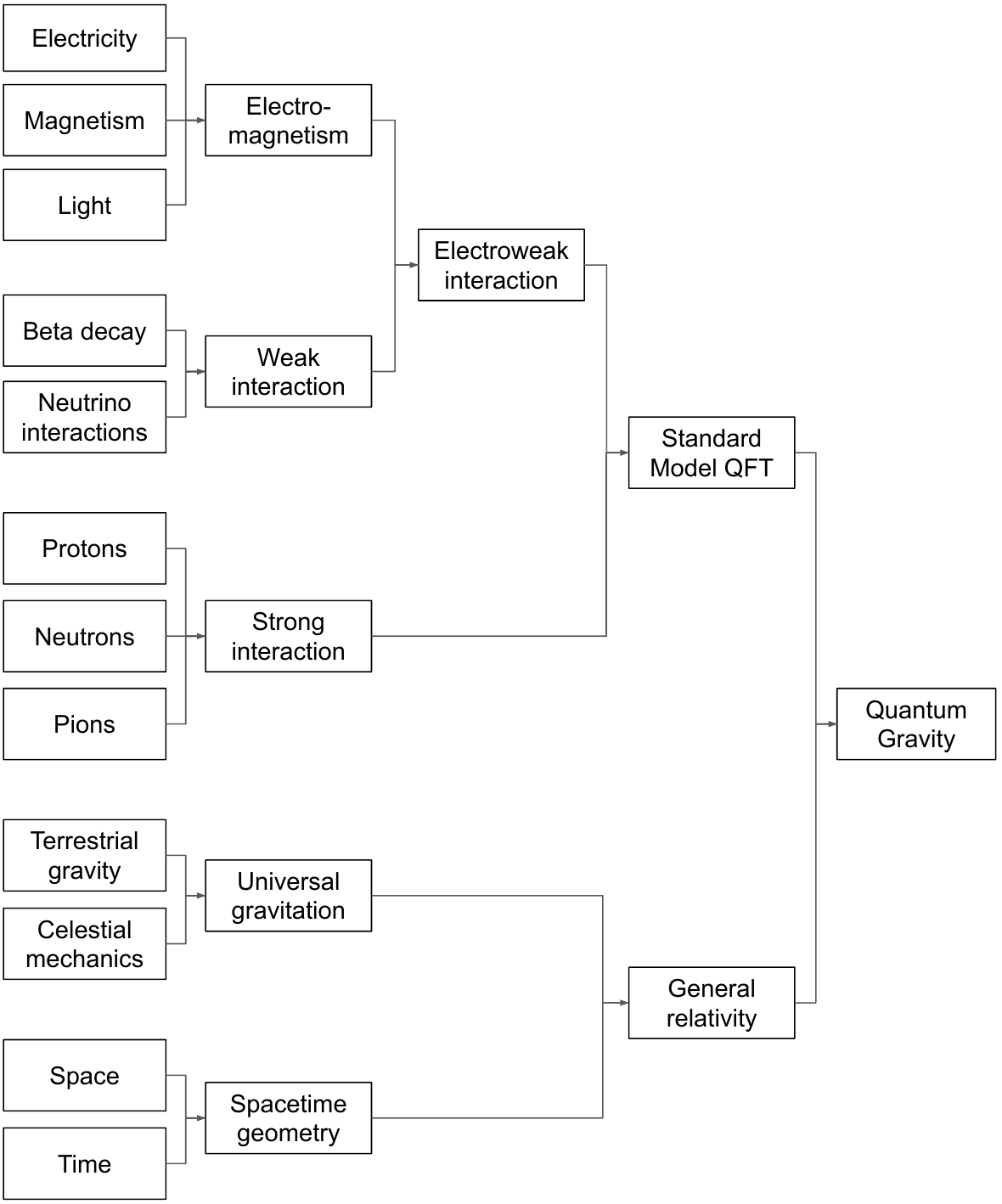}
    \captionof{figure}{The `physicist's tale' of unification.}
  \label{fig:unif}
\end{minipage}
}

\medskip

Unification is a traditional guiding principle in physics, and is often viewed as means of producing successful theories. Familiar examples (representing various different ideas, and degrees, of unification) include Maxwell’s theory of electromagnetism, which unified light as well as the electric and magnetic forces; the electroweak theory, which unified the electromagnetic force and the weak force; and even GR, with its identification of inertial mass with gravitational mass, and spacetime with gravity. There is a tendency to view the history of physics as a history of unification, and the path forward as one of continuing this trajectory to its ultimate end in a final, unified theory: \cite{Salimkhani2018} calls this the ``physicist's tale", following \cite{Maudlin1996}, who states that it ``has become so pervasive as to rank almost as dogma''. It is illustrated, for instance, in Fig. \ref{fig:unif}. 

As shown in Fig. \ref{fig:unif}, unification is a way of motivating a theory of QG, here illustrated as a final ToE (but see Fig. \ref{Flowchart} for an alternative picture, illustrating QG as part of a non-unified fundamental level). For those inclined towards unification, the current situation in physics---the split picture of the world it presents---is unsettling, and calls us to question the fundamental nature of both GR as well as the framework of quantum theory and the Standard Model of particle physics. This may be the case \textit{even if we accept} that there is a usable theory that treats gravity in the same framework as the (other) fundamental forces, namely LEQG (introduced in \S\ref{primary}). In other words, the problem is not that we cannot combine GR and QM in a single theory: the problem, on this view, is that the theory is not properly \textit{unified}.

What is meant by unification? \cite{Maudlin1996} argues that there are several degrees (or levels) of unification, which exist between a ``lower bound'' that falls short of unification, and an ``upper bound'' that represents \textit{perfect unification}. The lower threshold states that unification is not merely that two (or more) theories are consistent with one another, or that they share a common dynamics, or that there exists a law-like connection (nomic correlation) between physical forces. The upper bound of perfect unification requires that there be a single theory that describes all phenomena as \textit{the same}---as fundamentally stemming from a single origin, e.g., as manifestations of a single entity or interaction (``All is One''). And,``it is this deeper sense of unification, the idea that all the physical forces are at base one and the same, which contemporary physicists invoke when they speculate on the theories to come'' \citep[][p. 132]{Maudlin1996}. \citet{Morrison2000} refers to this type of unification---where two phenomena hitherto thought to be distinct are identified---as \textit{reductive} unification.\footnote{It is just one of 13 different forms of unification described in \citet{Morrison2000}.} 

Unification is typically seen as an external motivation for QG (the desire to unify two separate frameworks), but it can also be considered as an internal motivation (within a single framework), if the lack of unification \textit{within} the Standard Model of particle physics is taken as a problem motivating QG. Although the Standard Model can be written as a single theory, it appears as a disjointed amalgam of separate (particle) fields, rather than a unified theory in Maudlin's sense, and this drives many physicists to seek a more unified theory beyond. But why unification? Unification is generally regarded as an epistemic virtue, conferring support for a theory, and being used as a means of justification of a theory (note that this can be the case even without requiring or implying the metaphysical assumption that the world itself is unified); several authors have given Bayesian analyses of unification as an epistemic virtue \citep{Myrvold2003}. Unified theories have also been argued as being more explanatory, testable, falsifiable, and successful than non-unified (or less-unified) ones \citep[][\S1.1.4]{Schindler2018}. (Additionally, unification can be used \textit{heuristically} in guiding theory-development, as described by \cite{Kao2019}.)

String theory is often promoted as being a unified theory, as well as being a ToE, since it treats gravity as on par with the fundamental forces of the Standard Model, and all stem from the same basic physics (the behaviour of strings). Yet, QG is not necessarily a unified theory, nor a ToE. In the first case, it may be a semiclassical or hybrid theory, which is a non-unified combination of GR and QFT, and in the second case, it may just be a quantum theory of gravity, and not a theory that combines gravity with the Standard Model forces. Requiring QG to be a unified ToE is only justified if we take QG as a \textit{final} theory (i.e., `nothing beyond' in the schema illustrated in Fig. \ref{Flowchart}). 

There may be reasons for not wanting to unify gravity with the forces of the Standard Model of QFT, however, depending on how one interprets GR. The typical interpretation of GR is a geometrical one, according to which gravity is not properly a force at all, but the curvature of spacetime. As \citet[][p. 133]{Maudlin1996} states, ``Objects do not couple to the gravitational field, they merely exist in space-time.''\footnote{Cf. \cite{Maudlin2012}.}

This interpretation of GR stands in contrast to the `particle physics perspective' which drives string theory. Instead, it---along with the principle of background independence (\S\ref{BGInd})---motivates approaches to QG which primarily aim at a consistent quantum description of (more fundamental physics responsible for the low-energy existence of) spacetime, such as LQG and causal set theory. In such approaches the principle of background independence is prioritised over the principle of unification.

But in general, perfect unification of GR and QM is not a standard motivation for QG. If it were, this would mean that both GR and quantum theory (as a framework) are not fundamental, but must somehow be recovered from QG---using appropriate limiting procedures and approximation techniques---in the regimes where they are known to be successful. If any approach to QG could not demonstrate that this can be achieved, then its acceptance would be unlikely. But, this is not what most approaches aim at; they aim to retain the standard framework of quantum theory, rather than reduce it to a more basic theory. Arguably, even unification in a weaker sense is not properly called the problem of QG, if we would be satisfied with a `hybrid' approach to QG, or asymptotic safety.

\cite{Salimkhani2018} adopts the particle physics perspective and argues that the problem of QG is not the need to unify---nor even to combine---GR and QM. He uses results by Weinberg to argue that GR can be reduced to (derived from) the combination of SR and QM (i.e., QFT), and that we have LEQG; thus that the problem is that we need to find the correct theory at high energy scales. More generally, \cite{Salimkhani2021} argues that unification need not be understood as a goal of physics at all. Unification is considered an external influence on physics--- \textit{external} here meaning that it is driven by philosophical assumptions, such as metaphysical, metatheoretical, or epistemological considerations that are imposed on physics. Against this, Salimkhani argues that, instead, unification naturally arises in physics as a consequence of the more basic (or genuine) aims and methods of physics, i.e., factors properly \textit{internal} to physics itself, such as empirical adequacy and theoretical consistency. 

While the primary motivation for QG is to find a theory that describes the domains where both GR and QFT are supposed to be necessary, and which somehow combines GR and QM, it seems we should not understand the problem---as it currently stands---as to find a unified theory. This is not to say, however, that the goal of unification may not be heuristically useful, or that unification could not be an epistemic (or explanatory) virtue, or a means of confirmation for a theory of QG. These are questions that are left open, and depend on how else we conceive of the problem of QG.

\pagebreak

\section{Quantization}\label{quantiz}

Quantization is the procedure by which a quantum theory is produced from a classical one. There are three main ways in which we can attempt to perform a quantization of gravity, and each of these faces its own challenges (\S\ref{approaches}). However, when we refer to the need for QG to be a quantum theory, or more specifically, a theory of quantized gravity, this does not imply that it must be a theory obtained through the quantization of GR. Recall the four different ways of constructing QG identified by \cite{Butterfield2001}, described above (\S\ref{primary}), wherein the quantization of GR was only one option.\footnote{\cite{Butterfield2001} also describe how different structures in GR may be quantized.} In other words, even on the Standard Perspective, we can have a theory of QG that is quantum, but not produced through the quantization of GR. And, further, if we adopt the recondite New Framework perspective, we may deny not only that QG need be obtained through a quantization of GR, but also that QG need even be a quantum theory as we know it.

Physicists---sharing the `Standard Perspective'---argue that quantization of GR is necessary, but many of these arguments apply more broadly against any `hybrid' theory of QG in which gravity is fundamentally classical. These arguments include: 

\begin{itemize}
    \item \textit{The universality/fundamentality of quantum theory}: Quantum theory says that every dynamical field must be quantized, and GR tells us that spacetime is a dynamical field, therefore spacetime must be quantized;\footnote{This argument is made by many authors, but prominently in \cite{Rovelli2001, Rovelli2004}.}
    \item \textit{Unification}: Through quantization, gravity (whether conceived of as the curvature of spacetime or as a force), is treated as quantum in the same way as matter and the (other) fundamental forces; (but note that although quantization is promoted as a way of producing a `full' quantum theory of gravity, rather than a `hybrid' semiclassical theory, it still may not properly represent unification,\S\ref{unif}); 
    \item \textit{Spacetime singularities}: The presence of particular singularities in GR could be resolved by quantization of gravity, and points to the need for quantization (this argument is discussed in \S\ref{sing});
    \item \textit{Heuristic arguments and thought experiments}: Various arguments in which the involvement of quantum matter with classical gravity leads to paradox, inconsistency, or otherwise difficult-to-accept conclusions, and are thus taken to indicate the necessity of the quantization of gravity;\footnote{Two of these have already been described in \S\ref{incom}, from Feynmann, and \cite{BohrRosenfeld}. Others include \cite{Eppley1977, Peres2001}. These have already been well-covered in the literature, and as such I do not go over them again here, \cite{Grossardt, Grossardt2, Huggett2001, Kent2018, Mattingly2005, Mattingly2006, Oppenheim2023, Rydving2021, Tilloy2018, Wuthrich2005}.}
\end{itemize}

While these arguments are widely appealed to, none conclusively establish that QG must be a theory of quantized GR. 

A comment on the first of these arguments, regarding the fundamentality of quantum theory: It is a key assumption behind the Standard Perspective on addressing the Primary Motivation for QG, and may be questioned if we take the New Framework Perspective, according to which neither GR or quantum theory is fundamental as they stand. Additionally, as noted earlier, quantization is also standardly taken as the main characteristic principle of QM to be featured in QG, and thus to act in the role of principle-correspondence, as part of fulfilling the Primary Motivation (this could be true on either the New Framework Perspective, or the Standard Perspective). But, even if we grant the fundamentality of quantum theory, it still does not follow that gravity must be quantized: it may depend on what \textit{interpretation} of QM we are working with. As \citet{Adlam2022} states, 

\begin{quote}
    [...] the relevant notion of unification here is the idea that there should ultimately be only one type of stuff---i.e. quantum stuff---and therefore since ordinary matter gravitates, and ordinary matter is assumed to be composed entirely of quantum stuff, it follows that quantum stuff must gravitate. But not all interpretations of quantum mechanics agree that there exists only `quantum stuff', and therefore this conclusion is by no means inevitable. \citep[][p. 115]{Adlam2022}
\end{quote}

To recap: while the quantization of gravity is overwhelmingly viewed as the correct way of approaching the problem of QG, it is not the only one. Quantization is taken as a motivation for QG, but is itself driven by assumptions about the fundamentality of quantum theory as well as desires for unification. These principles also underlie the Standard Perspective on addressing the Primary Motivation for QG. Still, neither quantization, nor the more-basic principles motivating it, are necessary constraints on a theory of QG. 

\subsection{Hybrid theories and emergent gravity approaches}

If gravity is not quantized, one possibility is that it remains fundamentally classical, as in semiclassical gravity or other hybrid approaches. Recall that LEQG (\S\ref{primary}) is an effective field theory for QG (it is taken to reproduce the results of QG at low energies), and comprises both a semiclassical treatment of gravity as well as a description of the quantum fluctuations of the LEQG field. The dominant attitude towards semiclassical gravity is that it is an approximation to LEQG in the situations where quantum fluctuations of the spacetime geometry can be ignored, i.e., semiclassical gravity is a \textit{mean field description} in LEQG. This attitude holds that gravity is fundamentally quantized (and will be described as such by the more fundamental, unknown, theory of QG), and is only effectively treated as classical (i.e., as an approximation in the `low-energy' domains where we know GR is successful).

As \cite{huggettQGLab} points out, though, there are two alternative attitudes towards semiclassical gravity. One of these holds that it is not the mean field description of LEQG, but some other theory. If the low-energy approximation of the more fundamental theory of QG were not LEQG, this would allow that QG is not necessarily a theory of quantized gravity. This view fits the various examples of `emergent gravity' approaches mentioned below, as well as proposals such as that of \cite{Oppenheim2023}. \cite{huggettQGLab} take this view as an ``epistemically careful'' one, where we accept that semiclassical gravity approximates some theory \textit{X}, which in turn approximates QG, but we take no stance on whether \textit{X} is LEQG, or some other alternative (one in which gravity may even be classical---or not). It allows us to remain neutral on the exact relationship between semiclassical gravity and QG. Alternatively, there is the possibility that semiclassical gravity (\S\ref{incom}) is the fundamental theory of QG. This attitude is not a popular one, being plagued with both conceptual and mathematical difficulties.\footnote{See fn. \ref{fnCuriel}.} 

The label `emergent gravity' refers to a broad and eclectic class of approaches to QG which treat gravity as a low-energy effective phenomenon that emerges from the collective behaviour of more fundamental microscopic degrees of freedom. The microphysics itself is not based on a discretisation or quantization of GR \citep{Linnemann2018}. Simple examples of this approach are analogue models of spacetime from Bose-Einstein condensates (BECs).\footnote{See \citet{Barcelo2001E, Hu2009, Liberati2009, Visser2002, Volovik2003}. Philosophical discussions: \citet{Bain2008, Crowther2016}.} Others include Sakarov's induced gravity \citep{Sakharov1967, Sakharov2000}, Jacobson's gravitational thermodynamics \citep{Jacobson1995}, Padmanabhan's gravitational thermodynamics \citep{Padmanabhan2010}, and Verlinde's entropic gravity \citep{Verlinde2011}. 

Typically, these approaches share the conviction that gravity is analogous to hydrodynamics (or thermodynamics), and the Einstein equations are akin to the wave equation for sound in a medium (i.e., they are higher-level equations of state). Accordingly, they hold that quantization of GR does not give us a theory of the fundamental micro degrees of freedom---by analogy, we would arrive at a theory of phonons rather than a description of the underlying `atoms' of spacetime. If approaches that quantize the metric tensor produce theories of particles analogous to phonons, then it is unsurprising that they should break down at high-energy. Their breakdown can motivate the search for a high-energy theory beyond (quantum) GR, but we cannot say that the degrees of freedom of the high-energy theory would themselves need to be quantized in order to produce a theory of QG.

\subsection{Tabletop experiments}\label{semi}

In spite of the issues discussed above, there is widespread agreement that the quantum nature of gravity cannot be established (or ruled out) by conceptual niceties, theoretical arguments or mathematical deduction: it must be confirmed empirically. Even Rosenfeld maintained that the \cite{BohrRosenfeld} argument against semiclassical gravity by analogy with `semiclassical electromagnetism' (described above, \S\ref{incom}) did not demonstrate the inconsistency of semiclassical approaches, because no formal arguments can establish the necessity of quantization. Rather, according to \cite{Rosenfeld1963}, only experiment can answer such questions.

And, indeed, since then, various experiments have been proposed and carried out. The Colella-Overhauser-Werner neutron interferometry experiment, touted as ``observation of gravitationally induced quantum interference'' demonstrated that the gravitational field affects the behaviour of quantum systems \citep{COW1, COW2}. The much-discussed Page and Geilker quantum Cavendish experiment \citep{PageGeilker} claimed to provide evidence supporting the hypothesis that a consistent theory of gravity coupled to quantized matter should also have the gravitational field quantized; however, their results were shown to be dependent on the particular interpretation of QM adopted.\footnote{By ``much-discussed'' here, I mean that there were many confused expressions by physicists as to why Page and Geilker thought it necessary to undertake such an experiment in the first place: the results being thoroughly unsurprising. For philosophical discussion see \cite{huggettQGLab, Wallace2022}. Also a nice summary in \cite{Kiefer2007a}.} These experiments, respectively, show that quantum systems both feel the effect of gravity, and that quantum systems can be sources of gravitational fields---in accordance with predictions of LEQG \citep{Adlam2022}. However, neither of these experiments probe states in which gravity remains in a coherent quantum superposition, and as such, we have no empirical observations in the regime of LEQG where superpositions of spacetimes are important. More recently, a new class of `tabletop QG experiments' have been proposed which claim to do exactly this \citep{Bose2017, MV2017}. \cite{huggettQGLab} argue that how we assess the empirical scope of such experiments, though, is ``a much more intricate affair than may be at first thought, and crucially depends on matters of physical interpretation that are not settled by rational argument'' (p. 74). Note that these experiments are---at this time---proposals, likely years away from being carried out.

\pagebreak
\section{Singularity resolution}\label{sing}

Widely-cited as a motivation for QG is the need to resolve particular spacetime singularities in GR. Spacetime singularities are pathologies of a spacetime, and there are various ways in which spacetimes can be singular.\footnote{See, e.g., \cite{Curiel1999, Earman1996} for a discussion of different types.} Below, I discuss the two most common categories of spacetime singularity: incomplete geodesics, and curvature singularities.\footnote{According to \cite{Earman1996}, curvature singularities lead to geodesic incompleteness, whereas the opposite is not true. \citet[][\S1.1]{Curiel1999} argues that the two notions are actually independent.} Theorems of Penrose and Hawking \citep{Penrose1965, Hawking1973}, show that singularities (incomplete geodesics) are unavoidable in GR under very general, physically reasonable conditions. Because a common interpretation of these singularities is as representing the `breakdown' of spacetime, there is the worry, typically expressed, that GR `contains the seeds of its own destruction'. Spacetime singularities are are thus taken as \textit{internal motivation} for seeking a theory of QG, that is, they come from the theory (in this case, GR) itself. The corresponding constraint on QG is that it be a theory that \textit{resolves} the spacetime singularities---meaning that it should both be non-singular (in any physically problematic way), as well explain `what happens' in those domains where GR is thought to break down. Yet, it is not clear exactly if, how, or why, the spacetime singularities in GR signal a `breakdown' or incompleteness of GR---and thus, whether, or how, they in fact serve as internal motivations, or constraints, for QG. 

Although (particular) spacetime singularities are the most commonly cited motivation for QG, they are not the only singularities which could be resolved by QG. In QFT, there are several different types of divergences, including the UV divergences associated with the perturbative nature of the theory (\S\ref{UVC}), and Landau poles. These may be taken as internal reasons for not treating the framework of QFT, or particular QFTs (such as the Standard Model), as fundamental, and for seeking QG.\footnote{See, e.g., \cite{EllisInfinity, Kiefer2007}.} The UV-divergences do not necessarily point to the need for new physics, given that they could just be artefacts of the misapplication of perturbation theory. While the presence of Landau poles in the Standard Model of particle physics presents a stronger internal motivation for new physics, it is one that is usually ignored given that new physics is expected before these scales for \textit{external} reasons (that is, not from within the theory itself).

Here, I follow \citet{CrowtherDeHaro} in exploring how singularity resolution can motivate QG, and what the implications are for the theory sought. We find that, in general, singularities in current theories do not automatically point to the need for a new theory. While there are at least two examples of singularities (curvature singularities, and Landau poles) that do arguably motivate a new theory, it is not certain that these point to the need for QG rather than a different theory formulated `at the level of current theory' (i.e., a new classical theory of gravity such as a modification of GR, or an axiomatic formulation of QFT that does not feature divergences). The choice between these alternatives depends, amongst other things, on one's disposition towards the internal versus external motivations for QG. 

In \S\ref{UV}-\ref{Land}, I introduce the relevant QFT singularities, and \S\ref{QFTsing} discuss various responses to these. In \S\ref{geo}-\ref{curv}, I introduce the relevant GR singularities, and \S\ref{horror}, discuss various responses to these.

\subsection{UV divergences in QFT}\label{UV}

In \S\ref{UVC}, I described the problem of divergences associated with the perturbative methods in `conventional QFT' (CQFT), and which lie at the heart of the formalism itself. Historically, these infinities were removed in particular theories, such as quantum electrodynamics (QED), via the procedure of renormalization which rendered the theory finite and (impressively) predictive. This procedure, however, was physically suspicious, and the perturbative approach to QFT itself remained intrinsically approximate and conceptually problematic.\footnote{The discussion in this section largely follows \cite{CrowtherDeHaro}, \S3.} 

One response was the development of \textit{axiomatic QFT} (AQFT): instead of introducing informal renormalization techniques to treat interactions, this approach attempts to put QFT on a firm, non-perturbative footing, by specifying a mathematically precise set of axioms at the outset. Then, models of the axioms are constructed (constructive QFT).  Importantly, this approach is not an attempt at QG, but simply a new formulation of QFT at the level of QFT---i.e., as a combination of QM and special relativity without any singularities in the theory.\footnote{See, e.g., \cite{Fraser2011}.} Unlike CQFT, it is not an intrinsically approximate theory, since it is supposed to be directly defined on Minkowski spacetime, and so remain well-defined at arbitrarily small and arbitrarily large length scales. Although there are various simplified toy models satisfying the axioms of AQFT, there have been no realistic models constructed. In particular, it has not been demonstrated that QED or any other successful theories in high-energy physics admit formulations that satisfy the axioms of AQFT.

Meanwhile, in mainstream high-energy physics, the development of RG techniques led to a framework for studying CQFT systems at different energy scales, and ultimately to the discovery of the Standard Model. The RG analysis demonstrates that the means by which the cutoff is implemented has no bearing on the low-energy physics; the only effects that are significant at these scales are changes in the coefficients of finitely many interaction terms (the renormalizable interactions). In other words, RG theory tells us that large-scale phenomenology gives us almost no information about the nature of the short-distance cutoff, if there is one.

The dominant philosophical interpretation of this CQFT picture is that the UV divergences are not a real physical problem, but rather indications of the limitations of the perturbative framework of QFT. The framework itself is taken to be inherently approximate,\footnote{Cf. \cite{Fraser2020}.} and its models are supposed to be effective theories: not to be valid to arbitrarily high energy scales. ``This, in essence, is how modern particle physics deals with the renormalization problem\footnote{Footnote in the original suppressed.}: it is taken to presage an ultimate failure of quantum field theory at some short length scale, and once the bare existence of that failure is appreciated, the whole of renormalization becomes unproblematic, and indeed predictively powerful in its own right'' \cite[][p. 120]{Wallace2011}. The idea is that, whatever the unknown physics of QG turns out to be, the success of the CQFT models at known energies is explained, thanks to the RG.

This interpretation of CQFT as \textit{effective} means that the theory is not supposed to be reliable at short length scales. In particular, the need to employ a short-distance cutoff is not taken to indicate \textit{anything} regarding the physics beyond. This is in contrast with the case in condensed matter physics, where the RG is also employed because the system is described by a theory which diverges in the UV, but in which case the divergences, and the need to employ a short length scale cutoff is consistent with the existence of something physical---we know that we cannot treat matter as continuous at arbitrarily short length scales, because matter has a discrete structure at the atomic scale. In CQFT, however, there is no empirical evidence for the existence of a real physical cutoff (e.g., a discrete structure for spacetime at extremely high energies). It is possible that the UV divergences in CQFT simply reflect limitations of the theory, rather than any new physics.

\subsection{Landau Poles}\label{Land}

A motivation for treating particular QFTs (rather than the framework of QFT itself) as effective field theories is the existence of Landau poles. This type of infinity is thought more concerning than the UV-divergences because it is not taken to be merely due to the limitations of perturbative analysis.

The most familiar example of Landau pole is in QED. This theory is renormalizable, so in principle it should be able to be extended to arbitrarily high energies. Yet, the renormalized coupling grows with energy scale, and becomes infinite at a finite (though extremely large) energy scale, estimated as $10^{286}$eV (the original result comes from \cite{Landau1954}). The existence of this `pole' could mean the theory is mathematically inconsistent. This is avoided if the renormalized charge is set to zero, i.e., if the theory has no interactions. There is indication that in QED, the renormalized charge goes to zero as the cutoff is taken to infinity (a physical interpretation of this is that the charge is completely screened by vacuum polarisation). This is a case of quantum \textit{triviality}, where quantum corrections completely suppress the interactions in the absence of a cutoff. Since the theory is supposed to represent physical interactions, the coupling constant should be non-zero, and so the Landau pole and the associated triviality might be interpreted as a symptom of the theory being effective, or incomplete (i.e., that it fails to take into account other fundamental interactions relevant at high energy scales).

QED and $\phi^4$ theory are thought to be trivial in the continuum limit in this way. In other words, RG analysis of QED and $\phi^4$ does not indicate that these theories possess a stable UV fixed point \citep{Lusher, QED}. This means that the Standard Model of QFT is thought to suffer Landau poles both for the electron charge, and the Higgs.\footnote{However, since the Landau pole in QED is normally identified through perturbative one-loop or two-loop calculations, it is possible that the pole is a sign that the perturbative approximation breaks down at strong coupling. Similarly, although the Higgs also is suspected to be trivial at high energy, this has been difficult to prove theoretically.} Although the Landau pole in QED is problematic for the theory, it is usually ignored because it concerns an energy scale where QED is not thought to be valid anyway, given that electroweak unification occurs at an energy scale lower than this. The Higgs triviality is thus more interesting for the Standard Model, as potentially indicative of new physics at high energy.\footnote{This issue is also related to the principle of \textit{naturalness}, which the Higgs mass conflicts with.} However, it also concerns an energy scale where \textit{QFT itself} is not thought to be valid, based on the \textit{external} arguments for the Planck length. 

In summary, the Landau pole divergences in CQFT are typically interpreted as part of the formal (mathematical) grounds---internal to the theory---for treating particular QFTs as effective. But there are also external grounds (not from the theory itself) for treating particular QFTs---and perhaps the framework of QFT itself---as effective, which come from the motivations for QG. These external motivations hold regardless of whether or not there are divergences inherent to QFT, but they are reasons why one might not be concerned with finding a singularity-free theory of QFT in order to describe the world at arbitrarily small length scales. 

\subsection{Responses to QFT singularities}\label{QFTsing}

There are three different types of divergences associated with QFT that may bear on the problem of QG: non-renormalizability of gravity (\S\ref{UVC}), UV-divergences in QFT, and Landau poles. The perturbative non-renormalizability of quantum GR (LEQG, \S\ref{UVC}) does indicate the need for a theory of QG valid at the Planck scale, though this theory may be non-perturbative quantum GR (i.e., the asymptotic safety scenario). The UV divergences of QFT may be artefacts of perturbation theory, and there is a danger of misinterpreting the need for a high energy cutoff as indicating a physical or operational minimal length. They do not point directly to the need for QG. Landau poles in the Standard Model of particle physics do (likely) represent an incompleteness of the theory and motivate the need for a new theory---though this new theory need not be QG. Landau poles, however, are typically ignored by physicists who already expect QG to replace QFT at the Planck scale for external reasons (\S\ref{domains}, also \citet[][\S1]{Held2019}).

From the preceding discussion, we can identify four different possibilities in response to these singularities (following \cite{CrowtherDeHaro}):
\begin{enumerate}
    \item[i.] \label{optionAQFT}\textit{AQFT view}: Singularities in CQFT motivate a different QFT framework, one whose theories are singularity-free, but which does not include gravity (as in AQFT);
    \item[ii.] \label{optionNewPhysics}\textit{New physics}: Singularities in CQFT motivate a new, more fundamental theory at high-energy, and motivate treating our current theories as effective (applicable only at low energy scales), consistent with external motivations for QG;
     \item[iii.] \label{optionEffective}\textit{Effective theory view}: Ignore the Landau poles in the Standard Model and perturbative non-renormalizability of gravity (we shouldn't worry about interpreting them), since we have external reasons for thinking of these as non-fundamental effective theories, to be replaced by QG at high-energy scales (i.e., we appeal only to the external motivations for new physics, and the singularities in current theories do not count as motivations for new physics);
    \item[iv.] \label{optionAsySafety}\textit{Asymptotic Safety view}: UV divergences due to perturbative non-renormalizability and Landau poles do not motivate new physics or a new theory according to the asymptotic safety scenario for gravity and the Standard Model; these singularities do not appear in the full (non-perturbative) theory. 
\end{enumerate}

There is one more response to singularities in CQFT, which has been expressed prominently by \cite{Jackiw1999, Jackiw2000} and \cite{Batterman2011}. 

\begin{enumerate}
    \item[v.]\label{optionEmergent}\textit{Emergent physics view}: Singularities in CQFT are of physical significance, but not motivation for new, more fundamental physics. The singularities are important for facilitating and understanding the emergent, low-energy physics.
\end{enumerate}

Following \cite{CrowtherDeHaro}, we can classify these positions under four different `attitudes' towards singularities . \textit{Attitude 1} means that the singularities are resolved `at the level of QFT', without pointing to a more-fundamental theory or framework. Here, (i) and (iv) fall under Attitude 1. \textit{Attitude 2} means that the singularities are to be resolved by a more fundamental theory of QG, and here (ii) is an example. \textit{Attitude 3} is a positive attitude towards singularities, which means not resolving them at any level, because we have reason to keep them (or reasons why they cannot be removed). Example (v) is a case of Attitude 3, where the singularities are seen as importantly explanatory, or `essential'. Finally, \textit{Attitude 4} means that we do not care about the singularities in question; they are of no significance. (iii) is an example of this attitude. Importantly, one can (and should) take a different attitude towards different types of singularities. This classification also holds when considering spacetime singularities: taking them as pointing to QG is only one possible option (depending on which singularities are being considered).

The QFT divergences, in contrast to the spacetime singularities to be discussed, are not typically appealed to as motivation for QG (though this argument occasionally appears); more commonly the resolution of these is taken as a means of confirmation, or indication of pursuit-worthiness of approaches to QG that resolve these divergences (though we still classify this falling under Attitude 2). The adherence to quantization as the means of solving the problem of QG is consilient with the resolution of QFT divergences: if we quantize gravity, the UV divergences are usually expected to be resolved `for free' (via a minimal length), and this is taken to support approaches to QG that quantize GR. As stated in \S\ref{UVC}, the (alleged) resolution of the UV-divergence of quantum GR by string theory is presented as one of the selling points of string theory.

Some believe that the singularities in GR and QFT are to be cured by the existence of a minimal length, even without quantization of gravity (note, too, that the minimal length need not represent an actual discretisation of spacetime, but may be an operational minimal length, e.g., due to an extended probe).\footnote{E.g., \cite{EllisInfinity} states there is a widespread sentiment among QG physicists, that the singularities in GR and QFT are due to the assumption of a spacetime continuum. For more on the minimal length, see \cite{Hossenfelder2013}.}  \cite{Henson2009} expresses this view, particularly in regards to the need for QG to resolve the divergences associated with the non-renormalizability of GR, and uses it as motivation for the discreteness postulated by an approach to QG known as causal set theory.

\subsection{Geodesic incompleteness}\label{geo}

The definition of spacetime singularities in terms of \textit{geodesic incompleteness} states that a spacetime is singular if and only if it contains an incomplete, inextendible timelike geodesic. Such a geodesic is the worldline of a freely falling test object; the property that makes it singular is that the worldline ends within finite proper time and cannot be further extended. While this definition forms the basis of the Penrose and Hawking singularity theorems, it is not without problems \citep[see][\S 1.1]{Curiel2021}. Geodesic incompleteness appears to be a genuine physical worry, because it means that ``particles could pop in and out of existence right in the middle of a singular spacetime, and spacetime itself could simply come to an end, though no fundamental physical mechanism or process is known that could produce such effects'' \cite[][p.~S140]{Curiel1999}. Geodesic incompleteness leads to a lack of predictability and determinism, and so could indicate that the theory is incomplete \cite[][\S2.6]{Earman1995}.\footnote{Basically, determinism is said to hold if specifying the state of the world at a time together with the laws of nature jointly determine the state of the world at all times. In GR, physicists typically take determinism to hold if the spacetime admits of a well-posed initial value formulation. Cf., \cite{Dob2019, SmeenkDeterminism} for recent work on the problems of defining determinism in GR.} If the breakdown of determinism were visible to external observers, ``then those observers would be sprayed by unpredictable influences emerging from the singularities'' \cite[][p. 171]{Earman1992}. This would represent a nasty form of inconsistency---as Earman puts it, the laws would ``perversely undermine themselves''. 

The breakdown of determinism inside black holes occurs beyond the \textit{Cauchy horizon} (inside the event horizon): beyond this surface, the Einstein equations no longer give a unique solution. In response, \cite{Penrose1979} proposed \textit{strong cosmic censorship} (SCC), which postulates that the appearance of the Cauchy horizon in Schwarzschild black holes is non-generic, and that the interior region of these black holes is in some way unstable (under small perturbation of initial data) in the vicinity of the Cauchy horizon. Any passing gravitational waves would prevent the formation of Cauchy horizons, meaning that instead, spacetime would terminate at a `spacelike singularity', across which the metric is inextendable. 

SCC ensures that no violations of predictability are detectable even by local observers (i.e., an astronaut on a geodesically incomplete worldline would detect nothing up until, and presumably after, her disappearance), and so, the truth of this conjecture would render any singularities (incomplete geodesics) harmless in regards to determinism. As \citet[][p. 5]{StrongCC} states, ``The singular behaviour of Schwarzschild, though fatal for reckless observers entering the black hole, can be thought of as epistemologically preferable for general relativity as a theory, since this ensures that the future, however bleak, is indeed determined''. Thus, SCC may be able to save GR from the charge of incompleteness.

There are two issues that have been raised in the philosophy literature that bear on the question of indeterminism and incompleteness in GR, however. The first regards the nature of the spacetime singularities themselves: (essential) singularities are not located at a \textit{point} of spacetime (if a singularity is located at a point in spacetime, then this is indicative of having a non-essential singularity, i.e., one that can be removed, in the case of incomplete geodesics, by extending the geodesics beyond that point). As \cite{Curiel1999, Curiel2021} emphasises, a singular spacetime does not have any ``missing points'' of spacetime. (For this reason, Curiel and others have stressed the \textit{global}, rather than local nature of singularities). \cite{Earman1995, Earman1996} also uses this argument against the claim that spacetime singularities indicate that GR (alternatively abbreviated as GTR) is incomplete,

\begin{quote}
    In GTR the metric field $g_{ab}$ is now a physical field whose singularity behavior we have to judge. And the judgment has to start from the fact that a general relativistic spacetime $M, g_{ab}$ is such that $g_{ab}$ is defined and differentiable at every point of $M$--there are no singular points of spacetime where the laws of GTR fail to apply.
    
This definitional move may seem facile. In the case of a closed Robertson-Walker spacetime, Brandenberger \textit{et al}. see a kind of incompleteness resulting from the initial and final singularities.``[T]he [final] singularity implies that we cannot answer the question what will happen after the `big crunch' or (in the case of an expanding universe) what was before the `big bang'.''\footnote{\citep[][p. 1629]{Brandenberger1993}.} In response I would pose a dilemma. Either the initial and final singularities are essential--in the sense that it is impossible to extend through them in a way that EFE make sense [...]---or not. If so, then by the lights of GTR, talk about ``before" the big bang and ``after" the big crunch is physically meaningless. If not, then GTR can say something about the before and after. In either case GTR does not stand convicted out of its own mouth of raising meaningful questions it cannot answer. \citep[][pp. 631--632]{Earman1996}.
\end{quote}

\citet[][]{Smeenk2013} makes the same argument in regards to the big bang singularity: that the laws of GR apply throughout the entire spacetime, and there is no obvious incompleteness. If GR is the correct final theory, ``then there is nothing more to be said in regards to singularities'' (p. 634). While Smeenk does not view spacetime singularities as motivation for seeking QG, he states that there are other reasons for doubting that GR is the correct final theory, and good reasons to expect that the successor to GR will have novel implications for singularities---i.e., he recognises \textit{external motivations} for QG, without counting the GR singularities as internal motivation for seeking a new theory.

Although Earman generally advocates a ``tolerance for spacetime singularities''---arguing that we can treat them as \textit{predictions}, rather than pathologies of GR---he nevertheless believes there is one way in which the charge of incompleteness may be justified. This is the idea, described above, that if SCC does not hold, then the determinism of GR is undermined \citep{Earman1995, Earman1996}. Earman ties the determinism of GR to spacetime models that are globally hyperbolic and thus admit of a globally well-posed initial value formulation. The problem with this, however, is a ``dirty open secret'', that determinism in GR fails without help by fiat (i.e., the imposition of ad hoc constraints that simply rule out those spacetimes that do not admit a globally well-posed initial value formulation). 

This leads to the second issue discussed in the philosophy literature---whether GR is a deterministic theory, and whether all models of the theory represent physically possible spacetimes. These questions are explored in \cite{SmeenkDeterminism}, which also highlights a tension between the ``philosopher's conception'' of determinism and the conditions needed for having a well-posed initial value formulation in GR physics. \cite{Dob2019, Dob2020} also discusses the problems of defining determinism in GR, with the former paper arguing for a pluralistic conception. These issues with determinism in GR may take the bite out of the worry that geodesic incompleteness without SCC is a problem motivating QG (though, of course, this depends on one's convictions regarding the need for a deterministic theory).

\subsection{Curvature singularities}\label{curv}

According to the definition of curvature singularities, a spacetime is singular if the curvature, especially the scalar quantities constructed by contracting powers of the Riemann tensor, grows without bound in some region of the spacetime.\footnote{There are also curvature singularities whereby some of the physical components of the Riemann tensor do not have a limit, see, \citet[p.~37]{Earman1995}.} This gives rise to various problems, such as unbounded tidal forces and the lack of consistency of the semiclassical approximation. The semiclassical approximation being referred to here treats GR as an effective field theory, with the Einstein-Hilbert action supplemented with additional higher-curvature terms that represent the quantum corrections to the theory.\footnote{See \cite{CrowtherDeHaro}, \S2.3 for full details.} If we take a model, \textit{M}, of GR with a curvature singularity, and then check whether it is an approximate solution of the equations with the quantum corrections, we find that in general, \textit{M} may be a good solution far away from the singularity, but as we approach the singularity, the higher-order (high curvature) terms will start to dominate over the lowest order term (Einstein tensor). Because of this, \textit{M} will not be a model of the quantum-corrected theory. This shows how the curvature singularities can be used to predict that GR `breaks down', since it reveals an inconsistency between GR and (expected) QG effects that manifest at high curvature in the region close to the curvature singularity.

This is a standard argument found in the physics literature, and physicists tend to find the curvature singularities more concerning than geodesic incompleteness, which has been the focus in the philosophy literature. Notice, however, that here it is not the singularity \textit{itself} that is the problem (depending on the outcome of the charge of indeterminism described above), but the \textit{expected quantum effects} in regimes of high curvature---it is this that motivates the need for a new theory. 

\subsection{Responses to spacetime singularities}\label{horror}

Above, I argued that geodesic incompleteness may not represent an incompleteness of GR, depending on the status of the SCC and/or the assessment of (in)determinism of GR. Curvature singularities, meanwhile, are arguably problematic insofar as they represent domains of extreme curvature where GR is no longer thought to be reliable, and thus do motivate the need for a new theory to replace GR. If we take it that GR is unreliable due to its neglect of quantum effects in these domains (as is the standard interpretation), then this new theory required is QG. 

The dominant attitude towards spacetime singularities---especially the Big Bang singularity and black hole singularities---is as motivation for QG. This attitude treats GR as a classical approximation, an `effective theory', and the singularities in GR signal its limitations as such. Accordingly, singularities should be resolved by a more fundamental theory of QG.\footnote{What exactly is meant by `resolution' in different contexts is an interesting philosophical question. \citet{Thebault2023} explores this question in the context of Big Bang singularity resolution in quantum cosmology.} This attitude is inspired by the history of physics, in particular the suggestion that singularities in classical theories signal the breakdown of these theories and the need for quantization. An oft-cited example in this context is the resolution of the instability of the hydrogen atom in classical electrodynamics, by quantum mechanics. Electrodynamics predicts that the electron radiates energy as it orbits the nucleus and eventually falls into the nucleus, at which point the force becomes infinite. QM solves this by, first, rejecting the classical picture of an electron `orbiting' around the nucleus. Second, by confining the electron to discrete energy levels, and allowing it to emit energy only in discrete packets of energy (photons). And finally, the lowest energy level allowed by the theory is where the electron is, on average, located at a finite distance from the nucleus (the `Bohr radius'), so that it can never fall in. In regards to the spacetime singularities, the idea is that, analogously, gravity should be quantized and that QG solutions have a discrete spectrum, similar to the discrete spectrum of QM---and that, accordingly, singularities can be avoided. 

While this argument by analogy appears often in the literature, it may be criticised on various grounds. \cite{Earman1996} argues against the applicability of this analogy on two points (1) the instability of the atom in this picture comes not from a single, well-confirmed theory, but by combining the Rutherford model of the atom with classical electrodynamics, (2) this combination leads to a prediction that is empirically ``falsified by the fact that the material world exists'' (p. 629). By contrast, Earman states that there is no such definitive conflict with observation in the case of GR. The predictions of GR regarding the formation of black holes in gravitational collapse are meeting with increasing empirical success (though there is more difficulty in assessing the empirical success of Standard Model of cosmology which gives us the the Big Bang singularity). Another line of criticism is expressed by \cite{Tilloy2018}: just because quantization might solve the problem of singularities in GR does not mean it is the only solution, and promoting it as such is, in a way, avoiding the problem by saying it will be dealt with by an unknown, more-fundamental theory. Indeed, by looking to QG rather than seeking to understand singularities in GR `at the level of GR', we may be missing key physical insights of the theory. This attitude is advocated by \cite{Earman1996} (following \cite{Misner1969}), who argues that singularities in GR should be treated as \textit{predictions} of the theory.

In \cite{CrowtherDeHaro}, we find there are four different attitudes one may take towards singularities in GR, the same as those mentioned above for the divergences in QFT. \textit{Attitude 1} for spacetime singularities means that they are to be resolved classically, or `at the level of GR', rather than pointing to QG. \textit{Attitude 2} means they are to be resolved by QG; this is the attitude that dominates the physics literature, and which I have focused mostly on here. \textit{Attitude 3} is a tolerance for spacetime singularities, which means not resolving them at any level, because we have reason to keep them. This is exemplified by Earman's attitude expressed above, where they are treated as predictions of GR, not pointing to QG. \textit{Attitude 4} is an indifference to spacetime singularities: they are not of any significance. We find examples of each of these attitudes in the literature.

The choice of which attitude to take depends on the singularity being considered. We can---and should---adopt different attitudes towards different singularities. The choice also depends, amongst other things, on one's disposition towards the internal versus external motivations for QG, and one's position in regards to scientific realism. For instance, if the spacetime singularities (geodesic incompleteness) are not necessarily problematic for GR, then these would not count as internal motivations for QG. By contrast, in QFT the Landau poles are more readily interpreted as signalling incompleteness of the theories; these are, however, typically dismissed because they occur in regimes where the theory is not thought to be applicable anyway \textit{for external reasons} (e.g., the existence of the Planck scale as a minimal length). The tension between the internal and external reasons for treating a theory as effective is reflected in differing possible attitudes towards singularities in current theories: whether they are to be resolved by a more fundamental theory of QG, whether we should instead look to resolve them `at the level of current theories' (i.e., through developing a new framework for QFT, or modifying GR, etc.). Those who put weight on the external motivations for QG may tend to disregard alternative, internal, possibilities for singularity resolution.

Even if singularity resolution is not, in all cases, a motivation necessitating QG, it can be used as a \textit{guiding principle} inspiring new work. And, if a new theory of QG naturally resolves the problematic spacetime singularities (i.e., if QG describes the `domains' where GR is thought to `break down', or be problematically in deterministic, as well as explains why these occur in GR, and the theory itself not feature any problematic singularities) then, just as in the case of the QFT divergences, this may be promoted as a \textit{means of confirmation}, or an indication of pursuit-worthiness. Making the case for a stronger role, where singularity resolution is taken as a \textit{criterion of acceptance} for any theory of QG, however, requires further argument.

\pagebreak
\section{Black hole thermodynamics}\label{BHT}

The area of research known as \textit{black hole thermodynamics} (BHT) is currently one of the most important in theoretical physics; it suggests, tantalisingly, at a deep connection between GR, QFT and thermodynamics (the `three pillars perspective', \S\ref{threepillars}). BHT lacks empirical access or support, but is the most widely accepted, deeply trusted set of purely theoretical propositions in physics.\footnote{Curiel, fn. \ref{fnCuriel}.} BHT stems from attempts to bring together GR and QFT, but motivates the Primary Motivation, in suggesting the need for a new theory for its explanation. In doing so, it leads to various desiderata and potential constraints on the new theory, which I discuss in this chapter.

BHT is formulated in the framework of semiclassical gravity, whose basis is the semiclassical Einstein equations, which couple classical relativistic spacetime to quantum matter fields (\S\ref{incom}, eq. \ref{eqsemiclass}). As mentioned above, the dominant attitude towards this framework is that it is an approximation to LEQG in the situations where quantum fluctuations of the spacetime geometry can be ignored. Yet, semiclassical gravity, while thought to provide the most secure theoretical clues towards QG through BHT, is itself bedeviled with a plethora of theoretical and conceptual difficulties (\textit{even when} considered as an approximation, rather than a fundamental theory of QG). Because of this, and the absence of empirical support, the amount of trust in BHT as offering insights into the nature of QG may seem a striking---perhaps even suspicious---situation.

Here, I first introduce the basics of BHT and black hole statistical mechanics; \S\ref{HR} discusses Hawking radiation, the theoretical discovery of which was key to accepting BHT; \S\ref{infoparadox} discusses the black hole information loss paradox, which remains a hotly debated topic in physics though its relevance for QG is unclear; and \S\ref{EntHolog} entropy bounds and the holographic principle, which, arguably, offer more secure avenues along which to seek insights into QG.

Classically, there exists a formal analogy between the laws of black hole mechanics and the familiar laws of thermodynamics. Stated simply, the laws of \textit{black hole mechanics} (BHM) are as follows,

\begin{enumerate}
    \item[0] \textbf{Zeroth Law of BHM:} For a stationary black hole, the surface gravity $\kappa$ of the event horizon is constant everywhere on the horizon.
    \item[1] \textbf{First Law of BHM:} Changes in the mass, \textit{M}, of a stationary black hole are related to changes in the horizon area, \textit{A}, angular momentum, \textit{J}, and electric charge \textit{Q} as
    \begin{equation}
        dM=\frac{1}{8\pi}\kappa dA+\Omega dJ + \Phi dQ
    \end{equation}
    \textbf{Second Law of BHM (\textit{aka} the Area Theorem):} The area of the event horizon, \textit{A}, will not decrease in any physical process:
    \begin{equation}
        dA \geq 0
    \end{equation}
    where a physical process is any process which satisfies the null energy condition ($T_{\nu\mu}n^{\mu}n^{\nu}>0$ for null $n^{\mu}$) and during which the event horizon already existing does not `vanish'.
    \item[3] \textbf{Third Law of BHM:} No finite series of operations may reduce the surface gravity of a black hole to zero.
\end{enumerate}

The formal analogy holds when the surface gravity is taken to stand in as thermodynamic temperature, and the horizon area as thermodynamic entropy. However, the formal analogy is now generally believed to be not an analogy at all: rather, it is thought that black holes \textit{are} thermodynamic systems.\footnote{For a much more systematic and thorough account, see \cite{Curiel2021, Wallace2018a, Wallace2018b}. Other useful recent reviews: \cite{Alm2021, Harlow2016}.} This change in attitude came following the (theoretical) discovery of the \textit{Generalised Second Law}, and of Hawking radiation (\S\ref{HR}), which gave reason to attribute non-zero entropy and temperature to black holes. 

Classically, black holes have temperature of absolute zero, and do not possess entropy; according to the \textit{No-Hair Theorem}, they are completely characterised by only three parameters: mass, angular momentum, and electric charge. This leads to the worry that the second law of thermodynamics could potentially be violated in black hole spacetimes. For example, if a package of entropy, such as a cup of coffee, is lost into a black hole, and there were no corresponding increase in entropy of the black hole to compensate for this loss, then the total entropy of the universe (or, more correctly, the total entropy of the black hole plus the relevant external systems) has decreased, and the law has been violated. (There is a second way in which we can imagine the second law being violated, via the \textit{Geroch process} \citep{Bekenstein1972}: see below, \S\ref{EntHolog}). This led \citet{Bekenstein1974, Bekenstein1973, Bekenstein1972} to state that the second law, in the presence of black holes, is ``transcended''.

In response, and by reference to the \textit{Area Theorem} \citep{Hawking1971}, Bekenstein suggested that black holes possess entropy, $S_{BH}$, and that this is proportional to $A$. Following the (theoretical) discovery of Hawking radiation, which states that black holes radiate with temperature $\kappa / 2\pi$ \citep{Hawking1975}, the constant of proportionality has been set, so that,
\begin{equation}\label{SBH}
    S_{BH}=\frac{kAc^3}{4G\hbar}
\end{equation}
(where $k$ is Boltzmann's constant). This is known as the Bekenstein-Hawking entropy formula. Bekenstein also proposed that the (original) second law of thermodynamics be replaced by the \textit{Generalised Second Law} (GSL), which states that the total entropy, $S_{tot}$, including $S_{BH}$ as well as matter entropy, $S_{matt}$ cannot decrease,

\begin{equation}\label{eq:GSL}
    dS_{tot} = (dS_{BH} + dS_{matt}) \geq 0
\end{equation}

A ``dizzying variety of proofs'' have since been produced for the GSL in different regimes, relying on very different assumptions about the nature of the systems described, as well as different mathematical techniques \citep{Wall2009}. As \citet[][\S5.4]{Curiel2021} points out, though, the contradictory assumptions being made across the different proofs may actually threaten the result as tautology, rather than strengthening it as an interesting principle of nature through the consilience of derivations (the derivations, he argues, are \textit{not} consilient).

Nevertheless, the GSL and Hawking radiation results complete the description of a black hole as an ordinary thermodynamic system, rather than merely being analogous to one. Classically, the analogy only worked so long as heat exchanges with other systems were neglected: according to GR, nothing can escape a black hole, it cannot radiate and cannot be assigned a non-zero temperature. Hawking's result relies on \textit{quantum effects} to show that black holes radiate and thus can be assigned non-zero temperature and entropy: it provides a method by which black holes can be put in thermal contact with each other and with other thermodynamic systems. 

Given that black holes can be described as thermodynamic systems, and that all other thermodynamic systems are known to possess an underlying statistical mechanical description, we may speculate that the thermodynamics of black holes, too, is underpinned by a statistical mechanical description. This would imply that a black hole of given mass, charge, and angular momentum, has $\approx exp (A/4G)$ microstates. The statistical mechanical entropy has been calculated in the framework of LEQG, and reproduces the entropy formula, providing evidence that the black hole thermodynamic entropy has an associated statistical mechanical description.\footnote{See \cite{Wallace2018b} for details and original references.} Since then, the recovery of the entropy formula by calculations of statistical mechanical entropy (counting of microstates) has been taken as a \textit{guiding principle}, a \textit{means of confirmation}, and---most conspicuously---a prominent \textit{criterion of theory acceptance} for QG. Various approaches to QG have managed to reproduce this result, for different types of black holes, and to varying levels of precision, including in loop quantum gravity \citep{Meissner2004, Rovelli1996}, and in the context of the AdS/CFT duality \citep{Witten1998, Aharony2004}. However, the most-discussed results are in string theory: \cite{SV1996} showed that the statistical mechanical entropy of a certain type of black hole (known as `extremal' black holes) can be calculated in string theory, and, to leading order, the result matches the area formula. This calculation quickly gained widespread acceptance and is viewed as one of string theory's main successes.\footnote{For details, see \cite{DEHARO2020a, Deharo2020b, Wallace2018b}.} 

The derivation of the Bekenstein-Hawking entropy formula by the counting of microstates is now held as one of the main criteria of theory acceptance for any theory of QG. This is particularly striking given that it is a purely theoretical result that lies far from empirical scrutiny. (Compare with the recovery of GR, the other main criterion of acceptance for QG, which is explicitly motivated by empirical concerns). This raises interesting philosophical questions regarding the scientific method, and the future trajectory of fundamental physics \citep[][\S5.3]{Curiel2021}. 

The results of BHT, however, are certainly tantalising, and, if we take them seriously, do seem to point to the need for a new, more fundamental, theory. This is because there is not a way in GR, nor in semiclassical gravity, to describe a black hole as system whose physical properties arise as gross statistical measures over underlying microstates. Additionally, while Hawking radiation is predicted in the framework of semiclassical gravity (more strictly, \textit{QFT in curved spacetime}, see below), this radiation is not generated by micro-degrees of freedom (the event horizon is treated as a simple, classical geometrical structure). A more fundamental theory appears to be required if we are to explain (the origin of) BHT. Because BHT relies on both GR and QFT effects, its explanation serves as an \textit{external} motivation for QG: it is, in essence a problem `of our own making', rather than one suggested by our best theories as they stand on their own. 

Nevertheless, \citet[][p. 116]{Wallace2018b} argues that the convergence of these various different calculations is ``overwhelmingly'' evidence in support of BHT as underpinned by statistical mechanics---and, furthermore, that this supports both the belief in the correctness of LEQG (``that general relativity can be analysed by QFT at low energies just like any other field theory''), as well as the correctness of the AdS/CFT duality (``so that black hole statistical mechanics is dual to the statistical mechanics of a conformal field theory in at least some important cases''). Wallace's attitude towards the consilience of these derivations thus stands in stark contrast with Curiel's worries in the case of the (``multiplicity and multifariousness'' of the derivations for both the) GSL and Hawking radiation. One reason for this may relate to the inter-theory relations of correspondence and reduction that are expected to hold between QG and LEQG: since QG is expected to be a UV completion of LEQG, it is expected that the two theories share approximately the same results in the domains where they overlap, and because QG is also supposed to be more fundamental (analogous to statistical mechanics as more fundamental than thermodynamics), it is also expected to \textit{explain} the BHT results in LEQG. Of course, as mentioned above, this idea of correspondence cannot---at this stage, however---be an application of the Generalised Correspondence Principle, given that neither of these theories is among our best accepted (empirically successful) physics. Thus, as suggested by \cite{huggettQGLab}, above (\S\ref{semi}), we might still choose the ``epistemically careful'' attitude towards LEQG, as well as the specific relationship between QG and semiclassical gravity.

\subsection{Hawking radiation}\label{HR}

\citet{Hawking1974, Hawking1975} showed that the QFT vacuum state becomes a state with (real) particles present if there is a black hole---i.e., that black holes \textit{radiate}. More precisely, Hawking considered the vacuum state of a scalar field in the presence of an event horizon and found that it evolves into a thermal state of temperature $T$. Because the thermal spectrum of the radiation encodes a Planckian temperature that is proportional to the surface gravity, and the surface gravity plays the role of temperature in the zeroth and first laws of BHT, the temperature is associated with the black hole. Whereas, classically, a black hole is thought to absorb everything and emit nothing (including light), Hawking's result shows that black holes emit thermal radiation with the Planckian power spectrum characteristic of a perfect blackbody at a fixed temperature. It results in energy being radiated away, leading to the eventual `evaporation' of the black hole. A primordial black hole of mass $10^{12}$kg, produced in the early universe, would evaporate around now. The Hawking temperature of a solar mass black hole is $10^{-7}$K and its lifetime is $10^{64}$ years. Figs. \ref{fig:stages} and \ref{fig:penrose}, from \cite{Alm2021}, depict this process.

A non-technical heuristic explanation attributes the Hawking effect to vacuum fluctuations leading to the separation of positive and negative modes through the black hole's event horizon. The QFT vacuum state swarms with pairs of particles that are being constantly created and annihilated. If such a pair forms near the event horizon, one of the entangled partners may fall into the black hole, while the other escapes to infinity---the separation of the pair precludes their otherwise typical recombination, and results in the exterior region having an effective excess energy (real particles). The escape of the excess particles to the asymptotic region is then what gives rise to outgoing radiation.\footnote{Note, however, that while this is the standard heuristic picture described, it is quite imprecise, and does not, for instance, allow for a distinction between the Hawking effect and the Unruh effect, cf.\ \cite{barbado2016hawking}.}

Hawking's original calculation was done using the framework of \textit{QFT in curved spacetime}.\footnote{See \cite{Wald1994}.} This is a non-trivial, but consistent, theory of quantum fields on the background of a classical relativistic spacetime, which differs from semiclassical gravity in its neglect of the backreaction of matter on the curved spacetime. This neglect is justified in general if the relevant length scales are large and curvature effects small. Thus, we expect this framework to naturally apply in the cases where the curvature of spacetime is well above the Planck length, and where we can assume (i.e., we have some theoretical grounds for expecting) that we can ignore any quantum properties of spacetime itself. Nevertheless, the empirical support for QFT in curved spacetime is both very indirect as well as highly nonspecific, since it comes not from indirect measurements concerning the predictions of particular QFTs in curved spacetime, but rather from claims that particular QFTs in curved spacetime show the right behaviour in \textit{various limits}.\footnote{This means that same-limit-behaviour might be reproducible by myriad theories other than QFTs in curved spacetime.} An example is the geometric optics limit of electrodynamics in curved spacetime, which is a limit of \textit{quantum} electrodynamics in curved spacetime, describes light rays as tracing out null geodesics given sufficiently high frequency relative to the curvature scale. That light indeed moves on null geodesics in curved spacetime has been observationally verified, e.g. in gravitational lensing effects (see \cite{DysonGravitationalLensing}). The prediction of Hawking radiation is one of only two main specific predictions that could serve as a test for this framework.\footnote{The other is the prediction of a specific primordial density perturbation spectrum associated with cosmic inflation scenarios. For criticisms of this as a possible confirmation of QFT in curved spacetime, see \cite{CLW}.}

\makebox[0pt][l]{%
\begin{minipage}{\textwidth}
\centering
    \includegraphics[width=\textwidth]{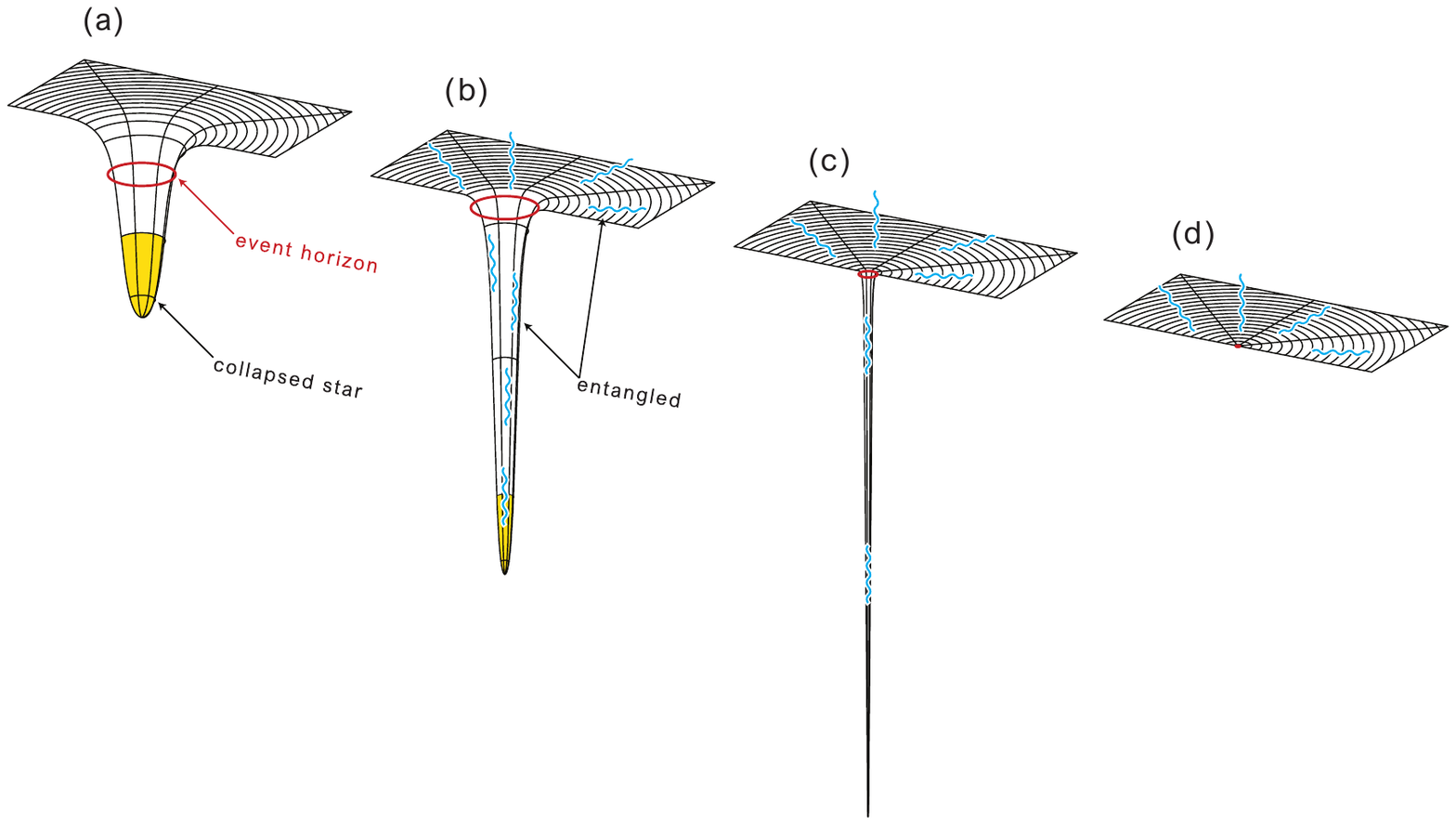}
    \captionsetup{width=0.8\textwidth}
 \captionof{figure}{\textit{The Hawking process of black hole evaporation}: (a) Stellar collapse: the interior geometry continues to elongate in one direction, while pinching towards zero size in the angular directions; (b) Hawking process creates entangled pairs of particles, with one partner trapped behind the horizon and the other escaping to infinity where it is observed as blackbody radiation. The black hole shrinks as its mass is slowly carried away; (c) Eventually, angular directions shrink to zero, as does the event horizon; (d) Final stage: a smooth spacetime containing thermal radiation but no black hole.}
 \label{fig:stages}
\end{minipage}
}

\medskip

\makebox[0pt][l]{%
\begin{minipage}{\textwidth}
\centering
    \includegraphics[width=\textwidth]{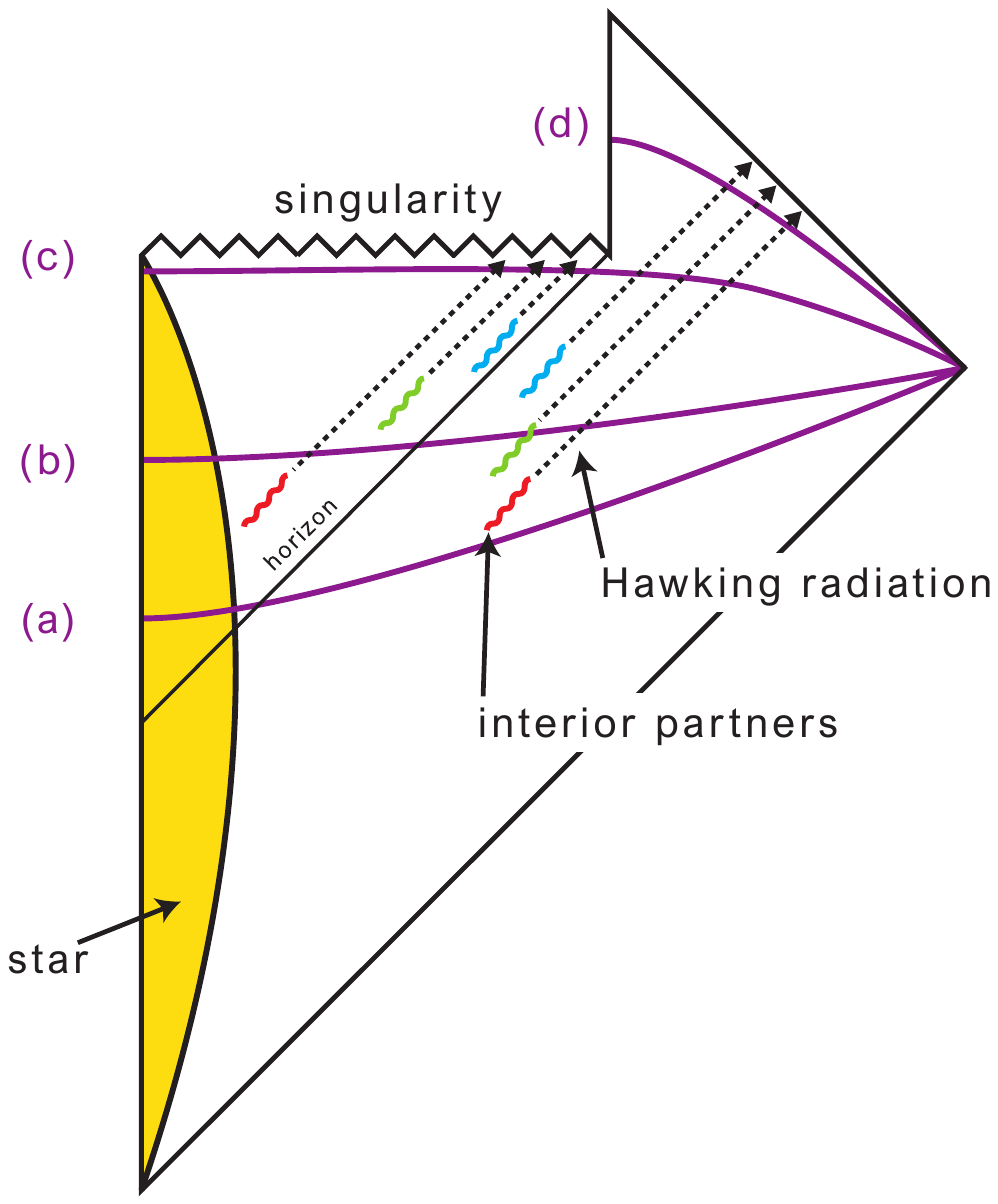}
     \captionsetup{width=0.8\textwidth}
 \captionof{figure}{Penrose diagram for the formation and evaporation of a black hole. Spatial slices (a)-(d) correspond to the slices drawn in Fig. \ref{fig:stages}.  \citep{Alm2021}.}
 \label{fig:penrose}
\end{minipage}
}

\medskip

 Another problem with Hawking's original derivation is that it involves field modes whose wavelengths become shorter than the Planck length near the black hole horizon---this `trans-Planckian' regime is beyond the domain of applicability of QFT in curved spacetime, and so we should not expect to trust its results there (instead, requiring a theory of QG to describe this regime). The problem arises due to an exponential gravitational redshift assumed to occur at the horizon, which means that the black hole radiation detected at late times (i.e., the outgoing particles) corresponds to extremely high (divergent) frequency when traced back to the horizon. 

Yet, there are at least four other derivations of the Hawking effect in addition to Hawking's original method \citep{Wallace2018a}. Together, Wallace claims, these strongly suggest that Hawking radiation is a real consequence of QFT in curved spacetime. \cite{Curiel2021}, on the other hand, argues that not all these derivations are consilient\footnote{Curiel also notes that this is not standard consilience, since it is not the prediction of the value for an experimentally observed quantity (say), by way of different mechanisms, but rather the prediction of the same (purely theoretical) phenomenon by different theoretical routes.}, given that some appear to rely on assumptions that contradict assumptions made by others: leaving the physical content of the prediction of  Hawking radiation obscure at the foundational level. Curiel argues against identifying Hawking radiation with blackbody radiation given that, in the semiclassical picture, it is not generated by micro-degrees of freedom on the event horizon---emphasising, instead, that we need further assumptions to bridge the derivation of the statistical mechanical entropy in LEQG (which does, at least naively, appear to treat the quantum microstates as located in a thin skin around the event horizon) with the derivation of Hawking radiation (which describes quantum fields external to the horizon).\footnote{Curiel calls this `the temperature decoupling problem'. Cf. fn. \ref{fnCuriel}.}

Similarly, while Wallace argues that we can understand QFT on curved spacetime as an application of the general machinery of modern QFT, \citet[][\S6.1]{Curiel2021} highlights not only a profound difference between the two  but also calls attention to the fact that QFT itself (even in the flat spacetime of special relativity) is fraught with difficulties.\footnote{See, e.g., \citet{FraserThesis, Kuhlmann2023}.}

While the detection of Hawking radiation may not reveal anything of its fundamental nature, it is nevertheless fair to say that an empirical test of Hawking radiation is required in order to entrench BHT. Yet, the Hawking radiation of typical astrophysical black holes has an extremely weak empirical signature, making its detection a near impossibility. As a black hole's temperature $T$ is proportional to its surface gravity $\kappa$---and, thereby, inversely proportional to its mass $M$---the temperature of the radiation becomes smaller the bigger the black hole considered. Consequently, black holes detectable in astronomical settings would emit radiation of an order several million times smaller than the temperature of the cosmic microwave background \citep{Thebault2019}.

\subsection{Black hole information paradox}\label{infoparadox}

If black holes radiate---if they are ordinary statistical mechanical systems obeying standard QM---then we would expect that their radiation and eventual evaporation be described as a fully unitary process, in accordance with QM. Yet, the calculations of the Hawking effect in QFT in curved spacetime describe a process that appears strongly non-unitary. This leads to what is known as the `information loss paradox'. In this context, there are two different problems that have been discussed: the `final state problem' (or `evaporation time problem') and the `Page time problem'; I briefly introduce each in turn.

Quantum systems are subject to \textit{unitary evolution}: the initial conditions together with the Schr\"{o}dinger equation fix the future state of the system. This means that unitary evolution `preserves information': if we start with a quantum system in a known state, then quantum theory ensures that the system will evolve in such a ways that we can infer the state of the system at some later time, and vice versa---the quantum evolution takes a pure state to a pure state, and this process is reversible. This is challenged by black hole evaporation. If we drop a cup of coffee into the black hole, information about its state still exists (though inaccessible) beyond the horizon, but once the black hole completely evaporates, all that is left is Hawking radiation (as shown in Fig. \ref{fig:stages} (d)). Hawking radiation is perfectly thermal, meaning that the `information' about what was dropped in cannot be recovered: we cannot infer the previous state of the system given the final state of the Hawking radiation. Another way of seeing the problem is to imagine forming a black hole through the collapse of a pure state (such as a large amplitude gravity wave), and then noting that at the final state of evaporation is a \textit{mixed state}: to an outside observer, the process of black hole formation and evaporation is a pure-to-mixed, irreversible transition. In other words: the evolution is \textit{non-unitary}.

To see why we end up with thermal (mixed state) radiation despite starting from an initial pure state, recall that the effect arises from considering the QFT vacuum in the presence of a horizon. In doing so, we are essentially splitting the original vacuum state into two parts: one interior and one exterior. While the QFT vacuum state is a pure state overall, its degrees of freedom are entangled at short distances, which means that when we consider only half of the space, we find a mixed state on that half. In the heuristic picture described above (\S\ref{HR}) involving pair creation, the two particles are entangled with one another, forming a pure state, but when we consider only the outgoing radiation, we find it in a mixed state looking like a thermal state at the Hawking temperature. [The presentation of the information loss paradox in terms of `information' thus risks being a bit misleading---the basic problem has really to do with the entanglement, or \textit{correlations}, between matter within the black hole and matter far outside the black hole].

One difficulty with this story, however, is that the final stages of black hole evaporation occur when the black hole's curvature is of the Planck scale, and so the semiclassical description that gives rise to the `paradox' is not itself valid here. It cannot tell us what actually happens in the final stages of evaporation. There are various proposals that attempt to exploit this `loophole' in order to save unitarity: for instance by having the `information' ejected in the final bursts of energy, or by having it encoded in a non-radiating Planck scale `remnant' black hole, or something even more exotic \citep{Hawking1993}. None of these proposals appear satisfactory \citep{Marolf2017, Unruh2017}.

For GR physicists, and most philosophers, this black hole `final state problem' is not really seen as a paradox---or, indeed, a problem.\footnote{Authors who argue this view include, e.g., \citet{Belot1999, maudlin2017, Unruh2017, Wald2001}. Cf.  \cite{Manchak2018}.} On this view, `the problem' is simply an argument to the conclusion that the black hole evaporation is a non-unitary process (this is often supported, for instance, by claiming that that the black hole evaporating spacetime is not globally hyperbolic, and we do not expect unitary evolution in such spacetimes). Early on, Hawking himself \citep{Hawking1976} stated that the full theory of QG will be one that violates the standard dynamical principle of quantum evolution. Hawking later changed his view, persuaded by arguments in favour of the holographic principle, that `information' is stored on the event horizon, \S\ref{EntHolog}. The opposing side in this debate holds that the evolution of an initial pure state to a final mixed state is in conflict with QM. Arguments to this effect are mainly found from within the `particle physics perspective' on QG (typically, string theorists), which treats unitary evolution as an inviolable principle. Why exactly this need be so, however, is debatable. 

The other main form of the `information loss paradox' is the `Page time problem'. Here, the problem occurs not in the final stages of evaporation where we expect the semiclassical theory to be invalid anyway, but rather in a regime where the black hole is still big, and where we have no reason to expect the semiclassical description to be incorrect. To understand this paradox, we need to first distinguish between two notions of entropy that we ordinarily use in physics.\footnote{This presentation follows \cite{Alm2021}.} The first is the \textit{von Neumann entropy}, $S_{vN}$, also called the \textit{fine-grained entropy}, or `entanglement entropy'. Given the density matrix, $\rho$, describing the quantum state of the system, it is defined as,

\begin{equation}\label{eq:VN}
    S_{vN}=-\text{Tr}[\rho \textbf{log} \rho]
\end{equation}

This equation quantifies our ignorance about the precise quantum state of the system. It vanishes for a pure state, indicating our complete knowledge of the quantum state. Importantly, it is invariant under unitary time evolution, $\rho\rightarrow U\rho U^{-1}$. 

The second notion of entropy is the \textit{coarse-grained entropy}, $S_T$, also known as the \textit{thermodynamic entropy}. In this case, we have the density matrix $\rho$, but instead of measuring all observables, we measure only a subset of simple, or coarse-grained, observables, $A_i$. We then consider all possible density matrices $\Tilde{\rho}$ that give the same result as our system for the observables we are tracking, and then compute the von Neumann entropy, $S(\Tilde{\rho})$. Finally, we maximise this over all choices of $\Tilde{\rho}$. A simple example is the ordinary entropy used in thermodynamics: here, $A_i$ are often chosen to be a few observables, e.g., the approximate energy and the volume, and then the thermodynamic entropy is obtained by maximising the von Neumann entropy among all states with that approximate energy and volume. The coarse-grained entropy obeys the second law of thermodynamics: it tends to increase under unitary time evolution. The generalised entropy, $S_{tot}$ in eq. \ref{eq:GSL}, comprising the entropy of the black hole plus the entropy outside---if it is to be either of these two entropies---would be the coarse grained, thermodynamic entropy \citep{Alm2021}.

The von Neumann entropy cannot be larger than the thermodynamic entropy: to see this, realise that we can always consider $\rho$ a candidate $\Tilde{\rho}$. Or, in other words, because the thermodynamic entropy provides a measure of the total number of degrees of freedom available to the system, it sets an upper bound on how much the system can be entangled with something else. 

\begin{equation}
    S_{vN} \geq S_T
\end{equation}

Now, consider the black hole as an ordinary thermodynamic system. The von Neumann entropy of the matter that formed the black hole will be extremely small compared to the black hole's initial thermodynamic entropy.\footnote{Although it's implausible to think that the stellar precursor to an astrophysical black hole is in a pure state, the thermodynamic entropy of such a precursor is typically negligible compared to the entropy of the black hole that forms from it, so actually the initial purity of the system state plays no essential role here provided that the initial state’s von Neumann entropy is much lower than its thermodynamic entropy \citep{Wallace2020}.} The system cools through the emission of thermal radiation---i.e., through the emission of quanta of radiation in highly mixed states. In the early stages of black hole evaporation, the von Neumann entropy will be almost exactly thermal because the radiation is entangled with the system. [If the original state of the black hole, considered as a thermodynamic system, is pure, and its dynamics are unitary, then the total state of the system plus radiation is pure. But if the radiation is thermal, this means each emitted particle is in a mixed state, and so must be entangled with some other system (in this case, the black hole) in order for the total state to be pure (no two emitted quanta can be entangled in thermal radiation)]. As the system cools, the von Neumann entropy of the radiation rises: the entanglement of the emitted particles increases. However, as the black hole radiates, its area shrinks, and at some point it will not be possible for the emitted radiation to be thermal anymore. This is because the number of degrees of freedom of the black hole is given by its thermodynamic entropy---i.e., the area of its horizon. When the area is sufficiently small, it will no longer be possible for the outgoing radiation to be entangled with the system, and at this point, the von Neumann entropy of the radiation will decrease along with the thermodynamic entropy of the black hole.

In summary: if, as in accordance with BHT, black holes are ordinary thermodynamic systems that can be described, by an external observer, as a quantum system with $e^{A/4G}$ microstates, that evolves unitarily, then---from the argument above---the von Neumann entropy of the Hawking radiation needs to decrease at some point in order that it not exceed the thermodynamic entropy of the black hole. This argument was made by \citet{Page1993}, who suggested that the entropy of the radiation would need to follow the curve (the \textit{Page curve}) indicated in Fig. \ref{fig:page}, rather than the Hawking curve. The entropy of the radiation at early stages increases until it equals the decreasing thermodynamic entropy of the black hole, and then the two remain equal for the rest of the evaporation process. The time at which the von Neumann entropy of the radiation is maximal (the turnover point), is known as the \textit{Page time}. The Page time for a Schwarzschild black hole is approximately half the total evaporation time: at this point, it is not possible for the black hole's radiation to be thermal, and instead it should be maximally entangled with the early-time radiation.

\makebox[0pt][l]{%
\begin{minipage}{\textwidth}
\centering
    \includegraphics[width=\textwidth]{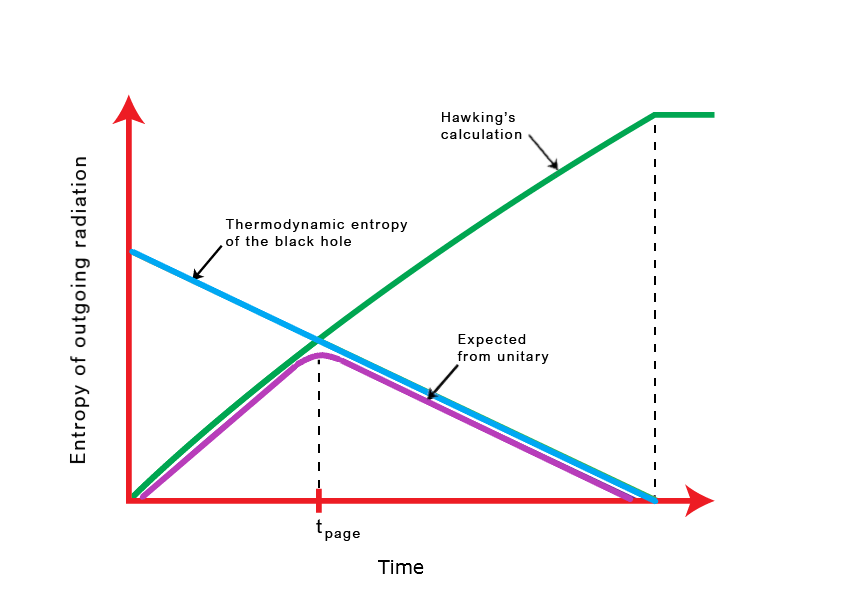}
    \captionsetup{width=0.8\textwidth}
 \captionof{figure}{The Page curve: Schematic behaviour of the entropy of the outgoing radiation. In Hawking's calculation (green), the entropy monotonically increases until \textit{t}\textsubscript{end}, when the black hole completely evaporates. If the process is unitary, the entropy of radiation must be less than the thermodynamic entropy of the black hole (blue), and follows the Page curve (purple) it saturates this maximum at \textit{t}\textsubscript{page}.}
 \label{fig:page}
\end{minipage}
}

\medskip

Now, the problem is that this result is in clear conflict with the predictions regarding Hawking radiation in QFT in curved spacetime. In these calculations, the Hawking radiation is \textit{exactly} thermal, with no entanglement between the early-time and late-time emitted particles. Indeed, there there is \textit{no possibility} of such entanglement on this picture, given that each mode of the Hawking radiation is maximally entangled with a partner mode inside the event horizon (a principle known as the \textit{monogamy of entanglement} ensures that these modes cannot be entangled with others), which has no means of escaping.

The Page time problem is indeed a paradox: a clash between the predictions of QFT in curved spacetime and the predictions of black hole statistical mechanics. As stated, it occurs when the black hole is still macroscopic in scale, a regime where \textit{both} these frameworks are expected to be valid, and so closes the `loophole' left open by the evaporation time problem (where we could attempt to appeal to unknown, exotic QG effects in the Planckian regime). \textit{Prima facie}, there are two ways to resolve the paradox:

\begin{enumerate}
    \item Accept that QFT in curved spacetime fails as a description of the entire spacetime of an evaporating black hole; retain the statistical-mechanical underpinnings of black hole thermodynamics; try to understand why and when the QFT description breaks down.

    \item Retain QFT in curved spacetime, reject black hole statistical mechanics, and find some non-statistical-mechanical understanding of BHT.
\end{enumerate}

\citet{Wallace2020} presents these two options as each packaged with their own explanatory burdens. Option 1, according to Wallace, necessitates explaining when and \textit{why} the QFT description breaks down, \textit{given that} the breakdown occurs in regimes ``well within the applicability domain of the theory''. Option 2 necessitates explaining not only \textit{why} we have such a striking theoretical analogy between BHT and ordinary thermodynamic  systems, but also \textit{why} we have a managed to reproduce the thermodynamic entropy of black holes in statistical-mechanical calculations in various different contexts. Failure to address the compatibility of this vast network of calculations, he says, would leave these results ``miraculous''. 

Again, though, we might follow Curiel in questioning the necessity of taking on these additional explanatory demands in order to reject one or the other framework, given the tenuous theoretical status of each. In the case of option 1, we might need only to appeal to the fact that ``semiclassical gravity is an effective field theory with no empirical support---deal with it''. Or, more seriously, it may be philosophically fruitful to question whether, or why, we should be compelled to believe that results developed in the framework of semiclassical gravity are revelatory of genuine insights into a more fundamental theory---given not only the problems associated with semiclassical gravity, but also due to the very nature of effective field theories in general (this is not to imply, of course, that Wallace does not provide sound arguments for his trust in LEQG in various non-Planckian regimes, see \cite{Wallace2022}). In the case of option 2, we might again wish to look more closely at the `consilience' of derivations and explore the inter-theory relations between the various frameworks, and/or bemoan the underavailability of physical assumptions bridging the frameworks generating the statistical-mechanical results with the BHT ones. 

Rather than fuel debate here, my goal is to encourage more philosophical interest in exploring these issues. Whichever of these routes one chooses to pursue, it is likely that interesting philosophical insights will be gained---whether concerning QG, BHT, semiclassical gravity, QFT, effective field theory, or more general issues in the philosophy of science such as inter-theory relations, scientific methodology, and scientific progress. In regards to the specific questions being considered in this \textit{Element}, we can ask how the information loss paradox should be used in motivating and constraining the search for QG. It is a theoretical puzzle that arises by taking seriously our best current theories, developing natural ways of combining them, and then pushing these until breaking point. Surely the paradox serves as an extremely stimulating heuristic and guiding principle, and any approach towards QG that indicates a solution could take this as an indication of pursuit-worthiness, or means of confirmation (given that it is an indication of \textit{consistency} of the approach with low-energy theoretical results). At this stage, however, the resolution of the paradox via new fundamental physics should not be taken as a constraint (criterion of acceptance) on a theory of QG: a \textit{dissolution} is not ruled out, neither a solution `at the level of current theories'.

On that note, I close this section by gesturing to the wide variety of different solutions to the paradox being explored in physics. Particularly noteworthy are recent results involving \textit{replica wormholes}, which show not only that entanglement entropy follows the Page curve,\footnote{This is consisistent with other results in the context of AdS/CFT, which has convinced most in the high-energy physics community that unitarity is not violated in black hole evaporation.} but also suggest a possible mechanism that allows the `information' to escape via wormholes\footnote{\cite{Alm2020, penington2020}.} (perhaps as a proxy for nonlocal effects\footnote{\cite{Marolf2020}.}). Exactly what this could mean physically, and what it could imply for QG, is still very much open to speculation.

\subsection{Entropy bounds}\label{EntHolog}

Various theoretical results in BHT suggest that there is an upper bound on how much entropy can be contained in a given region of spacetime. There are several different proposals for the entropy bound, with varying degrees of generality. These results are striking, and in conflict with the picture of the world presented by current theories---the explanation of the entropy bound is motivation for a new, more fundamental, theory.

The idea of an entropy bound originally came from considering the \textit{Geroch process}, which is a mechanism via which the GSL can apparently be violated.\footnote{Here following \citet[][\S5.4.2]{Curiel2021}, but see, e.g., \cite{Bekenstein1972, Bousso2002} for more details.} Imagine we prepare a massless box full of energetic radiation with high entropy, far away from the black hole. The mass of the radiation will be attracted by the black hole's gravitational force. We could then use this weight to drive an engine to produce energy while slowly lowering the box towards the event horizon of the black hole---this process extracts energy (work), but not entropy, from the radiation in the box. If we keep slowly lowering the box towards the event horizon, we can arrange for all the mass-energy of the radiation to have been exhausted when the box reaches the event horizon. Dropping the box into the black hole, then, will not increase the size of the event horizon (because the mass-energy of the black hole does not increase), but will mean that the thermodynamic entropy outside the black hole has decreased.

In response, Bekenstein proposed that there must be a limit to how much entropy can be contained in a given region of spacetime \citep{Bekenstein1981}. The \textit{Bekenstein bound} states that maximum entropy contained in a spherical system of radius \textit{R} and total gravitating energy \textit{E} is $S \leq \frac{2\pi ER}{\hbar}$. For a weakly gravitating system, the Bekenstein bound reduces to $S \leq \frac{A}{4\hbar G}$, where \textit{A} is the area of a spherical surface fully containing the system. 

If we assume, however, that the Bekenstein bound holds also for strongly gravitating systems, we can arrive at a more general conjecture known as the \textit{spherical entropy bound}, using the GSL and considering the \textit{Susskind process}, where, instead of lowering a thermodynamic system into a black hole, we \textit{convert} the system into a black hole \citep{Susskind1995}. This bound states that the maximum entropy contained in a region of space is $S \leq \frac{A}{4\hbar G}$, where \textit{A} is the area of any surface bounding the region. To see why this is, imagine that we find a region with entropy greater than $\frac{A}{4\hbar G}$, then we could add energy to it adiabatically without changing its entropy, and thereby transform it into a black hole of area \textit{A}. Such a black hole would then have entropy greater than the Bekenstein-Hawking entropy, (eq. \ref{SBH}). Thus, if the Bekenstein-Hawking entropy is correct, the entropy of a black hole of surface area \textit{A} must be the maximum possible entropy of any region of space with surface area \textit{A}. By definition, then, the entropy of a single Schwarszchild black hole precisely saturates the spherical entropy bound: a black hole is the most entropic object we can put inside a given spherical surface \citep{tHooft1993}. This form of the entropy bound is associated with the \textit{holographic principle}, discussed below.

The heuristic derivation of the spherical entropy bound rests on a large number of fairly strong assumptions (e.g., suitable asymptotic conditions, spherical surface, and gravitational stability of the region). Additionally, the bound suffers several counter examples, so does not have general validity. \citet{Bousso1999} proposed a successful generalisation of this bound, known as the \textit{covariant entropy bound}. Heuristically, we can imagine a spherical star with nonzero entropy, $S_0$ that burns out and undergoes catastrophic gravitational collapse. For an outside observer, the star will form a black hole whose surface area will be at least $4S_0$ in accordance with the GSL. But we can imagine following the star as it falls through its own horizon, and watch it shrink to zero radius as it crushes to a singularity: $A\rightarrow0$. But, the entropy of the star must still be at least $S_0$. This is a case where the spherical entropy bound does not apply, being a regime of dominant gravity. But we can imagine taking a snapshot of the star just before it crunches to a point: light does not have time to cover the entire area of the star before it completely crunches, and so the entropy of the area traversed by the light will always be less than the area of the star's surface. The covariant entropy bound is formulated in terms of \textit{light sheets}, where a light sheet is a way of taking a `snapshot' of a matter system in light-cone time (so the entropy on a light sheet is given by the entropy of the matter system). More technically, as illustrated in Fig. \ref{fig:lightsheet}, a light sheet associated with a surface is a set of null geodesics leaving the surface orthogonally such that the expansion of the set in the direction going away from the surface is zero or negative, i.e., the geodesics are remaining parallel or coming closer together as they get further from the surface. The light sheet continues up until the geodesics intersect or encounter a singularity of spacetime.

\medskip 
\makebox[0pt][l]{%
\begin{minipage}{\textwidth}
\centering
    \includegraphics[width=\textwidth]{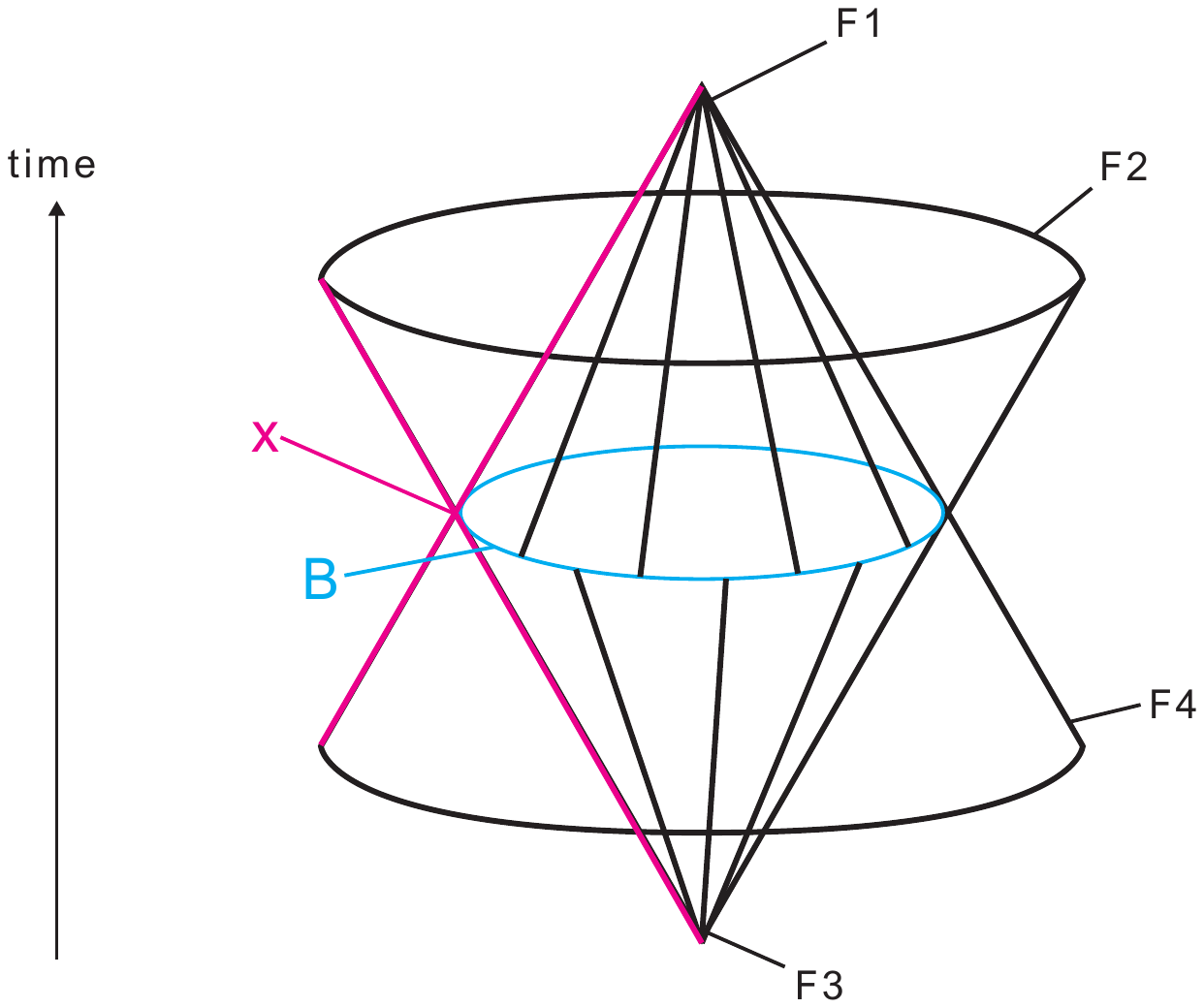}
     \captionsetup{width=0.8\textwidth}
 \captionof{figure}{Bousso's light sheet construction, depicting four null hypersurfaces orthogonal to a spherical surface, \textit{B} (also, the four null geodesics orthogonal to a point \textit{x} on \textit{B}, to make the illustration clearer). The two cones $F_1$, $F_3$ have negative expansion and hence are light sheets. The covariant entropy bound states that the entropy of each light sheet will not exceed the area of \textit{B}. The other two families of light rays, $F_2$, $F_4$ give the `skirts'; these have increasing cross-sectional area, and so are not light sheets and their entropy does not relate to the area of \textit{B}. \citep{Bousso2002}.}
 \label{fig:lightsheet}
\end{minipage}
}

\medskip

The covariant entropy bound simply states that the entropy on a light sheet associated with a surface \textit{B} will not exceed the area of \textit{B}, i.e., $S\leq A/4G$. This bound appears to hold universally in nature, at least at the semiclassical level---it holds for arbitrary physically realistic matter systems and arbitrary surfaces in any spacetime that is a solution of Einstein's equations. This bound reduces to the spherical entropy bound in special cases where the latter is known to hold. In addition, \cite{Bousso2002} argues that the covariant entropy bound \textit{implies} the GSL, i.e., that the GSL can be regarded as a consequence of the covariant entropy bound.

Yet, as \cite{Bousso2002} emphasises, the origin of the bound remains mysterious: aside from its success, little motivation for a light-like bound can be offered. Although the discovery of the covariant entropy bound was initially inspired by BHT (via the need for a generalisation of the Bekenstein bound, which itself was proposed in order to save the GSL), the covariant entropy bound is no longer connected, via derivation, to these discoveries, and stands on its own. Essentially, the principle has outstripped its original motivations. [Here, it is interesting to recall, as well, that BHT has strange motivations itself. The analogy of the area theorem with the second law of thermodynamics gives perhaps unreasonable weight to the second law itself, which is known not to be fundamental, but open to statistical violations: its appearance is due to perhaps contingent features of our universe].  We cannot look to its mode of discovery in order to justify the covariant entropy bound: instead, its justification is that it holds true in all calculations.

That there could be a maximal bound on entropy inside a region of spacetime is itself a striking result that conflicts with our current picture of the world: according to QFT, the number of degrees of freedom in any given region of spacetime is infinite, so this result calls into question the fundamental status of field theory. But the entropy bound is more incredible given its form: it implies that the number of degrees of freedom grows in proportion to the area surrounding the region, rather than the volume of the region. This clashes with the standard dynamics of all other known types of physical systems, where entropy increases as the volume of any spatial region. Notably, even if QG describes a minimal length (e.g., through a discretisation of spacetime), which could serve as a short distance cutoff for QFT, the information content of a given region would still be expected to grow with the volume.

As with the other results coming from BHT, however, it is precisely the \textit{inexplicability} of the entropy bound that excites researchers: its origin is expected to be found in a new theory. Because the bound relates aspects of spacetime geometry to the number of quantum states of matter (it involves the fundamental constants of both \textit{G} and \textit{h}), the suggestion is that any theory that explains the bound will be a theory of QG: a theory that `combines' GR and QFT. The theory will also need to explain why it is that field theory (QFT) gives the wrong answer. Still, however, it is possible that established physics is correct, and that the entropy bound result is an artefact of the semiclassical approximation. While the entropy bound is promoted as holding true in all calculations, we have to keep in mind that these calculations are still remote from empirical tests. 

We can ask how the entropy bound can, or should, be used in the search for QG. \citet{Bousso2002} suggests that the bound may be useful as a \textit{guiding principle} in the context of string theory. This is because the bound holds in situations where string theory suffers limitations, for instance in highly dynamical geometries, as well as in the non-perturbative regimes of string theory: the hope is that the bound can help aid the development of string theory in these areas. (The AdS/CFT duality is an example of a non-perturbative formulation of string theory where the entropy bound, as well as the holographic principle is satisfied; the idea is to find others, which could physically represent our universe). It seems possible that the covariant entropy bound (or the holographic principle) be used as a \textit{postulate} for a candidate theory of QG, similar to how approaches such as causal set theory take \textit{discreteness} as a postulate, for instance. The explanation of the bound can, and has, been appealed to as a \textit{means of confirmation} for QG. Calculations in causal set theory, for instance, give evidence for the existence of an entropy bound which leads to Susskind's spherical entropy bound in the continuum \citep{Rideout2006}. Finally, at this stage, it seems difficult that the explanation of the entropy bound be taken as a \textit{criterion of acceptance} for QG, given its status in semiclassical gravity. The use of the entropy bound in various roles may also be influenced by the attitude we take towards it: \citet{Adlam2023} discusses arguments for and against the epistemic and ontological interpretations of the entropy bound, highlighting the different commitments, challenges, and implications of each.

\subsection{Holographic principle}\label{Holog}

The most common understanding of the holographic principle is based on the idea that the description of a volume---`bulk'---of space can be equivalently defined on a surface bounding the volume of space. In other words, a physical theory defined only on the (\textit{N}-1)-dimensional boundary of the region it encloses completely describes the \textit{N}-dimensional physics of the bulk. The principle is then the claim that the full theory of QG can be reformulated as a theory all of whose degrees of freedom are defined on the boundary---a concrete example is the AdS/CFT duality, mentioned above (p. \ref{AdS}). The idea, originally from \citet{tHooft1993}, was popularised, and given a precise interpretation in string theory, by \citet{Susskind1995}. The picture is that ``The three-dimensional world of ordinary experience––-the universe filled with galaxies, stars, planets, houses, boulders, and people–-–is a hologram, an image of reality coded on a distant two-dimensional surface'' \citep[][p. 410]{Susskind2008}. Susskind's presentation, following t'Hooft, is based on a response to the information loss paradox, where the suggestion is that the `information' is stored on the black hole horizon. 

There are slightly different statements of the holographic principle, and its relation to entropy bounds. \citet{Bousso2002}, for instance, takes the holographic principle as simply the claim that the origin of the entropy bound is to be found in a new theory of QG. \citet{Smolin2001}, states that, while entropy bounds provide a limit on the number of degrees of freedom in a region of space, the holographic principle is supposed to go further in telling us about the nature of those degrees of freedom, by postulating a form of dynamics in which the quantum evolution of the spacetime and matter fields is described in terms of observables measurable on the screens (surfaces). ``The holographic principle is not a relationship between two independent sets of concepts: bulk theories and measures of geometry vs. boundary theories and measures of information. Instead, it is the assertion that in a fundamental theory the first set of concepts must be completely reduced to the second'' \citep{Smolin2001}. These claims suggest, however, that the more fundamental principle is the entropy bound itself, rather than the holographic duality, given that the latter is supposed to explain the former. 

\citet{JakslandLinnemann} usefully distinguish between the holographic duality and entropy bounds, in that the former is \textit{inter-representational}, while the latter are \textit{intra-representational}. The duality is inter-representational because it describes a relationship between different equivalent representations of the same physics, i.e., it relates two \textit{separate descriptions} (different theories): one on the `bulk' and one on the `boundary'. In the case of the AdS/CFT duality, for instance, the duality relates a string theory in \textit{N}-dimensional spacetime `bulk', to a gauge theory (CFT) formulated on the (\textit{N-1})-dimensional `boundary'. On the other hand, the entropy bounds apply only in a single theory, and are thus intra-representational. For this reason (though as should already be evident from the characterisation of the holographic principle, above), the existence of an entropy bound does not imply the holographic principle. Nevertheless, the holographic principle does suggest an entropy bound in the bulk theory, and its aim is to motivate or explain the existence of the bound. The main question is, then, how the holographic duality can be used to develop a theory of QG.

One suggestion is made by \citet{deHaro2018}, which utilises an account of dualities as isomorphic representations of a single ``common-core'' theory.\footnote{Cf. \citet{deHaro2021, Rickles2017}.} We can then distinguish between different functions of dualities in theory-construction. The \textit{theoretical function} of dualities is their use in developing the common-core theory that is represented on both sides of the duality: the idea is to utilise the dual theories, and exploit the relationship between them, to extract the content of this common physics that is being represented differently by each. The \textit{heuristic function}, rather than discovering and describing the `equivalent physics' (the existing physics of the underlying common-core theory), aims at discovering \textit{new physics} beyond the common-core theory: constructing a new theory that supersedes the common-core theory. While the theoretical function relies on the exactness of the duality, and it being a non-perturbative description of equivalent physics (in order to identify the physics of the common-core), the heuristic function requires breaking the (exactness of the) duality, and showing that it is recovered only approximately (e.g., in some appropriate limit, or perturbatively) from a more fundamental theory beyond the common-core theory. The heuristic function thus treats the duality as an `intermediate step' in discovering new physics. 

\citet{JakslandLinnemann} argue that, given that holography is inter-representational, it cannot, as it stands, tell us anything `outside' of itself (it only tells us about the relationship between features on the two different sides of the duality). In order for the duality to be useful for discovering new physics, we need to render it intra-representational---so that we are working within a single description, which, they argue, need not be the common-core theory. The heuristic use of the duality, for \citet{JakslandLinnemann}, is to get at the intra-representational description. On this account, it does not make sense to recover the duality, whether approximately or not. As a simple example of the heuristic impotency of duality, take the \textit{Ryu-Takayanagi formula} \citep{Ryu}. If we consider a spatial slice, $\Sigma$ of an AdS spacetime on whose boundary we define the dual CFT, the formula is a conjecture relating the entanglement entropy $S$ of the CFT (boundary theory) in some spatial subregion, with the area $A$ of a `Ryu-Takayanagi surface' in the bulk theory, as $S=A/(4G)$. This formula is clearly inter-representational. As such, \citet{JakslandLinnemann} argue that it cannot testify to the origin of the area in terms of entropy or vice versa: in the representation where the area exists, the entropy is not there, and where the entropy exists, the area is not there.

So, the holographic duality can be used as a \textit{guiding principle} in the search for QG, but this involves either the \textit{inexactness} of the duality, following \citet{deHaro2018}, or its transformation into a non-dual description, according to \citet{JakslandLinnemann}. The holographic principle can also be used as a \textit{postulate}, as for instance \citet{Smolin2017}, who presents it as one of four principles of QG, in the attempt to build QG as a principle theory, in Einstein's sense. The holographic principle could potentially function as a constitutive principle---indeed, it seems the most likely candidate for a coordinating principle defining a new physical framework. For instance, \citet[][p. 861]{Bousso2002} suggests that the holographic principle may not only aid the search for a non-perturbative version of string theory, but ``could also contribute to a background-independent formulation that would illuminate the conceptual foundation of string theory''. \citet{Smolin2001} also presents his weak holographic principle as playing something like a coordinating role. The holographic principle can also be appealed to as a \textit{means of confirmation}: for instance the AdS/CFT duality as embodying the holographic principle is presented as an argument in favour of string theory, especially in its background independent, non-perturbative form. Additionally, if we follow \citet{deHaro2018}, the approximate recovery of the duality from a more-fundamental theory of QG (a theory of new physics, beyond the common-core theory embodied by the duality), could serve as a means of confirmation (though \citet{JakslandLinnemann} argue against this). Finally, considering taking the holographic principle as a \textit{criterion of acceptance} for QG requires further argument. This may be prompted, perhaps, if the entropy bound is taken as a fundamental result motivating QG---the holographic principle would then be required in order to explain the origin of the entropy bound.

\pagebreak

\section*{Acknowledgements}
Thanks to Jim Weatherall for letting me do this; the referees for suggestions for improvements; Niels Linnemann and Kian Salimkhani for motivation and support; Eric Curiel for comments on Chapter 7; Sebastiano Orioli for making (most of) the figures.

\bibliography{thesis}

\begin{thebibliography}{}

\bibitem[\protect\citeauthoryear{Adlam}{Adlam}{2022}]{Adlam2022}
Adlam, E. (2022).
\newblock Tabletop experiments for quantum gravity are also tests of the
  interpretation of quantum mechanics.
\newblock {\em Foundations of Physics}, {\em 52\/}(115).

\bibitem[\protect\citeauthoryear{Adlam}{Adlam}{2023}]{Adlam2023}
Adlam, E. (2023).
\newblock Are entropy bounds epistemic?
\newblock {\em arXiv:2303.10781v3 [physics.hist-ph]}.

\bibitem[\protect\citeauthoryear{Aharony, Marsano, Minwalla, Papadodimas \&
  Van~Raamsdonk}{Aharony et~al.}{2004}]{Aharony2004}
Aharony, O., Marsano, J., Minwalla, S., Papadodimas, K., \& Van~Raamsdonk, M.
  (2004).
\newblock {The Hagedorn - deconfinement phase transition in weakly coupled
  large N gauge theories}.
\newblock {\em Adv. Theor. Math. Phys.}, {\em 8}, 603--696.

\bibitem[\protect\citeauthoryear{Almheiri, Hartman, Maldacena, Shaghoulian \&
  Tajdini}{Almheiri et~al.}{2020}]{Alm2020}
Almheiri, A., Hartman, T., Maldacena, J., Shaghoulian, E., \& Tajdini, A.
  (2020).
\newblock Replica wormholes and the entropy of hawking radiation.
\newblock {\em Journal of High Energy Physics}, {\em 2020\/}(5).

\bibitem[\protect\citeauthoryear{Almheiri, Hartman, Maldacena, Shaghoulian \&
  Tajdini}{Almheiri et~al.}{2021}]{Alm2021}
Almheiri, A., Hartman, T., Maldacena, J., Shaghoulian, E., \& Tajdini, A.
  (2021).
\newblock The entropy of hawking radiation.
\newblock {\em Rev. Mod. Phys.}, {\em 93}, 035002.

\bibitem[\protect\citeauthoryear{Ambj\o{}rn, Jurkiewicz \& Loll}{Ambj\o{}rn
  et~al.}{2006}]{Ambjorn2006}
Ambj\o{}rn, J., Jurkiewicz, J., \& Loll, R. (2006).
\newblock The universe from scratch.
\newblock {\em Contemporary Physics}, {\em 47\/}(2), 103--117.

\bibitem[\protect\citeauthoryear{Ambj\o{}rn, Jurkiewicz \& Loll}{Ambj\o{}rn
  et~al.}{2009}]{Ambjorn2009}
Ambj\o{}rn, J., Jurkiewicz, J., \& Loll, R. (2009).
\newblock Quantum gravity: the art of building spacetime.
\newblock In D.~Oriti (Ed.), {\em Approaches to Quantum Gravity}  (pp.\
  341--359). Cambridge: Cambridge University Press.

\bibitem[\protect\citeauthoryear{Anderson}{Anderson}{2017}]{AndersonTime}
Anderson, E. (2017).
\newblock {\em The Problem of Time: Quantum Mechanics versus Special
  Relativity}.
\newblock Springer Cham.

\bibitem[\protect\citeauthoryear{Ashtekar}{Ashtekar}{1995}]{Ashtekar1995}
Ashtekar, A. (1995).
\newblock Mathematical problems of non-perturbative quantum general relativity.
\newblock In B.~Julia \& J.~Zinn-Justin (Eds.), {\em Les Houches, Session LVII,
  1992: Gravitation and Quantizations}  (pp.\ 181--283). Elsevier Science
  Publishing Co.

\bibitem[\protect\citeauthoryear{Ashtekar}{Ashtekar}{2005}]{Ashtekar2005}
Ashtekar, A. (2005).
\newblock The winding road to quantum gravity.
\newblock {\em Current Science}, {\em 89}, 2064--2074.

\bibitem[\protect\citeauthoryear{Ashtekar \& Geroch}{Ashtekar \&
  Geroch}{1974}]{Ashtekar1974}
Ashtekar, A. \& Geroch, R. (1974).
\newblock Quantum theory and gravitation.
\newblock {\em Reports on Progress in Physics}, {\em 37}, 1211--1256.

\bibitem[\protect\citeauthoryear{Bain}{Bain}{2008}]{Bain2008}
Bain, J. (2008).
\newblock Condensed matter physics and the nature of spacetime.
\newblock In D.~Dieks (Ed.), {\em The Ontology of Spacetime II}  chapter~16,
  (pp.\ 301--329). Oxford: Elsevier.

\bibitem[\protect\citeauthoryear{Barbado, Barcel{\'o}, Garay \& Jannes}{Barbado
  et~al.}{2016}]{barbado2016hawking}
Barbado, L.~C., Barcel{\'o}, C., Garay, L.~J., \& Jannes, G. (2016).
\newblock Hawking versus {U}nruh effects, or the difficulty of slowly crossing
  a black hole horizon.
\newblock {\em Journal of High Energy Physics}, {\em 2016\/}(10), 161--174.

\bibitem[\protect\citeauthoryear{Barbour}{Barbour}{2000}]{Barbour2000}
Barbour, J. (2000).
\newblock {\em The end of time}.
\newblock Oxford University Press.

\bibitem[\protect\citeauthoryear{Barcel\'{o}, Liberati \& Visser}{Barcel\'{o}
  et~al.}{2011}]{Barcelo2011}
Barcel\'{o}, C., Liberati, S., \& Visser, M. (2011).
\newblock Analogue gravity.
\newblock {\em Living Reviews in Relativity}.

\bibitem[\protect\citeauthoryear{Barcel\'{o}, Visser \& Liberati}{Barcel\'{o}
  et~al.}{2001}]{Barcelo2001E}
Barcel\'{o}, C., Visser, M., \& Liberati, S. (2001).
\newblock Einstein gravity as an emergent phenomenon?
\newblock {\em International Journal of Modern Physics D}, {\em 10\/}(6),
  799--806.

\bibitem[\protect\citeauthoryear{Batterman}{Batterman}{2017}]{Batterman2017}
Batterman, R. (2017).
\newblock Autonomy of theories: An explanatory problem.
\newblock {\em Nous}.

\bibitem[\protect\citeauthoryear{Batterman}{Batterman}{2011}]{Batterman2011}
Batterman, R.~W. (2011).
\newblock Emergence, singularities, and symmetry breaking.
\newblock {\em Foundations of Physics}, {\em 41}, 1031--1050.

\bibitem[\protect\citeauthoryear{Bekenstein}{Bekenstein}{1981}]{Bekenstein1981}
Bekenstein, J. (1981).
\newblock A universal upper bound on the entropy to energy ratio for bounded
  systems.
\newblock {\em Phys. Rev. D}, {\em 23\/}(2), 287.

\bibitem[\protect\citeauthoryear{Bekenstein}{Bekenstein}{1972}]{Bekenstein1972}
Bekenstein, J.~D. (1972).
\newblock Black holes and the second law.
\newblock {\em Lettere al Nuovo Cimento}, {\em 4}, 737--740.

\bibitem[\protect\citeauthoryear{Bekenstein}{Bekenstein}{1973}]{Bekenstein1973}
Bekenstein, J.~D. (1973).
\newblock Black holes and entropy.
\newblock {\em Phys. Rev. D}, {\em 7\/}(8), 2333--2346.

\bibitem[\protect\citeauthoryear{Bekenstein}{Bekenstein}{1974}]{Bekenstein1974}
Bekenstein, J.~D. (1974).
\newblock Generalized second law of thermodynamics in black-hole physics.
\newblock {\em Phys. Rev. D.}, {\em 9}, 3292--3300.

\bibitem[\protect\citeauthoryear{Belot}{Belot}{2011}]{Belot2011}
Belot, G. (2011).
\newblock Background-independence.
\newblock {\em General Relativity and Gravitation}, {\em 43\/}(10), 2865--2884.

\bibitem[\protect\citeauthoryear{Belot, Earman \& Ruetsche}{Belot
  et~al.}{1999}]{Belot1999}
Belot, G., Earman, J., \& Ruetsche, L. (1999).
\newblock The hawking information loss paradox: The anatomy of a controversy.
\newblock {\em British Journal for the Philosophy of Science}, {\em 50\/}(2),
  189--229.

\bibitem[\protect\citeauthoryear{Bern}{Bern}{2002}]{Bern2002}
Bern, Z. (2002).
\newblock Perturbative quantum gravity and its relation to gauge theory.
\newblock {\em Living Reviews in Relativity}, {\em 5\/}(5).

\bibitem[\protect\citeauthoryear{Bohr \& Rosenfeld}{Bohr \&
  Rosenfeld}{1933}]{BohrRosenfeld}
Bohr, N. \& Rosenfeld, L. (1933).
\newblock Zur frage der messbarkeit der elektromagnetischen feldgr\"{o}ssern.
\newblock {\em Kong. Dan. Vid. Sel. Mat. Fys. Med.}, (8).

\bibitem[\protect\citeauthoryear{Bonanno, Eichhorn, Gies, Pawlowski, Percacci,
  Reuter, Saueressig \& Vacca}{Bonanno et~al.}{2020}]{Criticalasysafety}
Bonanno, A., Eichhorn, A., Gies, H., Pawlowski, J.~M., Percacci, R., Reuter,
  M., Saueressig, F., \& Vacca, G.~P. (2020).
\newblock Critical reflections on asymptotically safe gravity.
\newblock {\em Frontiers in Physics}, {\em 8}.

\bibitem[\protect\citeauthoryear{Bose, Mazumdar, Morley, Ulbricht,
  Toro\ifmmode~\check{s}\else \v{s}\fi{}, Paternostro, Geraci, Barker, Kim \&
  Milburn}{Bose et~al.}{2017}]{Bose2017}
Bose, S., Mazumdar, A., Morley, G.~W., Ulbricht, H.,
  Toro\ifmmode~\check{s}\else \v{s}\fi{}, M., Paternostro, M., Geraci, A.~A.,
  Barker, P.~F., Kim, M.~S., \& Milburn, G. (2017).
\newblock Spin entanglement witness for quantum gravity.
\newblock {\em Phys. Rev. Lett.}, {\em 119}, 240401.

\bibitem[\protect\citeauthoryear{Bousso}{Bousso}{1999}]{Bousso1999}
Bousso, R. (1999).
\newblock A covariant entropy conjecture.
\newblock {\em J. High Energy Phys.}, {\em 07}.

\bibitem[\protect\citeauthoryear{Bousso}{Bousso}{2002}]{Bousso2002}
Bousso, R. (2002).
\newblock The holographic principle.
\newblock {\em Reviews of Modern Physics}, {\em 74}, 825--874.

\bibitem[\protect\citeauthoryear{Brandenberger, Mukhanov \&
  Sornborger}{Brandenberger et~al.}{1993}]{Brandenberger1993}
Brandenberger, R., Mukhanov, V., \& Sornborger, A. (1993).
\newblock Cosmological theory without singularities.
\newblock {\em Phys. Rev. D}, {\em 48}, 1629--1642.

\bibitem[\protect\citeauthoryear{Brown}{Brown}{2005}]{Brown2005}
Brown, H. (2005).
\newblock {\em Physical Relativity: Space-time Structure from a Dynamical
  Perspective}.
\newblock Oxford: Oxford University Press.

\bibitem[\protect\citeauthoryear{Burgess}{Burgess}{2004}]{Burgess2004}
Burgess, C.~P. (2004).
\newblock Quantum gravity in everyday life: General relativity as an effective
  field theory.
\newblock {\em Living Reviews in Relativity}.

\bibitem[\protect\citeauthoryear{Butterfield \& Bouatta}{Butterfield \&
  Bouatta}{2015}]{Butterfield2015}
Butterfield, J. \& Bouatta, N. (2015).
\newblock Renormalization for philosophers.
\newblock In T.~Bigaj \& C.~W\"{u}thrich (Eds.), {\em Metaphysics in
  Contemporary Physics}, Poznan Studies in Philosophy of Science  (pp.\
  437--485).

\bibitem[\protect\citeauthoryear{Butterfield \& Isham}{Butterfield \&
  Isham}{1999}]{Butterfield1999}
Butterfield, J. \& Isham, C. (1999).
\newblock On the emergence of time in quantum gravity.
\newblock In J.~Butterfield (Ed.), {\em The Arguments of Time}  (pp.\
  116--168). Oxford: Oxford University Press.

\bibitem[\protect\citeauthoryear{Butterfield \& Isham}{Butterfield \&
  Isham}{2001}]{Butterfield2001}
Butterfield, J. \& Isham, C. (2001).
\newblock Spacetime and the philosophical challenge of quantum gravity.
\newblock In C.~Callender \& N.~Huggett (Eds.), {\em Physics Meets Philosophy
  at the Planck Scale}  (pp.\ 33--89). Cambridge: Cambridge University Press.

\bibitem[\protect\citeauthoryear{Calmet \& Latosh}{Calmet \&
  Latosh}{1998}]{Calmet2018}
Calmet, X. \& Latosh, B. (1998).
\newblock Dark matter in quantum gravity.
\newblock {\em European Physical Journal C}, 520.

\bibitem[\protect\citeauthoryear{Cao \& Schweber}{Cao \&
  Schweber}{1993}]{Cao1993}
Cao, T.~Y. \& Schweber, S.~S. (1993).
\newblock The conceptual foundations and the philosophical aspects of
  renormalization theory.
\newblock {\em Synthese}, {\em 97\/}(1), 33--108.

\bibitem[\protect\citeauthoryear{Caravelli \& Markopoulou}{Caravelli \&
  Markopoulou}{2011}]{Caravelli2011}
Caravelli, F. \& Markopoulou, F. (2011).
\newblock Properties of quantum graphity at low temperature.
\newblock {\em Physical Review D}, {\em 84\/}(2), 024002.

\bibitem[\protect\citeauthoryear{Carlip}{Carlip}{2001}]{Carlip2001}
Carlip, S. (2001).
\newblock Quantum gravity: a progress report.
\newblock {\em Reports on Progress in Physics}, {\em 64\/}(8), 885.

\bibitem[\protect\citeauthoryear{Chiribella}{Chiribella}{2020}]{Chiribella2020}
Chiribella, G. (2020).
\newblock Quantum superpositions of causal structures.
\newblock {\em Critical Hermeneutics}, {\em 4\/}(special II), 1--24.

\bibitem[\protect\citeauthoryear{Colella, Overhauser \& Werner}{Colella
  et~al.}{1975}]{COW2}
Colella, R., Overhauser, A.~W., \& Werner, S.~A. (1975).
\newblock Observation of gravitationally induced quantum interference.
\newblock {\em Phys. Rev. Lett.}, {\em 34}, 1472--1474.

\bibitem[\protect\citeauthoryear{Cotler \& Strominger}{Cotler \&
  Strominger}{2022}]{cotler2022}
Cotler, J. \& Strominger, A. (2022).
\newblock The universe as a quantum encoder.
\newblock {\em arXiv:2201.11658 [hep-th]}.

\bibitem[\protect\citeauthoryear{Crowther}{Crowther}{2015}]{Crowther2015}
Crowther, K. (2015).
\newblock Decoupling emergence and reduction in physics.
\newblock {\em European Journal for Philosophy of Science}, {\em 5\/}(3),
  419--445.

\bibitem[\protect\citeauthoryear{Crowther}{Crowther}{2016}]{Crowther2016}
Crowther, K. (2016).
\newblock {\em Effective Spacetime: Understanding Emergence in Effective Field
  Theory and Quantum Gravity}.
\newblock Heidelberg: Springer.

\bibitem[\protect\citeauthoryear{Crowther}{Crowther}{2018}]{Crowther2018}
Crowther, K. (2018).
\newblock Inter-theory relations in quantum gravity: Correspondence, reduction,
  and emergence.
\newblock {\em Studies in History and Philosophy of Modern Physics}, {\em 63},
  74--85.

\bibitem[\protect\citeauthoryear{Crowther}{Crowther}{2019}]{CrowtherDigging}
Crowther, K. (2019).
\newblock When do we stop digging? {C}onditions on a fundamental theory of
  physics.
\newblock In A.~Aguirre, B.~Foster, \& Z.~Merali (Eds.), {\em What is
  `Fundamental'?} Springer.

\bibitem[\protect\citeauthoryear{Crowther}{Crowther}{2020}]{Crowther2020}
Crowther, K. (2020).
\newblock What is the point of reduction in science?
\newblock {\em Erkenntnis}, {\em 85\/}(6), 1437--1460.

\bibitem[\protect\citeauthoryear{Crowther}{Crowther}{2021}]{Crowther2021}
Crowther, K. (2021).
\newblock Defining a crisis: The roles of principles in the search for a theory
  of quantum gravity.
\newblock {\em Synthese}, {\em 198\/}(Suppl 14), 3489--3516.

\bibitem[\protect\citeauthoryear{Crowther \& De~Haro}{Crowther \&
  De~Haro}{2022}]{CrowtherDeHaro}
Crowther, K. \& De~Haro, S. (2022).
\newblock Four attitudes towards singularities in the search for a theory of
  quantum gravity.
\newblock In A.~Vassallo (Ed.), {\em The Foundations of Spacetime Physics:
  Philosophical Perspectives}  (pp.\ 223--250). Routledge.

\bibitem[\protect\citeauthoryear{Crowther \& Linnemann}{Crowther \&
  Linnemann}{2019}]{CrowtherLinnemann}
Crowther, K. \& Linnemann, N. (2019).
\newblock Renormalizability, fundamentality, and a final theory: The role of
  {UV}-completion in the search for quantum gravity.
\newblock {\em British Journal for the Philosophy of Science}, {\em 70\/}(2),
  377--406.

\bibitem[\protect\citeauthoryear{Crowther, Linnemann \& W\"{u}thrich}{Crowther
  et~al.}{2019}]{CLW}
Crowther, K., Linnemann, N.~S., \& W\"{u}thrich, C. (2019).
\newblock What we cannot learn from analogue experiments.
\newblock {\em Synthese}, (Suppl 16), 1--26.

\bibitem[\protect\citeauthoryear{Curiel}{Curiel}{1999}]{Curiel1999}
Curiel, E. (1999).
\newblock The analysis of singular spacetimes.
\newblock {\em Philosophy of Science}, {\em 66\/}(3), 145.

\bibitem[\protect\citeauthoryear{Curiel}{Curiel}{2023}]{Curiel2021}
Curiel, E. (2023).
\newblock Singularities and black holes.
\newblock {\em Stanford Encyclopedia of Philosophy},
  https://plato.stanford.edu/archives/sum2023/entries/spacetime--singularities/.

\bibitem[\protect\citeauthoryear{Dafermos \& Luk}{Dafermos \&
  Luk}{2017}]{StrongCC}
Dafermos, M. \& Luk, J. (2017).
\newblock The interior of dynamical vacuum black holes {I}: The
  $c^{\circ}$-stability of the {K}err {C}auchy horizon.
\newblock {\em arXiv:1710.01722}.

\bibitem[\protect\citeauthoryear{Dardashti, Hartmann, Th\'{e}bault \&
  Winsberg}{Dardashti et~al.}{2019}]{Dardashti2019}
Dardashti, R., Hartmann, S., Th\'{e}bault, K., \& Winsberg, E. (2019).
\newblock Hawking radiation and analogue experiments: A bayesian analysis.
\newblock {\em Studies in History and Philosophy of Science Part B: Studies in
  History and Philosophy of Modern Physics}, {\em 67}, 1--11.

\bibitem[\protect\citeauthoryear{Dardashti, Th\'{e}bault \& Winsberg}{Dardashti
  et~al.}{2017}]{Dardashti2017}
Dardashti, R., Th\'{e}bault, K., \& Winsberg, E. (2017).
\newblock Confirmation via analogue simulation: what dumb holes could tell us
  about gravity.
\newblock {\em British Journal for the Philosophy of Science}, (68), 55--89.

\bibitem[\protect\citeauthoryear{Dawid}{Dawid}{2013}]{Dawid2013a}
Dawid, R. (2013).
\newblock {\em String Theory and the Scientific Method}.
\newblock Cambridge University Press.

\bibitem[\protect\citeauthoryear{{De~Haro}}{{De~Haro}}{2018}]{deHaro2018}
{De~Haro}, S. (2018).
\newblock The heuristic function of duality.
\newblock {\em Synthese}, https://doi.org/10.1007/s11229--018--1708--9.

\bibitem[\protect\citeauthoryear{{De Haro} \& Butterfield}{{De Haro} \&
  Butterfield}{2021}]{deHaro2021}
{De Haro}, S. \& Butterfield, J. (2021).
\newblock On symmetry and duality.
\newblock {\em Synthese}, {\em 198\/}(4), 2973--3013.

\bibitem[\protect\citeauthoryear{{De Haro}, {van Dongen}, Visser \&
  Butterfield}{{De Haro} et~al.}{2020}]{DEHARO2020a}
{De Haro}, S., {van Dongen}, J., Visser, M., \& Butterfield, J. (2020).
\newblock Conceptual analysis of black hole entropy in string theory.
\newblock {\em Studies in History and Philosophy of Modern Physics}, {\em 69},
  82--111.

\bibitem[\protect\citeauthoryear{Dewitt \& Rickles}{Dewitt \&
  Rickles}{2011}]{DewittRickles}
Dewitt, C.~M. \& Rickles, D. (2011).
\newblock {\em The role of gravitation in physics: report from the 1957 Chapel
  Hill Conference}.

\bibitem[\protect\citeauthoryear{Doboszewski}{Doboszewski}{2019}]{Dob2019}
Doboszewski, J. (2019).
\newblock Relativistic spacetimes and definitions of determinism.
\newblock {\em European Journal for Philosophy of Science}, {\em 9\/}(2), 24.

\bibitem[\protect\citeauthoryear{Doboszewski}{Doboszewski}{2020}]{Dob2020}
Doboszewski, J. (2020).
\newblock Epistemic holes and determinism in classical general relativity.
\newblock {\em British Journal for the Philosophy of Science}, {\em 71\/}(3),
  1093--1111.

\bibitem[\protect\citeauthoryear{Donoghue}{Donoghue}{2020}]{Donoghue2020}
Donoghue, J.~F. (2020).
\newblock A critique of the asymptotic safety program.
\newblock {\em Frontiers in Physics}, {\em 8}, 56.

\bibitem[\protect\citeauthoryear{Dowker}{Dowker}{2020}]{Dowker2020}
Dowker, F. (2020).
\newblock Being and becoming on the road to quantum gravity: or, the birth of a
  baby is not a baby.
\newblock In {\em Beyond Spacetime: The Foundations of Quantum Gravity}  (pp.\
  133--142). Cambridge University Press.

\bibitem[\protect\citeauthoryear{Dvali, Giudice, Gomez \& Kehagias}{Dvali
  et~al.}{2011}]{Dvali}
Dvali, G., Giudice, G.~F., Gomez, C., \& Kehagias, A. (2011).
\newblock Uv-completion by classicalization.
\newblock {\em Journal of High Energy Physics}, {\em 2011\/}(8), 1--31.

\bibitem[\protect\citeauthoryear{{Dyson}, {Eddington} \& {Davidson}}{{Dyson}
  et~al.}{1920}]{DysonGravitationalLensing}
{Dyson}, F.~W., {Eddington}, A.~S., \& {Davidson}, C. (1920).
\newblock {A Determination of the Deflection of Light by the Sun's
  Gravitational Field, from Observations Made at the Total Eclipse of May 29,
  1919}.
\newblock {\em Philosophical Transactions of the Royal Society of London Series
  A}, {\em 220}, 291--333.

\bibitem[\protect\citeauthoryear{Earman}{Earman}{1992}]{Earman1992}
Earman, J. (1992).
\newblock Cosmic censorship.
\newblock {\em PSA: Proceedings}, {\em 2}, 171--180.

\bibitem[\protect\citeauthoryear{Earman}{Earman}{1995}]{Earman1995}
Earman, J. (1995).
\newblock {\em Bangs, Crunches, Whimpers, and Shrieks: Singularities and
  Acausalities in Relativistic Spacetimes}.
\newblock New York: Oxford University Press.

\bibitem[\protect\citeauthoryear{Earman}{Earman}{1996}]{Earman1996}
Earman, J. (1996).
\newblock Tolerance for spacetime singularities.
\newblock {\em Foundations of Physics}, {\em 26\/}(5), 623--640.

\bibitem[\protect\citeauthoryear{Earman}{Earman}{2006}]{Earman2006}
Earman, J. (2006).
\newblock The implications of general covariance for the ontology and ideology
  of spacetime.
\newblock In D.~Dieks (Ed.), {\em The Ontology of Spacetime}  (pp.\ 3--23).
  Amsterdam: Elsevier.

\bibitem[\protect\citeauthoryear{Eichhorn}{Eichhorn}{2019}]{Eichhorn2019}
Eichhorn, A. (2019).
\newblock An asymptotically safe guide to quantum gravity and matter.
\newblock {\em Frontiers in Astronomy and Space Sciences}, {\em 5}, 47.

\bibitem[\protect\citeauthoryear{Ellis, Meissner \& Nicolai}{Ellis
  et~al.}{2018}]{EllisInfinity}
Ellis, G.~F., Meissner, K., \& Nicolai, H. (2018).
\newblock The physics of infinity.
\newblock {\em Nature Physics}, {\em 14}, 770--772.

\bibitem[\protect\citeauthoryear{Eppley \& Hannah}{Eppley \&
  Hannah}{1977}]{Eppley1977}
Eppley, K. \& Hannah, E. (1977).
\newblock The necessity of quantizing the gravitational field.
\newblock {\em Foundations of Physics}, {\em 7\/}(1), 51--68.

\bibitem[\protect\citeauthoryear{Fraser}{Fraser}{2011}]{Fraser2011}
Fraser, D. (2011).
\newblock How to take particle physics seriously: A further defence of
  axiomatic quantum field theory.
\newblock {\em Studies In History and Philosophy of Modern Physics}, {\em
  42\/}(2), 126--135.

\bibitem[\protect\citeauthoryear{Fraser}{Fraser}{2016}]{FraserThesis}
Fraser, J.~D. (2016).
\newblock {\em What is Quantum Field Theory? {I}dealisation, Explanation and
  Realism in High Energy Physics}.
\newblock PhD thesis, University of Leeds.

\bibitem[\protect\citeauthoryear{Fraser}{Fraser}{2020}]{Fraser2020}
Fraser, J.~D. (2020).
\newblock The real problem with perturbative quantum field theory.
\newblock {\em British Journal for the Philosophy of Science}, {\em 71\/}(2),
  391--413.

\bibitem[\protect\citeauthoryear{Gies \& Jaeckel}{Gies \& Jaeckel}{2004}]{QED}
Gies, H. \& Jaeckel, J. (2004).
\newblock Renormalization flow of {QED}.
\newblock {\em Phys. Rev. Lett.}, {\em 93\/}(11), 110405.

\bibitem[\protect\citeauthoryear{Giulini}{Giulini}{2007}]{Giulini2007}
Giulini, D. (2007).
\newblock Remarks on the notions of general covariance and background
  independence.
\newblock In I.-O. Stamatescu \& E.~Seiler (Eds.), {\em Lecture Notes in
  Physics}, volume 721  (pp.\ 105--120). Berlin: Springer.

\bibitem[\protect\citeauthoryear{Gomes}{Gomes}{2020}]{Gomes2020}
Gomes, H. (2020).
\newblock Back to parmenides.
\newblock In {\em Beyond Spacetime: The Foundations of Quantum Gravity}  (pp.\
  176--206). Cambridge University Press.

\bibitem[\protect\citeauthoryear{Goswami, Cao, Paz-Silva, Romero \&
  White}{Goswami et~al.}{2020}]{Goswami2020}
Goswami, K., Cao, Y., Paz-Silva, G.~A., Romero, J., \& White, A.~G. (2020).
\newblock Increasing communication capacity via superposition of order.
\newblock {\em Phys. Rev. Research}, {\em 2}, 033292.

\bibitem[\protect\citeauthoryear{Gro{\ss}ardt}{Gro{\ss}ardt}{2021}]{Grossardt}
Gro{\ss}ardt, A. (2021).
\newblock Comment on ``do gedankenexperiments compel quantization of gravity".
\newblock {\em arXiv:2107.14666 [gr-qc]}.

\bibitem[\protect\citeauthoryear{Gro{\ss}ardt}{Gro{\ss}ardt}{2022}]{Grossardt2}
Gro{\ss}ardt, A. (2022).
\newblock Three little paradoxes: Making sense of semiclassical gravity.
\newblock {\em AVS Quantum Science}, {\em 4\/}(1), 010502.

\bibitem[\protect\citeauthoryear{Hagar}{Hagar}{2014}]{Hagar2014}
Hagar, A. (2014).
\newblock {\em Discrete or continuous? The quest for fundamental length in
  modern physics}.
\newblock Cambridge University Press.

\bibitem[\protect\citeauthoryear{Harlow}{Harlow}{2016}]{Harlow2016}
Harlow, D. (2016).
\newblock Jerusalem lectures on black holes and quantum information.
\newblock {\em Reviews of Modern Physics}, {\em 88\/}(1).

\bibitem[\protect\citeauthoryear{Hartmann}{Hartmann}{2002}]{Hartmann2002}
Hartmann, S. (2002).
\newblock On correspondence.
\newblock {\em Studies In History and Philosophy of Modern Physics}, {\em
  33\/}(1), 79--94.

\bibitem[\protect\citeauthoryear{Hawking}{Hawking}{1971}]{Hawking1971}
Hawking, S. (1971).
\newblock Gravitational radiation from colliding black holes.
\newblock {\em Phys. Rev. Lett.}, {\em 26}, 1344.

\bibitem[\protect\citeauthoryear{Hawking}{Hawking}{1974}]{Hawking1974}
Hawking, S. (1974).
\newblock Black hole explosions?
\newblock {\em Nature}, {\em 248}, 30--31.

\bibitem[\protect\citeauthoryear{Hawking}{Hawking}{1975}]{Hawking1975}
Hawking, S. (1975).
\newblock Particle creation by black holes.
\newblock {\em Communications in Mathematical Physics}, {\em 43}, 199--220.

\bibitem[\protect\citeauthoryear{Hawking, King \& McCarthy}{Hawking
  et~al.}{1976}]{Hawking1976}
Hawking, S., King, A., \& McCarthy, P. (1976).
\newblock A new topology for curved space-time which incorporates the causal,
  differential, and conformal structures.
\newblock {\em Journal of Mathematical Physics}, {\em 17\/}(2), 174--181.

\bibitem[\protect\citeauthoryear{Hawking \& Stewart}{Hawking \&
  Stewart}{1993}]{Hawking1993}
Hawking, S. \& Stewart, J. (1993).
\newblock Naked and thunderbolt singularities in black hole evaporation.
\newblock {\em Nuclear Physics B}, {\em 400\/}(1-3), 393--415.

\bibitem[\protect\citeauthoryear{Hawking \& Ellis}{Hawking \&
  Ellis}{1973}]{Hawking1973}
Hawking, S.~W. \& Ellis, G.~F. (1973).
\newblock {\em The Large-Scale Structure of Space-Time}.
\newblock Cambridge: Cambridge University Press.

\bibitem[\protect\citeauthoryear{Held}{Held}{2019}]{Held2019}
Held, A. (2019).
\newblock {\em From particle physics to black holes: The predictive power of
  asymptotic safety}.
\newblock PhD thesis, University of Heidelberg, Heidelberg.

\bibitem[\protect\citeauthoryear{Henson}{Henson}{2009}]{Henson2009}
Henson, J. (2009).
\newblock The causal set approach to quantum gravity.
\newblock In D.~Oriti (Ed.), {\em Approaches to Quantum Gravity: Toward a New
  Understanding of Space, Time and Matter}  (pp.\ 393--413). Cambridge:
  Cambridge University Press.

\bibitem[\protect\citeauthoryear{Hossenfelder}{Hossenfelder}{2013}]{Hossenfelder2013}
Hossenfelder, S. (2013).
\newblock Minimal length scale scenarios for quantum gravity.
\newblock {\em Living Reviews in Relativity}, (2).

\bibitem[\protect\citeauthoryear{Hu}{Hu}{2009}]{Hu2009}
Hu, B.-L. (2009).
\newblock Emergent/quantum gravity: macro/micro structures of spacetime.
\newblock In H.~T. Elze, L.~Diosi, L.~Fronzoni, J.~Halliwell, \& G.~Vitiello
  (Eds.), {\em Fourth International Workshop Dice 2008: From Quantum Mechanics
  through Complexity to Spacetime: The Role of Emergent Dynamical Structures},
  volume 174 of {\em Journal of Physics Conference Series}  (pp.\
  12015--12015). Bristol: Iop Publishing Ltd.

\bibitem[\protect\citeauthoryear{Huggett \& Callender}{Huggett \&
  Callender}{2001}]{Huggett2001}
Huggett, N. \& Callender, C. (2001).
\newblock Why quantize gravity (or any other field for that matter)?
\newblock {\em Philosophy of Science}, {\em 68\/}(3), S382--S394.

\bibitem[\protect\citeauthoryear{Huggett, Linnemann \& Schneider}{Huggett
  et~al.}{2023}]{huggettQGLab}
Huggett, N., Linnemann, N., \& Schneider, M.~D. (2023).
\newblock {\em Quantum Gravity in a Laboratory?}
\newblock Elements in the Foundations of Contemporary Physics. Cambridge
  University Press.

\bibitem[\protect\citeauthoryear{Huggett \& Vistarini}{Huggett \&
  Vistarini}{2015}]{HuggettVistarini}
Huggett, N. \& Vistarini, T. (2015).
\newblock Deriving general relativity from string theory.
\newblock {\em Philosophy of Science}, {\em 82\/}(5), 1163--1174.

\bibitem[\protect\citeauthoryear{Huggett \& W\"{u}thrich}{Huggett \&
  W\"{u}thrich}{2013}]{Huggett2013}
Huggett, N. \& W\"{u}thrich, C. (2013).
\newblock Emergent spacetime and empirical (in)coherence.
\newblock {\em Studies in History and Philosophy of Modern Physics}, {\em
  44\/}(3), 276--285.

\bibitem[\protect\citeauthoryear{Isham}{Isham}{1993}]{Isham1993}
Isham, C. (1993).
\newblock Canonical quantum gravity and the problem of time.
\newblock In L.~Ibort \& M.~Rodriguez (Eds.), {\em Integrable Systems, Quantum
  Groups, and Quantum Field Theories}  (pp.\ 157--288). Dordrecht: Kluwer.

\bibitem[\protect\citeauthoryear{Jackiw}{Jackiw}{1999}]{Jackiw1999}
Jackiw, R. (1999).
\newblock The unreasonable effectiveness of quantum field theory.
\newblock In T.~Cao (Ed.), {\em Conceptual Foundations of Quantum Field Theory}
   (pp.\ 148--159). Cambridge: Cambridge University Press.

\bibitem[\protect\citeauthoryear{Jackiw}{Jackiw}{2000}]{Jackiw2000}
Jackiw, R. (2000).
\newblock What good are quantum field theory infinities?
\newblock In A.~Fokas, A.~Grigoryan, T.~Kibble, \& B.~Zegarlinski (Eds.), {\em
  Mathematical Physics 2000}  (pp.\ 101--110). World Scientific.

\bibitem[\protect\citeauthoryear{Jacobson}{Jacobson}{1995}]{Jacobson1995}
Jacobson, T. (1995).
\newblock Thermodynamics of spacetime: The {E}instein equation of state.
\newblock {\em Phys. Rev. Lett.}, {\em 75}, 1260--1263.

\bibitem[\protect\citeauthoryear{Jaksland \& Linnemann}{Jaksland \&
  Linnemann}{2020}]{JakslandLinnemann}
Jaksland, R. \& Linnemann, N.~S. (2020).
\newblock Holography without holography: How to turn inter-representational
  into intra-theoretical relations in ads/cft.
\newblock {\em Studies in History and Philosophy of Modern Physics}, {\em 71},
  101--117.

\bibitem[\protect\citeauthoryear{Kao}{Kao}{2019}]{Kao2019}
Kao, M. (2019).
\newblock Unification beyond justification: A strategy for theory development.
\newblock {\em Synthese}, {\em 196\/}(8), 3263--3278.

\bibitem[\protect\citeauthoryear{Kaplan}{Kaplan}{2005}]{Kaplan2005}
Kaplan, D.~B. (2005).
\newblock Five lectures on effective field theory.
\newblock {\em arXiv:nucl-th/0510023}.

\bibitem[\protect\citeauthoryear{Kastner \& Kauffman}{Kastner \&
  Kauffman}{2018}]{Kastner2018}
Kastner, R.~E. \& Kauffman, S. (2018).
\newblock Are dark energy and dark matter different aspects of the same
  physical process?
\newblock {\em Frontiers in Physics}, {\em 6}.

\bibitem[\protect\citeauthoryear{Kent}{Kent}{2018}]{Kent2018}
Kent, A. (2018).
\newblock Simple refutation of the {E}ppley–{H}annah argument.
\newblock {\em Classical and Quantum Gravity}, {\em 35\/}(24), 245008.

\bibitem[\protect\citeauthoryear{Kiefer}{Kiefer}{2007a}]{Kiefer2007a}
Kiefer, C. (2007a).
\newblock {\em Quantum Gravity\/} (2nd ed.).
\newblock Oxford: Oxford University Press.

\bibitem[\protect\citeauthoryear{Kiefer}{Kiefer}{2007b}]{Kiefer2007}
Kiefer, C. (2007b).
\newblock Why quantum gravity?
\newblock In I.-O. Stamatescu \& E.~Seiler (Eds.), {\em Approaches to
  Fundamental Physics}, volume 721 of {\em Lecture Notes in Physics}  (pp.\
  123--130). Berlin: Springer.

\bibitem[\protect\citeauthoryear{Kiefer}{Kiefer}{2013}]{Kiefer2013}
Kiefer, C. (2013).
\newblock Conceptual problems in quantum gravity and quantum cosmology.
\newblock {\em ISRN Mathematical Physics}, {\em 2013}, 509316.

\bibitem[\protect\citeauthoryear{Kretschmann}{Kretschmann}{1917}]{Kretschmann1917}
Kretschmann, E. (1917).
\newblock Über den physikalischen sinn der relativitätspostulate: A.
  einsteins neue und seine ursprüngliche relativitätstheorie.
\newblock {\em Annalen der Physik}, {\em 53}, 575--614.

\bibitem[\protect\citeauthoryear{Kucha\v{r}}{Kucha\v{r}}{1999}]{Kuchar1999}
Kucha\v{r}, K. (1999).
\newblock The problem of time in quantum geometrodynamics.
\newblock In J.~Butterfield (Ed.), {\em The Arguments of Time}  (pp.\
  169--196). Oxford: Oxford University Press.

\bibitem[\protect\citeauthoryear{Kuhlmann}{Kuhlmann}{2023}]{Kuhlmann2023}
Kuhlmann, M. (2023).
\newblock Quantum field theory.
\newblock In E.~N. Zalta (Ed.), {\em The Stanford Encyclopedia of Philosophy}.

\bibitem[\protect\citeauthoryear{Landau, Abrikosov \& Khalatnikov}{Landau
  et~al.}{1954}]{Landau1954}
Landau, L., Abrikosov, A., \& Khalatnikov, I. (1954).
\newblock The removal of infinities in quantum electrodynamics.
\newblock {\em Dokl. Akad. Nauk SSSR}.

\bibitem[\protect\citeauthoryear{Liberati, Girelli \& Sindoni}{Liberati
  et~al.}{2009}]{Liberati2009}
Liberati, S., Girelli, F., \& Sindoni, L. (2009).
\newblock Analogue models for emergent gravity.

\bibitem[\protect\citeauthoryear{Linnemann \& Visser}{Linnemann \&
  Visser}{2018}]{Linnemann2018}
Linnemann, N.~S. \& Visser, M.~R. (2018).
\newblock Hints towards the emergent nature of gravity.
\newblock {\em Studies in History and Philosophy of Modern Physics}, {\em 64},
  1 -- 13.

\bibitem[\protect\citeauthoryear{Loll}{Loll}{1998}]{Loll1998}
Loll, R. (1998).
\newblock Discrete approaches to quantum gravity in four dimensions.
\newblock {\em Living Reviews in Relativity}.

\bibitem[\protect\citeauthoryear{Loll}{Loll}{2019}]{Loll2019}
Loll, R. (2019).
\newblock Quantum gravity from causal dynamical triangulations: a review.
\newblock {\em Classical and Quantum Gravity}, {\em 37\/}(1), 013002.

\bibitem[\protect\citeauthoryear{Loll, Fabiano, Frattulillo \& Wagner}{Loll
  et~al.}{2022}]{QG30questions}
Loll, R., Fabiano, G., Frattulillo, D., \& Wagner, F. (2022).
\newblock Quantum gravity in 30 questions.
\newblock {\em arXiv:2206.06762 [hep-th]}.

\bibitem[\protect\citeauthoryear{L\"{u}sher \& Weisz}{L\"{u}sher \&
  Weisz}{1987}]{Lusher}
L\"{u}sher, M. \& Weisz, P. (1987).
\newblock Scaling laws and triviality bounds in the lattice $\phi^4$ theory:
  {(I)}. one-component model in the symmetric phase.
\newblock {\em Nuclear Physics B}, {\em 290}, 25--60.

\bibitem[\protect\citeauthoryear{Maldacena}{Maldacena}{1998}]{Maldacena1998}
Maldacena, J. (1998).
\newblock The large n limit of superconformal field theories and supergravity.
\newblock {\em Advances in Theoretical and Mathematical Physics}, {\em 2},
  231--252.

\bibitem[\protect\citeauthoryear{Manchak \& Weatherall}{Manchak \&
  Weatherall}{2018}]{Manchak2018}
Manchak, J.~B. \& Weatherall, J.~O. (2018).
\newblock Paradox regained? a brief comment on maudlin on black hole
  information loss.
\newblock {\em Foundations of Physics}, {\em 48\/}(6), 611--627.

\bibitem[\protect\citeauthoryear{Manohar}{Manohar}{1997}]{Manohar1997}
Manohar, A. (1997).
\newblock Effective field theories.
\newblock In {\em Perturbative and Nonperturbative Aspects of Quantum Field
  Theory}, volume 479 of {\em Lecture Notes in Physics}  (pp.\ 311--362).
  Berlin: Springer.

\bibitem[\protect\citeauthoryear{Markopoulou}{Markopoulou}{2009a}]{Markopoulou2009}
Markopoulou, F. (2009a).
\newblock New directions in background independent quantum gravity.
\newblock In D.~Oriti (Ed.), {\em Approaches to Quantum Gravity}  (pp.\
  129--149). Cambridge: Cambridge University Press.

\bibitem[\protect\citeauthoryear{Markopoulou}{Markopoulou}{2009b}]{MarkopoulouSpace}
Markopoulou, F. (2009b).
\newblock Space does not exist, so time can.
\newblock {\em arXiv:0909.1861 [gr-qc]}, 1--9.

\bibitem[\protect\citeauthoryear{Marletto \& Vedral}{Marletto \&
  Vedral}{2017}]{MV2017}
Marletto, C. \& Vedral, V. (2017).
\newblock Gravitationally induced entanglement between two massive particles is
  sufficient evidence of quantum effects in gravity.
\newblock {\em Phys. Rev. Lett.}, {\em 119}, 240402.

\bibitem[\protect\citeauthoryear{Marolf}{Marolf}{2017}]{Marolf2017}
Marolf, D. (2017).
\newblock The black hole information problem: past, present, and future.
\newblock {\em Reports on Progress in Physics}, {\em 80\/}(9), 092001.

\bibitem[\protect\citeauthoryear{Marolf \& Maxfield}{Marolf \&
  Maxfield}{2020}]{Marolf2020}
Marolf, D. \& Maxfield, H. (2020).
\newblock Transcending the ensemble: baby universes, spacetime wormholes, and
  the order and disorder of black hole information.
\newblock {\em Journal of High Energy Physics}, {\em 2020\/}(8).

\bibitem[\protect\citeauthoryear{Mattingly}{Mattingly}{2005}]{Mattingly2005}
Mattingly, J. (2005).
\newblock Is quantum gravity necessary?
\newblock In A.~Kox \& J.~Eisenstaedt (Eds.), {\em The Universe of General
  Relativity}  (pp.\ 327--338). Birkh\"{a}user.

\bibitem[\protect\citeauthoryear{Mattingly}{Mattingly}{2006}]{Mattingly2006}
Mattingly, J. (2006).
\newblock Why {E}ppley and {H}annah's thought experiment fails.
\newblock {\em Phys. Rev. D}, {\em 73}, 062025.

\bibitem[\protect\citeauthoryear{Mattingly}{Mattingly}{2009}]{Mattingly2009}
Mattingly, J. (2009).
\newblock Mongrel gravity.
\newblock {\em Erkenntnis}, {\em 70\/}(3), 379--395.

\bibitem[\protect\citeauthoryear{Maudlin}{Maudlin}{1996}]{Maudlin1996}
Maudlin, T. (1996).
\newblock On the unification of physics.
\newblock {\em Journal of Philosophy}, {\em 93\/}(3), 129--144.

\bibitem[\protect\citeauthoryear{Maudlin}{Maudlin}{2012}]{Maudlin2012}
Maudlin, T. (2012).
\newblock {\em Philosophy of Physics: Space and Time}.
\newblock Princeton University Press.

\bibitem[\protect\citeauthoryear{Maudlin}{Maudlin}{2017}]{maudlin2017}
Maudlin, T. (2017).
\newblock (information) paradox lost.

\bibitem[\protect\citeauthoryear{Meissner}{Meissner}{2004}]{Meissner2004}
Meissner, K.~A. (2004).
\newblock Black-hole entropy in loop quantum gravity.
\newblock {\em Classical and Quantum Gravity}, {\em 21\/}(22), 5245--5251.

\bibitem[\protect\citeauthoryear{Misner}{Misner}{1969}]{Misner1969}
Misner, C. (1969).
\newblock Absolute zero of time.
\newblock {\em Physical Review}, {\em 186\/}(5), 1328--1333.

\bibitem[\protect\citeauthoryear{Misner, Thorne \& Wheeler}{Misner
  et~al.}{2017}]{MTW2017}
Misner, C.~W., Thorne, K.~S., \& Wheeler, J.~A. (2017).
\newblock {\em Gravitation}.
\newblock Princeton University Press.

\bibitem[\protect\citeauthoryear{Morganti}{Morganti}{2020}]{Morganti2020b}
Morganti, M. (2020).
\newblock Fundamentality in metaphysics and the philosophy of physics. part ii:
  The philosophy of physics.
\newblock {\em Philosophy Compass}, {\em 15\/}(10), 1--14.

\bibitem[\protect\citeauthoryear{Morrison}{Morrison}{2000}]{Morrison2000}
Morrison, M. (2000).
\newblock {\em Unifying Scientific Theories: Physical Concepts and Mathematical
  Structures}.
\newblock Cambridge: Cambridge University Press.

\bibitem[\protect\citeauthoryear{Myrvold}{Myrvold}{2003}]{Myrvold2003}
Myrvold, W.~C. (2003).
\newblock A {B}ayesian account of the virtue of unification.
\newblock {\em Philosophy of Science}, {\em 70\/}(2), pp. 399--423.

\bibitem[\protect\citeauthoryear{Nickles}{Nickles}{1973}]{Nickles1973}
Nickles, T. (1973).
\newblock Two concepts of intertheoretic reduction.
\newblock {\em The Journal of Philosophy}, {\em 70\/}(7), 181--201.

\bibitem[\protect\citeauthoryear{Niedermaier \& Reuter}{Niedermaier \&
  Reuter}{2006}]{Niedermaier2006}
Niedermaier, M. \& Reuter, M. (2006).
\newblock The asymptotic safety scenario in quantum gravity.
\newblock {\em Living Reviews in Relativity}, {\em 9},
  http://www.livingreviews.org/lrr--2006--5.

\bibitem[\protect\citeauthoryear{Norton}{Norton}{2003}]{Norton2003}
Norton, J. (2003).
\newblock General covariance, gauge theories and the kretschmann objection.
\newblock In K.~Brading \& E.~Castellani (Eds.), {\em Symmetries in Physics:
  Philosophical Reflections}  (pp.\ 110--123). Cambridge University Press.

\bibitem[\protect\citeauthoryear{Oppenheim}{Oppenheim}{2023}]{Oppenheim2023}
Oppenheim, J. (2023).
\newblock A postquantum theory of classical gravity?
\newblock {\em Phys. Rev. X}, {\em 13}, 041040.

\bibitem[\protect\citeauthoryear{Oriti}{Oriti}{2009}]{Oriti2009b}
Oriti, D. (2009).
\newblock The group field theory approach to quantum gravity.
\newblock In {\em Approaches to quantum gravity: Toward a new understanding of
  space time and matter}  (pp.\ 310--331). Cambridge: Cambridge University
  Press.

\bibitem[\protect\citeauthoryear{Oriti \& Pang}{Oriti \&
  Pang}{2021}]{Oriti2021}
Oriti, D. \& Pang, X. (2021).
\newblock Phantom-like dark energy from quantum gravity.
\newblock {\em Journal of Cosmology and Astroparticle Physics}, {\em
  2021\/}(12), 040.

\bibitem[\protect\citeauthoryear{Overhauser \& Colella}{Overhauser \&
  Colella}{1974}]{COW1}
Overhauser, A.~W. \& Colella, R. (1974).
\newblock Experimental test of gravitationally induced quantum interference.
\newblock {\em Phys. Rev. Lett.}, {\em 33}, 1237--1239.

\bibitem[\protect\citeauthoryear{Padmanabhan}{Padmanabhan}{2004}]{Padmanabhan2004}
Padmanabhan, T. (2004).
\newblock Equipartition of energy in the horizon degrees of freedom and the
  emergence of gravity.
\newblock {\em Class. Quant. Grav.}, {\em 21}, 1129--1136.

\bibitem[\protect\citeauthoryear{Padmanabhan}{Padmanabhan}{2010}]{Padmanabhan2010}
Padmanabhan, T. (2010).
\newblock Thermodynamical aspects of gravity: new insights.
\newblock {\em Reports on Progress in Physics}, {\em 73\/}(4), 046901.

\bibitem[\protect\citeauthoryear{Page}{Page}{1993}]{Page1993}
Page, D.~N. (1993).
\newblock Information in black hole radiation.
\newblock {\em Phys. Rev. Lett.}, {\em 71}, 3743--3746.

\bibitem[\protect\citeauthoryear{Page \& Geilker}{Page \&
  Geilker}{1981}]{PageGeilker}
Page, D.~N. \& Geilker, C.~D. (1981).
\newblock Indirect evidence for quantum gravity.
\newblock {\em Phys. Rev. Lett.}, {\em 47}, 979--982.

\bibitem[\protect\citeauthoryear{Penington, Shenker, Stanford \&
  Yang}{Penington et~al.}{2020}]{penington2020}
Penington, G., Shenker, S.~H., Stanford, D., \& Yang, Z. (2020).
\newblock Replica wormholes and the black hole interior.

\bibitem[\protect\citeauthoryear{Penrose}{Penrose}{1965}]{Penrose1965}
Penrose, R. (1965).
\newblock Gravitational collapse and space-time singularities.
\newblock {\em Phys. Rev. Lett.}, {\em 14}, 57--59.

\bibitem[\protect\citeauthoryear{Penrose}{Penrose}{1979}]{Penrose1979}
Penrose, R. (1979).
\newblock Singularities and time-asymmetry.
\newblock In S.~Hawking \& W.~Israel (Eds.), {\em General Relativity: An
  {E}instein Centenary Survey}  (pp.\ 581--638). Cambridge University Press.

\bibitem[\protect\citeauthoryear{Penrose}{Penrose}{2002}]{Penrose2002}
Penrose, R. (2002).
\newblock Gravitational collapse: The role of general relativity.
\newblock {\em General Relativity and Gravitation}, {\em 34\/}(7), 1141--1165.

\bibitem[\protect\citeauthoryear{Penrose}{Penrose}{2004}]{Penrose2004}
Penrose, R. (2004).
\newblock {\em The road to reality}.
\newblock Jonathan Cape.

\bibitem[\protect\citeauthoryear{Penrose}{Penrose}{2014}]{Penrose2014}
Penrose, R. (2014).
\newblock On the gravitization of quantum mechanics 1: Quantum state reduction.
\newblock {\em Foundations of Physics}, {\em 44}, 557--–575.

\bibitem[\protect\citeauthoryear{Peres \& Terno}{Peres \&
  Terno}{2001}]{Peres2001}
Peres, A. \& Terno, D.~R. (2001).
\newblock Hybrid classical-quantum dynamics.
\newblock {\em Phys. Rev. A}, {\em 63}, 022101.

\bibitem[\protect\citeauthoryear{Pitts}{Pitts}{2006}]{Pitts2006}
Pitts, J.~B. (2006).
\newblock Absolute objects and counterexamples: Jones-geroch dust, torretti
  constant curvature, tetrad-spinor, and scalar density.
\newblock {\em Studies In History and Philosophy of Modern Physics}, {\em 37},
  347--371.

\bibitem[\protect\citeauthoryear{Polchinski}{Polchinski}{2017}]{Polchinski2017}
Polchinski, J. (2017).
\newblock Dualities of fields and strings.
\newblock {\em Studies In History and Philosophy of Modern Physics}, {\em 59},
  6--20.

\bibitem[\protect\citeauthoryear{Pooley}{Pooley}{2010}]{Pooley2010}
Pooley, O. (2010).
\newblock Substantive general covariance: Another decade of dispute.
\newblock In M.~Su'{a}rez, M.~Dorato, \& M.~R'{e}dei (Eds.), {\em EPSA
  Philosophical Issues in the Sciences: Launch of the European Philosophy of
  Science Association}  (pp.\ 197--210). Springer.

\bibitem[\protect\citeauthoryear{Pooley}{Pooley}{2017}]{Pooley2017}
Pooley, O. (2017).
\newblock Background independence, diffeomorphism invariance, and the meaning
  of coordinates.
\newblock In {\em Towards a Theory of Spacetime Theories}  (pp.\ 105--144).
  Birkh\"{a}user.

\bibitem[\protect\citeauthoryear{Post}{Post}{1971}]{Post1971}
Post, H. (1971).
\newblock Correspondence, invariance and heuristics: In praise of conservative
  induction.
\newblock {\em Studies in History and Philosophy of Science Part A}, {\em
  2\/}(3), 213--255.

\bibitem[\protect\citeauthoryear{Radder}{Radder}{1991}]{Radder1991}
Radder, H. (1991).
\newblock Heuristics and the generalized correspondence principle.
\newblock {\em British Journal for the Philosophy of Science}, {\em 42},
  195--226.

\bibitem[\protect\citeauthoryear{Read}{Read}{2023}]{Read2023}
Read, J. (2023).
\newblock {\em Background Independence in Classical and Quantum Gravity}.
\newblock Oxford University Press.

\bibitem[\protect\citeauthoryear{Rickles}{Rickles}{2006}]{Rickles2006}
Rickles, D. (2006).
\newblock Time and structure in canonical gravity.
\newblock In D.~Rickles, S.~French, \& J.~Saatsi (Eds.), {\em The Structural
  Foundations of Quantum Gravity}  (pp.\ 152--196). Oxford: Oxford University
  Press.

\bibitem[\protect\citeauthoryear{Rickles}{Rickles}{2008}]{Rickles2008}
Rickles, D. (2008).
\newblock Who's afraid of background independence?
\newblock In D.~Deiks (Ed.), {\em The Ontology of Spacetime II}  chapter~7,
  (pp.\ 133--152). Amsterdam: Elsevier.

\bibitem[\protect\citeauthoryear{Rickles}{Rickles}{2017}]{Rickles2017}
Rickles, D. (2017).
\newblock Dual theories: Same but different or different but same?
\newblock {\em Studies in History and Philosophy of Science Part B: Studies in
  History and Philosophy of Modern Physics}, 61--67.

\bibitem[\protect\citeauthoryear{Rideout \& Zohren}{Rideout \&
  Zohren}{2006}]{Rideout2006}
Rideout, D. \& Zohren, S. (2006).
\newblock Evidence for an entropy bound from fundamentally discrete gravity.
\newblock {\em Classical and Quantum Gravity}, {\em 23\/}(22), 6195.

\bibitem[\protect\citeauthoryear{Rosenfeld}{Rosenfeld}{1963}]{Rosenfeld1963}
Rosenfeld, L. (1963).
\newblock On quantization of fields.
\newblock {\em Nuclear Physics}, {\em 40}, 353--356.

\bibitem[\protect\citeauthoryear{Rovelli}{Rovelli}{1996}]{Rovelli1996}
Rovelli, C. (1996).
\newblock Black hole entropy from loop quantum gravity.
\newblock {\em Physical Review Letters}, {\em 77\/}(16), 3288--3291.

\bibitem[\protect\citeauthoryear{Rovelli}{Rovelli}{2000}]{Rovelli2000}
Rovelli, C. (2000).
\newblock Notes for a brief history of quantum gravity.
\newblock In {\em 9th Marcel Grossmann Meeting on Recent Developments in
  Theoretical and Experimental General Relativity, Gravitation and Relativistic
  Field Theories (MG 9)}, (pp.\ 742--768).

\bibitem[\protect\citeauthoryear{Rovelli}{Rovelli}{2001}]{Rovelli2001}
Rovelli, C. (2001).
\newblock Quantum spacetime: What do we know?
\newblock In C.~Callender \& N.~Huggett (Eds.), {\em Physics Meets Philosophy
  at the Planck Scale: Contemporary Theories in Quantum Gravity}  (pp.\
  101--124). Cambridge: Cambridge University Press.

\bibitem[\protect\citeauthoryear{Rovelli}{Rovelli}{2004}]{Rovelli2004}
Rovelli, C. (2004).
\newblock {\em Quantum Gravity}.
\newblock Cambridge: Cambridge University Press.

\bibitem[\protect\citeauthoryear{Rovelli}{Rovelli}{2020}]{Rovelli2020}
Rovelli, C. (2020).
\newblock Space and time in loop quantum gravity.
\newblock In {\em Beyond Spacetime: The Foundations of Quantum Gravity}  (pp.\
  117--132). Cambridge University Press.

\bibitem[\protect\citeauthoryear{Rovelli \& Vidotto}{Rovelli \&
  Vidotto}{2014}]{Rovelli2014}
Rovelli, C. \& Vidotto, F. (2014).
\newblock {\em Covariant Loop Quantum Gravity: An Elementary Introduction to
  Quantum Gravity and Spinfoam Theory}.
\newblock Cambridge University Press.

\bibitem[\protect\citeauthoryear{Rydving, Aurell \& Pikovski}{Rydving
  et~al.}{2021}]{Rydving2021}
Rydving, E., Aurell, E., \& Pikovski, I. (2021).
\newblock Do gedanken experiments compel quantization of gravity?
\newblock {\em Phys. Rev. D}, {\em 104}, 086024.

\bibitem[\protect\citeauthoryear{Ryu \& Takayanagi}{Ryu \&
  Takayanagi}{2006}]{Ryu}
Ryu, S. \& Takayanagi, T. (2006).
\newblock Holographic derivation of entanglement entropy from the anti--de
  sitter space/conformal field theory correspondence.
\newblock {\em Phys. Rev. Lett.}, {\em 96}, 181602.

\bibitem[\protect\citeauthoryear{Sakharov}{Sakharov}{1967}]{Sakharov1967}
Sakharov, A. (1967).
\newblock Vacuum quantum fluctuations in curved space and the theory of
  gravitation.
\newblock {\em Doklady Akadmii Nauk SSSR}, {\em 177}, 70--71.

\bibitem[\protect\citeauthoryear{Sakharov}{Sakharov}{2000}]{Sakharov2000}
Sakharov, A.~D. (2000).
\newblock Vacuum quantum fluctuations in curved space and the theory of
  gravitation.
\newblock {\em General Relativity and Gravitation}, {\em 32}, 365--367.

\bibitem[\protect\citeauthoryear{Salimkhani}{Salimkhani}{2018}]{Salimkhani2018}
Salimkhani, K. (2018).
\newblock Quantum gravity: A dogma of unification?
\newblock In A.~Christian, D.~Hommen, N.~Retzlaff, \& G.~Schurz (Eds.), {\em
  Philosophy of Science. European Studies in Philosophy of Science, Vol 9.}
  (pp.\ 23--41). Cham: Springer.

\bibitem[\protect\citeauthoryear{Salimkhani}{Salimkhani}{2021}]{Salimkhani2021}
Salimkhani, K. (2021).
\newblock Explaining unification in physics internally.
\newblock {\em Synthese}, {\em 198\/}(6), 5861--5882.

\bibitem[\protect\citeauthoryear{Schindler}{Schindler}{2018}]{Schindler2018}
Schindler, S. (2018).
\newblock {\em Theoretical Virtues in Science: Uncovering Reality through
  Theory}.
\newblock Cambridge: Cambridge University Press.

\bibitem[\protect\citeauthoryear{Smeenk}{Smeenk}{2013}]{Smeenk2013}
Smeenk, C. (2013).
\newblock Philosophy of cosmology.
\newblock In R.~Batterman (Ed.), {\em Oxford Handbook of Philosophy of Physics}
   (pp.\ 607--652). Oxford: Oxford University Press.

\bibitem[\protect\citeauthoryear{Smeenk \& W\"{u}thrich}{Smeenk \&
  W\"{u}thrich}{2021}]{SmeenkDeterminism}
Smeenk, C. \& W\"{u}thrich, C. (2021).
\newblock Determinism and general relativity.
\newblock {\em Philosophy of Science}, {\em 88\/}(4), 638--664.

\bibitem[\protect\citeauthoryear{Smolin}{Smolin}{2001}]{Smolin2001}
Smolin, L. (2001).
\newblock The strong and weak holographic principles.
\newblock {\em Nuclear Physics B}, {\em 601\/}(1), 209 -- 247.

\bibitem[\protect\citeauthoryear{Smolin}{Smolin}{2006}]{Smolin2006}
Smolin, L. (2006).
\newblock The case for background independence.
\newblock In D.~Rickles, S.~French, \& J.~Saatsi (Eds.), {\em The Structural
  Foundations of Quantum Gravity}  (pp.\ 196--239). Oxford: Oxford University
  Press.

\bibitem[\protect\citeauthoryear{Smolin}{Smolin}{2013}]{Smolin2013}
Smolin, L. (2013).
\newblock {\em Time Reborn: From the Crisis of Physics to the Future of the
  Universe}.
\newblock Allen Lane.

\bibitem[\protect\citeauthoryear{Smolin}{Smolin}{2017}]{Smolin2017}
Smolin, L. (2017).
\newblock Four principles for quantum gravity.
\newblock In J.~Bagla \& S.~Engineer (Eds.), {\em Gravity and the Quantum},
  volume 187 of {\em Fundamental Theories of Physics}  (pp.\ 427--450).
  Springer.

\bibitem[\protect\citeauthoryear{Smolin}{Smolin}{2020}]{Smolin2020}
Smolin, L. (2020).
\newblock Temporal relationalism.
\newblock In {\em Beyond Spacetime: The Foundations of Quantum Gravity}  (pp.\
  143--175). Cambridge University Press.

\bibitem[\protect\citeauthoryear{Stachel}{Stachel}{2006}]{Stachel2006}
Stachel, J. (2006).
\newblock Structure, individuality and quantum gravity.
\newblock In D.~Rickles \& S.~French (Eds.), {\em The Structural Foundations of
  Quantum Gravity}  (pp.\ 53--82). Oxford: Oxford University Press.

\bibitem[\protect\citeauthoryear{Steinhauer}{Steinhauer}{2016}]{Steinhauer2016}
Steinhauer, J. (2016).
\newblock Observation of quantum {H}awking radiation and its entanglement in an
  analogue black hole.
\newblock {\em Nature Physics}, {\em 12\/}(10), 959--965.

\bibitem[\protect\citeauthoryear{Strominger \& Vafa}{Strominger \&
  Vafa}{1996}]{SV1996}
Strominger, A. \& Vafa, C. (1996).
\newblock Microscopic origin of the bekenstein-hawking entropy.
\newblock {\em Physics Letters B}, {\em 379\/}(1-4), 99--104.

\bibitem[\protect\citeauthoryear{Surya}{Surya}{2019}]{Surya2019}
Surya, S. (2019).
\newblock The causal set approach to quantum gravity.
\newblock {\em Living Reviews in Relativity}, https://arxiv.org/abs/1903.11544.

\bibitem[\protect\citeauthoryear{Susskind}{Susskind}{1995}]{Susskind1995}
Susskind, L. (1995).
\newblock The world as a hologram.
\newblock {\em Journal of Mathematical Physics}, {\em 36\/}(11), 6377--6396.

\bibitem[\protect\citeauthoryear{Susskind}{Susskind}{2008}]{Susskind2008}
Susskind, L. (2008).
\newblock {\em The Black Hole War: My Battle with Stephen Hawking to Make the
  World Safe for Quantum Mechanics}.
\newblock Little, Brown and Company.

\bibitem[\protect\citeauthoryear{'t~Hooft}{'t~Hooft}{1993}]{tHooft1993}
't~Hooft, G. (1993).
\newblock Dimensional reduction in quantum gravity.
\newblock {\em https://arxiv.org/abs/gr-qc/9310026}.

\bibitem[\protect\citeauthoryear{'t~Hooft \& Veltman}{'t~Hooft \&
  Veltman}{1974}]{Oneloop}
't~Hooft, G. \& Veltman, M. (1974).
\newblock One-loop divergencies in the theory of gravitation.
\newblock {\em Annales de l'IHP Physique theorique}, {\em 20\/}(1), 69--94.

\bibitem[\protect\citeauthoryear{Th\'{e}bault}{Th\'{e}bault}{2019}]{Thebault2019}
Th\'{e}bault, K. (2019).
\newblock What can we learn from analogue experiments?
\newblock In R.~Dardashti, R.~Dawid, \& K.~Th\'{e}bault (Eds.), {\em Why Trust
  a Theory?: Epistemology of Fundamental Physics}  (pp.\ 184–201). Cambridge
  University Press.

\bibitem[\protect\citeauthoryear{Th\'{e}bault}{Th\'{e}bault}{2021}]{Thebault2021}
Th\'{e}bault, K. (2021).
\newblock The problem of time.
\newblock In E.~Knox \& A.~Wilson (Eds.), {\em The Routledge Companion to
  Philosophy of Physics}. Routledge.

\bibitem[\protect\citeauthoryear{Th\'{e}bault}{Th\'{e}bault}{2023}]{Thebault2023}
Th\'{e}bault, K. (2023).
\newblock Big bang singularity resolution in quantum cosmology.
\newblock {\em Classical and Quantum Gravity}, {\em 40\/}(5), 055007.

\bibitem[\protect\citeauthoryear{Tilloy}{Tilloy}{2018}]{Tilloy2018}
Tilloy, A. (2018).
\newblock Binding quantum matter and space-time, without romanticism.
\newblock {\em Foundations of Physics}, {\em 48}, 1572--9516.

\bibitem[\protect\citeauthoryear{Unruh}{Unruh}{1981}]{Unruh1981}
Unruh, W. (1981).
\newblock Experimental black-hole evaporation?
\newblock {\em Physical Review Letters}, {\em 46}, 1351--1358.

\bibitem[\protect\citeauthoryear{Unruh \& Wald}{Unruh \&
  Wald}{2017}]{Unruh2017}
Unruh, W.~G. \& Wald, R.~M. (2017).
\newblock Information loss.
\newblock {\em Reports on Progress in Physics}, {\em 80\/}(9), 092002.

\bibitem[\protect\citeauthoryear{{van Dongen}, {De Haro}, Visser \&
  Butterfield}{{van Dongen} et~al.}{2020}]{Deharo2020b}
{van Dongen}, J., {De Haro}, S., Visser, M., \& Butterfield, J. (2020).
\newblock Emergence and correspondence for string theory black holes.
\newblock {\em Studies in History and Philosophy of Modern Physics}, {\em 69},
  112--127.

\bibitem[\protect\citeauthoryear{Verlinde}{Verlinde}{2011}]{Verlinde2011}
Verlinde, E. (2011).
\newblock On the origin of gravity and the laws of {N}ewton.
\newblock {\em Journal of High Energy Physics}, 1--27.

\bibitem[\protect\citeauthoryear{Verlinde}{Verlinde}{2017}]{Verlinde2017}
Verlinde, E. (2017).
\newblock Emergent gravity and the dark universe.
\newblock {\em SciPost Phys.}, {\em 2}, 016.

\bibitem[\protect\citeauthoryear{Visser, Barcel\'{o} \& Liberati}{Visser
  et~al.}{2002}]{Visser2002}
Visser, M., Barcel\'{o}, C., \& Liberati, S. (2002).
\newblock Analogue models of and for gravity.
\newblock {\em General Relativity and Gravitation}, {\em 34\/}(10), 1719--1734.
\newblock 3rd Australasian Conference on General Relativity and Graviation Jul
  11-13, 2001 Perth, australia.

\bibitem[\protect\citeauthoryear{Volovik}{Volovik}{2003}]{Volovik2003}
Volovik, G. (2003).
\newblock {\em The Universe in a Helium Droplet}.
\newblock Oxford: Oxford University Press.

\bibitem[\protect\citeauthoryear{Wald}{Wald}{1994}]{Wald1994}
Wald, R.~M. (1994).
\newblock {\em Quantum Field Theory in Curved Spacetime and Black Hole
  Thermodynamics}.
\newblock University of Chicago Press.

\bibitem[\protect\citeauthoryear{Wald}{Wald}{2001}]{Wald2001}
Wald, R.~M. (2001).
\newblock The thermodynamics of black holes.
\newblock {\em Living Reviews in Relativity}, {\em 4\/}(1), 6.

\bibitem[\protect\citeauthoryear{Wall}{Wall}{2009}]{Wall2009}
Wall, A.~C. (2009).
\newblock Ten proofs of the generalized second law.
\newblock {\em Journal of High Energy Physics}, {\em 2009\/}(06), 021.

\bibitem[\protect\citeauthoryear{Wallace}{Wallace}{2006}]{Wallace2006}
Wallace, D. (2006).
\newblock In defence of naivete: The conceptual status of lagrangian quantum
  field theory.
\newblock {\em Synthese}, {\em 151\/}(1), 33.

\bibitem[\protect\citeauthoryear{Wallace}{Wallace}{2011}]{Wallace2011}
Wallace, D. (2011).
\newblock Taking particle physics seriously: A critique of the algebraic
  approach to quantum field theory.
\newblock {\em Studies In History and Philosophy of Modern Physics}, {\em
  42\/}(2), 116--125.

\bibitem[\protect\citeauthoryear{Wallace}{Wallace}{2018a}]{Wallace2018a}
Wallace, D. (2018a).
\newblock The case for black hole thermodynamics part i: Phenomenological
  thermodynamics.
\newblock {\em Studies in History and Philosophy of Modern Physics}, {\em 64},
  52--67.

\bibitem[\protect\citeauthoryear{Wallace}{Wallace}{2018b}]{Wallace2018b}
Wallace, D. (2018b).
\newblock The case for black hole thermodynamics part ii: Statistical
  mechanics.
\newblock {\em Studies in History and Philosophy of Modern Physics}, {\em
  66\/}(C), 103--117.

\bibitem[\protect\citeauthoryear{Wallace}{Wallace}{2020}]{Wallace2020}
Wallace, D. (2020).
\newblock Why black hole information loss is paradoxical.
\newblock In {\em Beyond Spacetime: The Foundations of Quantum Gravity}  (pp.\
  209--236). Cambridge University Press.

\bibitem[\protect\citeauthoryear{Wallace}{Wallace}{2022}]{Wallace2022}
Wallace, D. (2022).
\newblock Quantum gravity at low energies.
\newblock {\em Studies in History and Philosophy of Science}, {\em 94}, 31--46.

\bibitem[\protect\citeauthoryear{Weinberg}{Weinberg}{1979}]{Weinberg1979}
Weinberg, S. (1979).
\newblock Ultraviolet divergencies in quantum theories of gravitation.
\newblock In S.~Hawking \& W.~Israel (Eds.), {\em General relativity, an
  Einstein Centenary survey}  (pp.\ 790--831). Cambridge: Cambridge University
  Press.

\bibitem[\protect\citeauthoryear{Weinberg}{Weinberg}{1995}]{Weinberg1995}
Weinberg, S. (1995).
\newblock {\em The quantum theory of fields, Vol. I (Foundations)}.
\newblock Cambridge University Press.

\bibitem[\protect\citeauthoryear{Weinberg}{Weinberg}{2009}]{Weinberg2009}
Weinberg, S. (2009).
\newblock Effective field theory, past and future.
\newblock {\em arXiv:hep-th/0908.1964v3}.

\bibitem[\protect\citeauthoryear{Weinfurtner, Tedford, Penrice, Unruh \&
  Lawrence}{Weinfurtner et~al.}{2013}]{Weinfurtner2013}
Weinfurtner, S., Tedford, E.~W., Penrice, M.~C., Unruh, W.~G., \& Lawrence,
  G.~A. (2013).
\newblock Classical aspects of hawking radiation verified in analogue gravity
  experiment.
\newblock In {\em Analogue Gravity Phenomenology}  (pp.\ 167--180). Springer.

\bibitem[\protect\citeauthoryear{Weinstein \& Rickles}{Weinstein \&
  Rickles}{2021}]{RicklesSEP}
Weinstein, S. \& Rickles, D. (2021).
\newblock Quantum gravity.
\newblock {\em The Stanford Encyclopedia of Philosophy},
  https://plato.stanford.edu/archives/fall2021/entries/quantum--gravity/.

\bibitem[\protect\citeauthoryear{Wheeler}{Wheeler}{2000}]{Wheeler2000}
Wheeler, J. (2000).
\newblock {\em Geons, Black Holes and Quantum Foam}.
\newblock W.W. Norton \& Company.

\bibitem[\protect\citeauthoryear{Williams}{Williams}{2021}]{Williams2021}
Williams, P. (2021).
\newblock Renormalization group methods.
\newblock In E.~Knox \& A.~Wilson (Eds.), {\em The Routledge Companion to
  Philosophy of Physics}. Routledge.

\bibitem[\protect\citeauthoryear{Williams}{Williams}{2006}]{Williams2006}
Williams, R. (2006).
\newblock Discrete quantum gravity.
\newblock {\em Journal of Physics: Conference Series}, {\em 33\/}(1), 38.

\bibitem[\protect\citeauthoryear{Witten}{Witten}{1998}]{Witten1998}
Witten, E. (1998).
\newblock {Anti-de Sitter space, thermal phase transition, and confinement in
  gauge theories}.
\newblock {\em Adv. Theor. Math. Phys.}, {\em 2}, 505--532.

\bibitem[\protect\citeauthoryear{W\"{u}thrich}{W\"{u}thrich}{2005}]{Wuthrich2005}
W\"{u}thrich, C. (2005).
\newblock To quantize or not to quantize: Fact and folklore in quantum gravity.
\newblock {\em Philosophy of Science}, {\em 72}, 777--788.

\bibitem[\protect\citeauthoryear{Zee}{Zee}{2010}]{Zee2010}
Zee, A. (2010).
\newblock {\em Quantum Field Theory in a Nutshell\/} (Second ed.).
\newblock Princeton: Princeton University Press.

\bibitem[\protect\citeauthoryear{Zeh}{Zeh}{2011}]{Zeh2011}
Zeh, H. (2011).
\newblock Feynman’s interpretation of quantum theory.
\newblock {\em The European Physical Journal H}, {\em 36}, 63--74.

\bibitem[\protect\citeauthoryear{Zwiebach}{Zwiebach}{2009}]{Zwiebach2009}
Zwiebach, B. (2009).
\newblock {\em A First Course in String Theory\/} (Second ed.).
\newblock Cambridge: Cambridge University Press.

\end{thebibliography}

\end{document}